\documentclass[preprintnumbers, floatfix, letterpaper, twocolumn,aps,prd,epsfig,nofootinbib,natbib,longbibliography]{revtex4-2}

\usepackage{graphicx}
\usepackage{epstopdf}
\usepackage{natbib}
\usepackage{latexsym}
\usepackage{amssymb}
\usepackage{amsmath}
\usepackage{color}
\usepackage{mathrsfs}
\usepackage{xparse}
\usepackage{bbding}
\usepackage{enumitem}
\usepackage{pifont}
\usepackage[center]{subfigure}
\usepackage[
            pdfstartview=FitH,
            bookmarksnumbered=true,
            bookmarksopen=true,
            colorlinks,
            linkcolor=blue,
            anchorcolor=green,
            citecolor=blue
            ]{hyperref}
            
            \usepackage{setspace}
\usepackage{amssymb}
\usepackage{graphicx,wrapfig}
\usepackage{enumerate} 
\usepackage{color}
\usepackage{mathrsfs}
\usepackage{subfigure}
\usepackage{comment}
\definecolor{mg}{rgb}{0.0, 0.5, 0.0}

\begin{document}

  \renewcommand\arraystretch{2}
 \newcommand{\bq}{\begin{equation}}
 \newcommand{\eq}{\end{equation}}
 \newcommand{\bqn}{\begin{eqnarray}}
 \newcommand{\eqn}{\end{eqnarray}}
 \newcommand{\nb}{\nonumber}
 \newcommand{\lb}{\label}
 
\newcommand{\La}{\Lambda}
\newcommand{\va}{\scriptscriptstyle}
\newcommand{\be}{\nopagebreak[3]\begin{equation}}
\newcommand{\ee}{\end{equation}}

\newcommand{\ba}{\nopagebreak[3]\begin{eqnarray}}
\newcommand{\ea}{\end{eqnarray}}

\newcommand{\la}{\label}
\newcommand{\n}{\nonumber}
\newcommand{\su}{\mathfrak{su}}
\newcommand{\SU}{\mathrm{SU}}
\newcommand{\U}{\mathrm{U}}

\def\be{\nopagebreak[3]\begin{equation}}
\def\ee{\end{equation}}
\def\ba{\nopagebreak[3]\begin{eqnarray}}
\def\ea{\end{eqnarray}}
\newcommand{\f}{\frac}
\def\rmd{\rm d}
\def\lp{\ell_{\rm Pl}}
\def\d{{\rm d}}
\def\fe{\mathring{e}^{\,i}_a}
\def\fw{\mathring{\omega}^{\,a}_i}
\def\fq{\mathring{q}_{ab}}
\def\t{\tilde}

\def\db{\delta_b}
\def\dc{\delta_c}
\def\T{\mathcal{T}}
\def\GammaE{\Gamma_{\rm ext}}
\def\GammaEb{\bar\Gamma_{\rm ext}}
\def\GammaEh{\hat\Gamma_{\rm ext}}
\def\Hee{H_{\rm eff}^{\rm ext}}
\def\H{\mathcal{H}}

\newcommand{\R}{\mathbb{R}}

 \newcommand{\cb}{\color{blue}}
    \newcommand{\cc}{\color{cyan}}
        \newcommand{\cm}{\color{magenta}}
\newcommand{\rc}{\rho^{\scriptscriptstyle{\mathrm{I}}}_c}
\newcommand{\rd}{\rho^{\scriptscriptstyle{\mathrm{II}}}_c} 
\NewDocumentCommand{\evalat}{sO{\big}mm}{%
  \IfBooleanTF{#1}
   {\mleft. #3 \mright|_{#4}}
   {#3#2|_{#4}}%
}
\newcommand{\PRL}{Phys. Rev. Lett.}
\newcommand{\PL}{Phys. Lett.}
\newcommand{\PR}{Phys. Rev.}
\newcommand{\CQG}{Class. Quantum Grav.}




\title {A new quantization scheme of  black holes in effective loop quantum gravity}

\author{Wen-Cong Gan${}^{a, b}$}
\email{ganwencong@jxnu.edu.cn}

\author{Anzhong Wang${}^{b}$ \thanks{Corresponding author}}
\email{Anzhong$\_$Wang@baylor.edu; Corresponding author}

 \affiliation{${}^{a}$ College of Physics and Communication Electronics, Jiangxi Normal University, Nanchang, 330022, China\\
 ${}^{b}$ GCAP-CASPER, Physics Department, Baylor University, Waco, Texas 76798-7316, USA}

\date{\today}

\begin{abstract}

 Loop quantum cosmology has achieved great successes, in which the polymerization plays a crucial role. In particular, the phase-space-variable dependent polymerization turns out to be the unique one that leads to consistent quantization of the homogeneous and isotropic universe. However, when applying the same scheme to the quantization of black holes, it meets resistances, when the Kantowski-Sachs (KS) gauge is adopted. In this paper, we continue to study the quantum effects of the polymerization near the location that a classical black hole horizon used to be, from the point of view of effective loop quantum gravity in the KS gauge. In particular,  we find a phase-space-variable dependent polymerization scheme that leads to negligible quantum effects near the location of the classical black hole horizon, but significantly alters the spacetime structure near the origin, so that the classical singularity is finally replaced by a finite and regular transition surface. The final geodesically-complete spacetime consists of the regular transition surface that connects a black hole in one side and an anti-trapped region in the other side.  In the anti-trapped region, no white hole horizons are found and the spacetime is extended to infinity, at which the geometric radius of the two-spheres becomes infinitely large.

\end{abstract}

\maketitle

\section{Introduction}
\label{sec:Intro}
 \renewcommand{\theequation}{1.\arabic{equation}}\setcounter{equation}{0}

 General relativity (GR) breaks down at the singularity due to the divergence of curvature, and corrections from quantum effects should dominate when curvature reaches the Plank scale \cite{Hawking:1973uf}. There are mainly two kinds of singularities in GR, i.e. the cosmological singularity and black (white) hole singularity. 
 Loop quantum gravity (LQG) \cite{Thiemann:2007pyv} solves the singularity problem in cosmology  \cite{Bojowald:2001xe,Bojowald:2002gz,Ashtekar:2003hd,Ashtekar:2006wn}. In loop quantum cosmology (LQC), the resolution of the big bang singularity is crucially related to the fact that in LQG the area operator has a minimal and non-zero area gap, after the  quantization of the spacial geometry. The quantum corrections of Einstein's equations due to this minimal area gap will produce an effective energy momentum tensor which violates the energy conditions and thus prevents the singularity  to be formed, once the quantum gravity effects are taken into account. Especially, the singularity is replaced by a quantum bounce, which connects the trapped region in the past of the bounce  to an anti-trapped region in the future of it.

In loop quantum black holes (LQBHs), the problem has been treated  similarly \cite{Olmedo:2016ddn,Ashtekar:2018cay,Ashtekar:2020ifw,Gambini:2022hxr,Ashtekar:2023cod}. In particular, the internal of the Schwarzschild black hole can be written in the form of the Kantowski-Sachs (KS) universe, and thus the technics of LQC can be applied to LQBHs. 
Quantum effective black hole models are constructed by the same procedure as in LQC: Hamiltonian constraint is re-written in terms of gravitational connections and triads. Quantum parameters are introduced in Hamiltonian constraint via replacing the gravitational connection by holonomies of Ashtekar's connection around some loops enclosed the minimal area. For  models with symmetry such as the KS model, the Hamiltonian can be greatly simplified and the symmetry reduced dynamic variables consist only four, that is, the connection components $(b, c)$ and the triad variables $(p_b,p_c)$, which satisfy the canonical relations
 \bqn
 \lb{eq1.1}
 \{c,p_c \}=2G \gamma,\quad  \{b,p_b \}=G \gamma,
 \eqn
 where $G$ is the Newton gravitational constant,  and $\gamma$  the Barbero-Immirzi parameter. Holonomies are simplified and consist of terms like 
$\exp(i\delta_b b)$ and $\exp(i\delta_c c)$,  where the two parameters $\delta_b$ and $\delta_c$ are related to the edge lengths of holonomies in different directions. The leading order quantum effects can be well captured by the effective Hamiltonian, $H^{\text{eff}}(b,p_b, c, p_c; \delta_b, \delta_c)$, obtained from the replacement of $b$ and $c$ by 
\bq
\lb{eq1.2}
b \rightarrow \frac{\sin(\delta_b b)}{\delta_b}, \quad 
c \rightarrow \frac{\sin(\delta_c c)}{\delta_c},
\eq
in the classical Hamiltonian, $H^{\text{GR}}(b,p_b, c, p_c)$. This is called the polymerization of the phase space variables with the two quantum polymerization parameters $(\delta_b,\delta_c)$ in the effective theory of LQG \cite{Olmedo:2016ddn,Ashtekar:2018cay,Ashtekar:2020ifw,Gambini:2022hxr,Ashtekar:2023cod}. Polymerized Hamiltonian leads to effective dynamic equations, and solutions to these equations show that the singularity is indeed avoided. The classical trajectories are obtained when $\delta_{b}, \; \delta_{c} \rightarrow 0$.
Different choices of $\delta_{b}$ and $\delta_{c}$ originate from different ways of enclosing the area gap and  lead to different quantization schemes with different effective dynamics of phase space variables \cite{Olmedo:2016ddn,Ashtekar:2018cay,Ashtekar:2020ifw,Gambini:2022hxr,Ashtekar:2023cod}.

In the literature, several schemes have been proposed. In particular, in the $\mu_o$-scheme \cite{Modesto:2004wm,Ashtekar:2005qt,Modesto:2005zm}, holonomies are eigenstates of the area operator with its eigenvalue being equal to the  minimal area gap. As a consequence, $(\delta_b,\delta_c)$ are constants in the $\mu_o$-scheme. However, in some models of  this scheme, the physical results depend on the fiducial cell \cite{Ashtekar:2018cay}. Lately, a comprehensive survey of the whole parameter phase space reveals that a large class of such LQBH solutions exists \cite{Ongole:2023pbs}, which is free of this  problem and meanwhile possesses all the desirable properties as LQBHs found so far. The latter includes the Ashtekar-Olmedo-Singh (AOS) model \cite{Ashtekar:2018cay} as a particular case.

The above mentioned drawbacks can be also overcome in the ``improved dynamics" scheme, i.e., the $\bar \mu$-scheme, in which the two quantum parameters are determined by the fact that the closed loops of holonomies have physical area equal to the minimal area gap. The physical area depends on phase space variables and then the quantum parameters are general functions of the phase space variables 
$(b, p_b; c, p_c)$, too.
This is the unique scheme in LQC that overcomes the dependence of fiducial cell and have consistent semiclassical behavior \cite{Ashtekar:2011ni,Joe:2014tca,Li:2023dwy}.
However, when the $\bar \mu$-scheme is applied to LQBHs \cite{Boehmer:2007ket}, it was found that the quantum effects can be very large near the location that the classical black hole horizon used to be present even  for very massive black holes.
 At the horizons of such classical black holes the spacetime curvature are very low,   and physically it is expected that the quantum effects should be very small, which rises doubts on the viability of the scheme when applied to LQBHs \cite{Ashtekar:2018cay,Ongole:2023pbs}.
 
 In addition, detailed analyses revealed that the resultant KS spacetime is geodesically complete \cite{Saini:2016vgo}, and the quantum effects actually are so strong that black/white hole horizons do not exist any longer, instead, they are replaced by an infinite number of transition surfaces, which always separate trapped regions from anti-trapped ones \cite{Zhang:2023noj,Gan:2022oiy}.  

 It must be noted that the above results are partially due to the  ``improved dynamics"  $\bar \mu$-scheme, and partially to the KS gauge, in which the metric takes the form of Eq.(\ref{metric}). In this gauge, the physical distance along the $x$-direction becomes zero at the singularity as well as at the black hole horizon. Then, the replacements (\ref{eq1.2}) lead to significant quantum gravity effects not only at the classical singularity (so that it is finally replaced by a regular transition surface) but also at the classical black horizon (so that finally such a horizon does not exist any more). In fact, with other gauge choices, black/white hole horizons indeed exist and quantum effects near these horizons are negligible \cite{Gambini:2022hxr,han2022improved}. 
 
In review of the above, to avoid large quantum effects near black hole horizons, there are at least two possibilities: (a) choose different gauges, or (b) choose different schemes. Such possibilities exist, because the operations of symmetry reduction and loop quantization do not commute. In particular, the properties of a classical system does not depend on the choice of the physical variables $(b, p_b; c, p_c)$,  but the polymerization of Eq.(\ref{eq1.2}) does. 

In this paper, we shall work with the second possibility, that is, we still choose the KS gauge but a different polymerization. In particular, since the physical distance along the $x$-direction depends on the gauge, we shall use its geometric distance, when calculating areas connected to this direction. For such a choice,  we find that the quantum effects  indeed become small near the locations that the classical horizons used to appear, and physical results are independent of the choice of the fudicial cell. On the other hand, curvatures remain finite and the classical Schwarzschild black hole singularity is replaced by a regular transition surface. So, the final geodesically-complete spacetime consists of the regular transition surface that connects a black hole in one side and an anti-trapped region in the other side.  In the anti-trapped region, no white hole horizons are found and the spacetime is extended to infinity, at which the geometric radius of the two-spheres becomes infinitely large.

It must be noted that in this paper we have no intention to claim that this is the only polymerization within the KS gauge that leads to the desirable properties, rather than to show the effects of different choices of the two quantum parameters $\delta_b$ and $\delta_c$ within the same gauge, so that they can shed lights on the nature and origin of the questions.  

The rest of the paper is organized as follows.
In Sec. \ref{bi}, we briefly review the KS spacetime,
while in Sec. \ref{sec-new}, we propose a new quantization scheme in the framework of LQBHs and investigate its properties. Our main conclusions are included in Sec. \ref{conclusion}.

\section{Internal Spacetimes of Loop Quantum Black Holes}
\lb{bi}
 \renewcommand{\theequation}{2.\arabic{equation}}\setcounter{equation}{0}

The internal spacetimes of a spherically symmetric black hole can always be written in
the KS form  
\bq
\lb{metric}
 ds^2 = - N^2 dT^2 + \frac{p_b^2}{L_o^2 \left|p_c\right|} dx^2 + \left|p_c\right| d\Omega^2, 
 \eq
 where $N, \; p_b, \; p_c$ are all functions of $T$ only, and $d\Omega^2 \equiv d\theta^2 + \sin^2\theta d\phi^2$, with $-\infty < x < \infty$, $0 \le \theta \le \pi$ and $0 \le \phi \le 2\pi$. Since the metric is independent of $x$, to keep the corresponding Hamiltonian finite, one usually first introduces a fiducial cell along $x$-direction with its length $L_o$, that is, $x \in \left[0, L_o\right]$, and finally takes the limit    $L_o \rightarrow \infty$ at the end of calculations.  Thus, the physics involved should not depend on the choice of $L_o$, which will be one of the main criteria for a model to be physically acceptable. 
 The function $N$ is often called the lapse function, and $p_b$ and $p_c$ are the dynamical variables, which satisfy the Poisson brackets (\ref{eq1.1}).

 It should be noted that the KS metric (\ref{metric}) is invariant under the gauge transformations
 \bq
 \lb{GTs}
 T = f(\hat T), \quad x = \alpha \hat x + x_o,
 \eq
via the redefinition of the lapse function and the length of the fiducial cell,
 \bq
 \lb{RDs}
 \hat N = Nf_{,\hat{T}}, \quad \hat{L}_o = \frac{L_o}{\alpha},
 \eq
 where $f(\hat{T})$ is an arbitrary function of $\hat{T}$, and $\alpha$ and $x_o$ are constants. 
 Using the above gauge freedom, we can always choose the lapse function
as  
\bqn 
  \lb{eq1}
  N^{\text{GR}}= \frac{\gamma \; {\text{sgn}}(p_c)\left|p_c\right|^{1/2}}{b},
  \eqn
  without loss of the generality.   Then, the corresponding classical Hamiltonian is given by 
\bqn
  \lb{eq2}
 H^{\text{GR}}[N^{\text{GR}}] &\equiv& N^{\text{GR}} \mathcal{H}^{\text{GR}}\nb\\
&=& -\frac{1}{2G \gamma}\left(2c\;p_c+\left(b+\frac{\gamma^2}{b}\right)p_b\right).~~~
\eqn
Then, the dynamical equations
\bq
\lb{eq2.8}
\frac{A^{\text{GR}}}{dT} = \left[A^{\text{GR}}, H^{\text{GR}}[N^{\text{GR}}]\right],
\eq
for any given  physical variable $A^{\text{GR}}$, have the classical Schwarzschild black hole internal solution, given by 
\bqn
 \lb{eq6d}
b^{\text{GR}}(T) &=&  \gamma \sqrt{2me^{-T} -1},  \nb\\
p^{\text{GR}}_b(T)&=&  e^{T} \sqrt{2me^{-T}-1},\nb\\
  c^{\text{GR}}(T)&=& - \gamma m e^{-2T},\nb\\
  p^{\text{GR}}_c(T)&=& e^{2T}, 
 \eqn
 where the parameter $m$ is related to the black hole mass $M$ via the relation $M = m/G$ \footnote{Note that $m$ has the dimension of length in the units adopted in this paper, $\left[m\right] = \left[M\right]\left[G\right] = M^{-1} = L$.}. 
 Note that the origin ($r = 0$) corresponds to $T = -\infty$ in the classical theory.

In LQC, the leading order quantum corrections are incorporated by the replacements of Eq.(\ref{eq1.2}) in the above classical Hamiltonian as well as the lapse function \cite{Ashtekar:2005qt}, so we find  
\bqn
\lb{eq8}
N^{\text{eff}} &=& \frac{\gamma\delta_b \sqrt{p_c}}{\sin(\delta_b b)},\\
\lb{eq9}
H^{\text{eff}}[N^{\text{eff}}] &=& - \frac{1}{2\gamma G}\Bigg[2 \frac{\sin(\delta_c c)}{ \delta_c} p_c\nb\\
&& + \left(\frac{\sin(\delta_b b)}{ \delta_b} + \frac{\gamma^2\delta_b}{\sin(\delta_b b)}\right)p_b\Bigg].
\eqn
Note that in writing the above expressions, we had assumed $p_c > 0$ without loss of the generality, as far as the effective theory is concerned. Clearly,
Eqs.\eqref{eq8} and \eqref{eq9} will reduce,  respectively, to the classical expressions of Eqs.\eqref{eq1} and \eqref{eq2} in the limit $\delta_b \rightarrow 0$ and $\delta_c \rightarrow 0$.
 
Assuming that  $\delta_b$ and $\delta_c$ depend only on $p_b$ and $p_c$,  the equations of motion (EoMs)  are given by \cite{Gan:2022oiy}
 \begin{widetext}
\bqn\lb{eq2.25m}
\dot b&=& G \gamma  \frac{\partial H^{\text{eff}}}{\partial p_b}=-\frac{1}{2}\Bigg\{2 \left(\frac{c\cos(\delta_c c)}{\delta_c}-\frac{\sin(\delta_c c)}{\delta_c^2}\right) \frac{\partial \delta_c}{\partial p_b} p_c+\left[\frac{\gamma^2 \delta_b}{\sin(\delta_b b)} +\frac{\sin(\delta_b b)}{\delta_b} \right]+p_b \frac{\partial}{\partial p_b} \left[\frac{\gamma^2 \delta_b}{\sin(\delta_b b)} +\frac{\sin(\delta_b b)}{\delta_b} \right]
\Bigg\},\nb\\ 
\\
\lb{eq2.25n}
\dot c&=&2G \gamma  \frac{\partial H^{\text{eff}}}{\partial p_c}=-\Bigg\{2 \left(\frac{c\cos(\delta_c c)}{\delta_c}-\frac{\sin(\delta_c c)}{\delta_c^2}\right) \frac{\partial \delta_c}{\partial p_c} p_c+2\frac{\sin(\delta_c c)}{\delta_c}+p_b \frac{\partial}{\partial p_c} \left[\frac{\gamma^2 \delta_b}{\sin(\delta_b b)} +\frac{\sin(\delta_b b)}{\delta_b} \right]
\Bigg\},\\
\lb{eq2.25o}
\dot p_c &=&-2G \gamma  \frac{\partial H^{\text{eff}}}{\partial c} = 2 p_c \cos(\delta_c c) , \\ 
\lb{eq2.25p}
\dot p_b &=&-G \gamma  \frac{\partial H^{\text{eff}}}{\partial b} = \frac{p_b}{2} \cos(\delta_b b) \left[1-\frac{\gamma^2 \delta_b^2}{\sin^2(\delta_b b)}\right].
\eqn
\end{widetext}

\section{ A Hybrid Scheme}\lb{sec-new}
 \renewcommand{\theequation}{3.\arabic{equation}}\setcounter{equation}{0}

To solve the above dynamical system of the first-order ordinary differential equations, we need first to fix the two parameters $\delta_b$ and $\delta_c$. To this goal, let us introduce the fiducial metric ${}^{o}q_{ab}$ on the slice $T =$ Constant \cite{Ashtekar:2018cay}
\bqn
\lb{eq3.1}
{}^{o}q_{ab} dx^a dx^b = dx^2 + r_o^2\left(d\theta^2
+ \sin^2\theta d\phi^2\right),
\eqn
where $r_o$ is a constant with a length dimension, 
$[r_o] = L$. Then, 
consider an infinitesimal rectangular plaquette in the $(x, \phi)$-plane  of the fiducial cell, $x \in [0, L_o]$, on the equator plane  $\theta = \pi/2$. The plaquette has two parallel links along the $z$-direction and two parallel links along $\theta = \pi/2$. Let $\delta_c$ ($\delta_b$) denotes the fractional length of the link along the $x$-direction (along the equator). In LQC, a consistent prescription for the polymerization parameters were obtained by requiring that the physical area $A_{(x,\phi)} \left(= 2\pi \delta_b\delta_c p_b\right)$ \cite{Ashtekar:2006wn} 
of the closed holonomy loop in the $(x,\phi)$-plane be equal to the minimum area gap, $\Delta \equiv 4\sqrt{3}\pi \gamma \ell_{pl}^2$, predicted by LQG.
However, in the black hole physics, the physical length along the $x$-direction becomes zero at the black hole horizon, so it seems improper to use the physical length here. In addition, as shown in Eq.(\ref{GTs}),  the $x$ coordinate has the rescaling symmetry, $x \rightarrow x' = \alpha x$, under which the physical length along the $x$-direction changes and can be assigned any value by properly choosing the parameter $\alpha$. However, the fraction length $L_o \delta_c$ with respect to ${}^{o}q_{ab}$ is metric independent. So, in this paper we shall adopt this length. Then, the area in the  $(x,\phi)$-plane
is given by 
\bqn
\lb{eq3.1a}
A_{(x,\phi)} =  \left(L_o \delta_c\right) \cdot \left(2\pi r\delta_b\right), 
\eqn
where $r \equiv \sqrt{p_c} = r_o \sqrt{\bar{p}_c}$ denotes the geometric radius of the two-spheres, and
$\left(2\pi  r\delta_b\right)$ is the fractional length along the equator. On the other hand,   the fractional
area of the two-spheres is given by
\bqn
\lb{eq3.1b}
A_{(\theta,\phi)} = \left(4\pi r^2\right) \delta_b^2. 
\eqn
Requiring both of them be equal to the  minimum area gap $\Delta$, we  find
\bq
\lb{eq3}
\delta_b = \sqrt{\frac{\Delta}{4\pi p_c}}, \quad
L_o \delta_c = \sqrt{\frac{\Delta}{\pi}}.
\eq
 To be distinguishable from the previous ones, we refer the above choice to as {\em the hybrid scheme}, in the sense that $\delta_b$ now depends on the phase space variable $p_c$, while $L_o \delta_c$ still remains a constant. Therefore, it is a hybrid of the $\mu_o$- and $\bar\mu$- schemes.  Hence, we have
  \bqn
 \lb{eq2.25k}
\frac{\partial \delta_b}{\partial p_b}&=&0, \quad \frac{\partial \delta_b}{\partial p_c}=-\frac{\delta_b}{2p_c},\\
\lb{eq2.25l}
\frac{\partial \delta_c}{\partial p_b}&=&0,  \quad  \frac{\partial \delta_c}{\partial p_c}=0.
\eqn
Hence, we find that
\bqn
\lb{eq2.25kka}
H^{\text{eff}}[N^{\text{eff}}] &=& - \left(\frac{p_c}{2\gamma G\delta_c \sin(\delta_b b)}\right) \hat H^{\text{eff}}[N^{\text{eff}}],\\
\lb{eq2.25kkb}
\hat H^{\text{eff}}[N^{\text{eff}}] &\equiv& 2\sin(\delta_b b)\sin(\delta_c c)
 \nb\\
&& +   \left(\sin^2(\delta_b b) + \gamma^2\delta_b^2\right) \frac{\delta_c p_b}{\delta_b p_c}.  
\eqn
Thus, the Hamiltonian constraint $H^{\text{eff}}[N^{\text{eff}}] \approx 0$ leads to 
\bq
\lb{eq2.25kkc}
\hat H^{\text{eff}}[N^{\text{eff}}] \approx 0.
\eq

On the other hand, inserting them into Eqs.(\ref{eq2.25m}) - (\ref{eq2.25p}), we obtain
\bqn
\lb{eq2.25s}
\dot b &=&-\frac{1}{2}\left(\frac{\sin(\delta_b b)}{ \delta_b} + \frac{\gamma^2\delta_b}{\sin(\delta_b b)}\right),\\
\lb{eq2.25t}
\dot c &=& -2\frac{\sin(\delta_c c)}{ \delta_c}+\frac{p_b}{2 p_c}\Bigg\{b\cos(\delta_b b)\left(1-\frac{\gamma^2 \delta_b^2}{\sin^2(\delta_b b)} \right)
\nb\\
&& - \left( \frac{\sin(\delta_b b)}{ \delta_b} - \frac{\gamma^2\delta_b}{\sin(\delta_b b)}\right)\Bigg\}, \\
\lb{eq2.25u}
\dot p_c &=& 2 p_c \cos(\delta_c c), \\ 
\lb{eq2.25v}
\dot p_b &=& \frac{p_b}{2} \cos(\delta_b b) \left(1-\frac{\gamma^2 \delta_b^2}{\sin^2(\delta_b b)}\right). 
\eqn

\begin{figure*}[htbp]
 \resizebox{\linewidth}{!}{\begin{tabular}{cc}
 \includegraphics[height=4.cm]{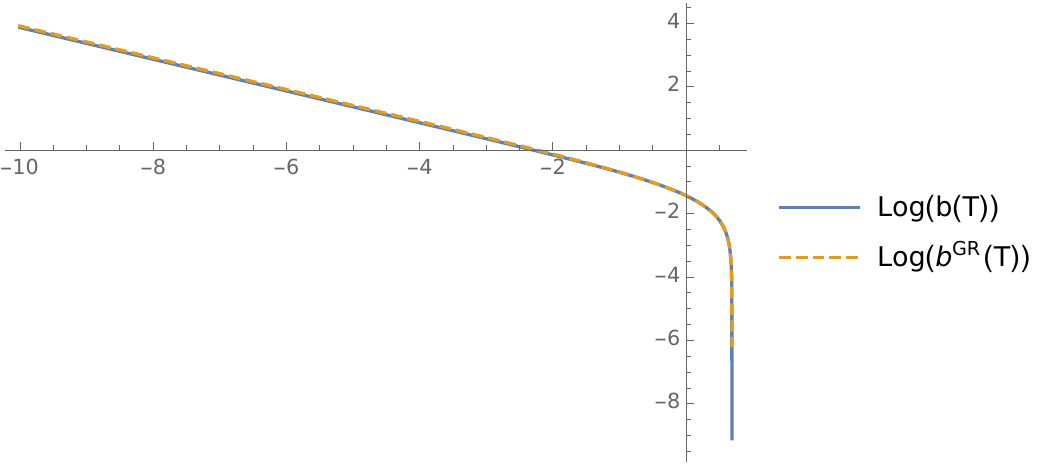}&
\includegraphics[height=4.cm]{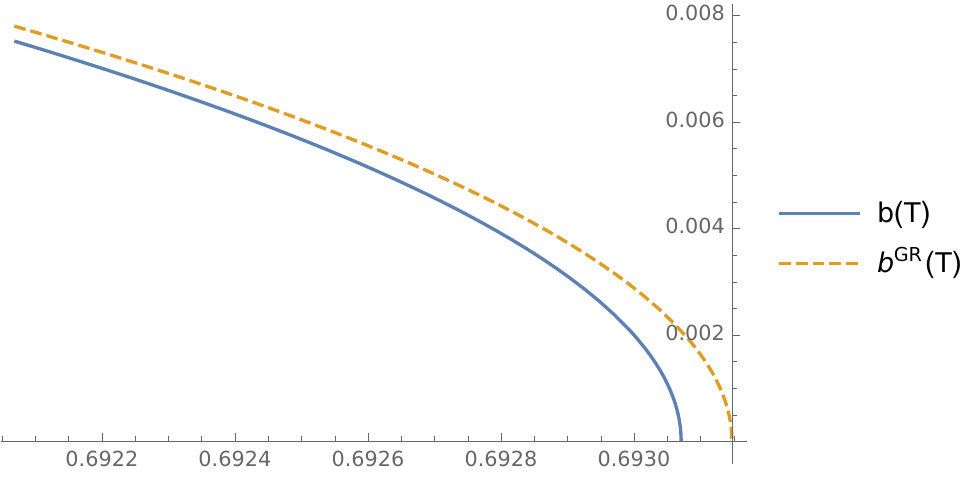}\\
 (a) & (b) \\[6pt]
\\
\includegraphics[height=4cm]{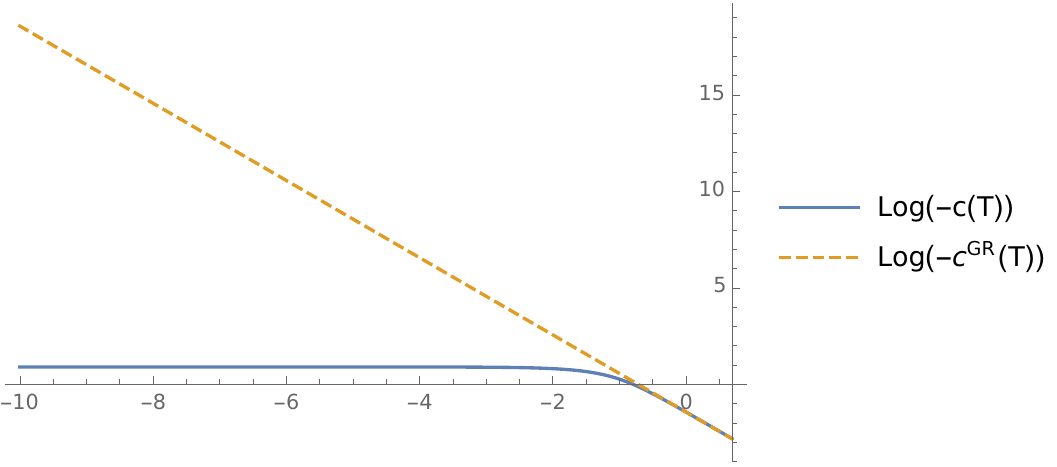}&
\includegraphics[height=4cm]{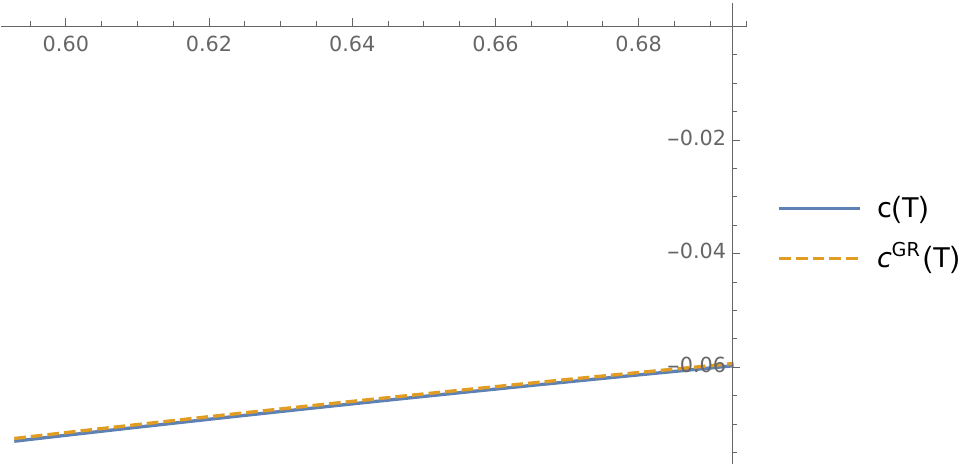}\\
(c) & (d) \\[6pt]
\\
\includegraphics[height=4cm]{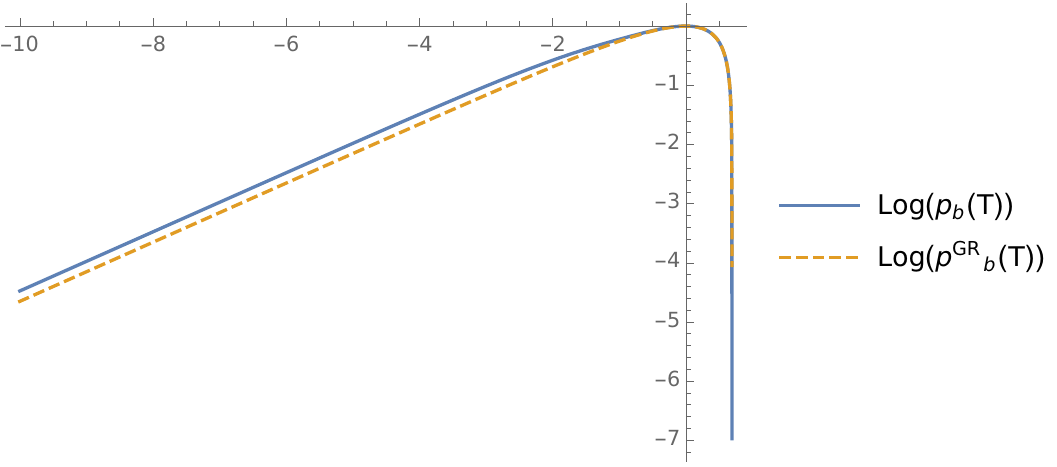}&
\includegraphics[height=4cm]{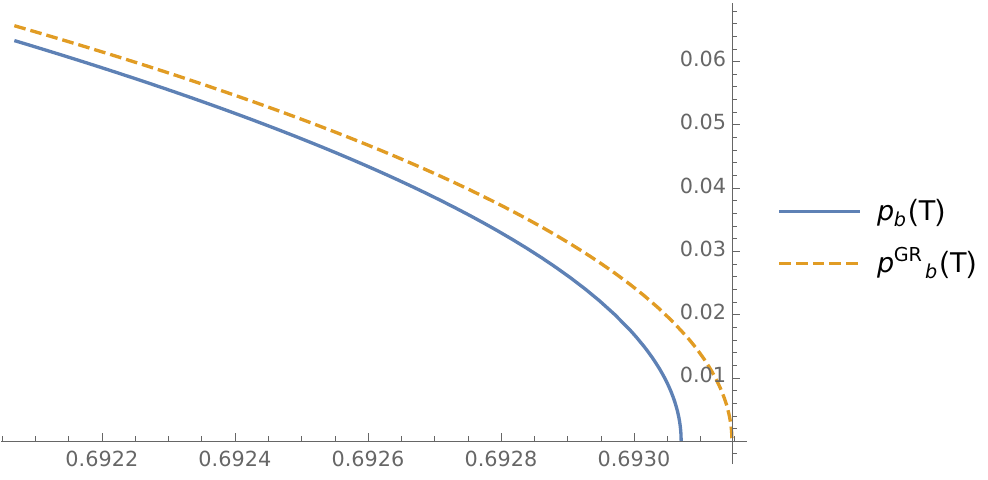}\\
(e) & (f)   \\[6pt]
\\
\includegraphics[height=4cm]{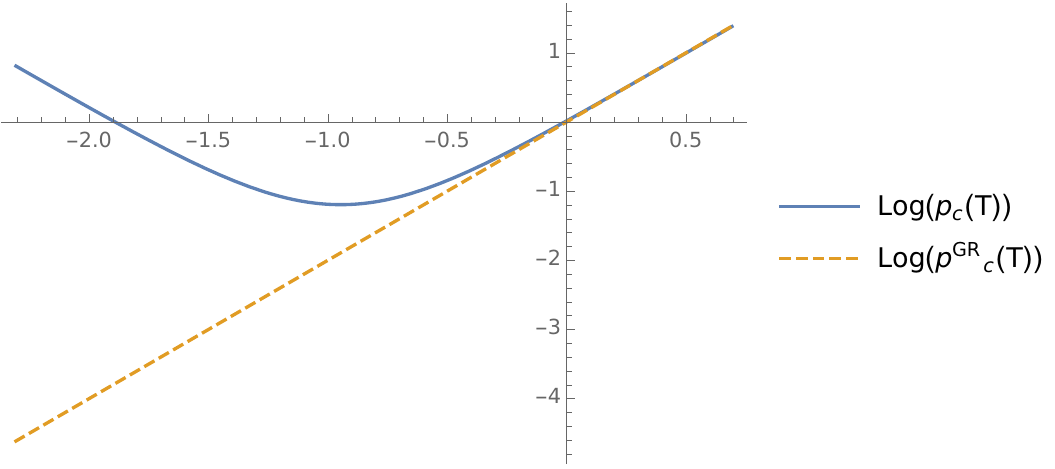}&
\includegraphics[height=4cm]{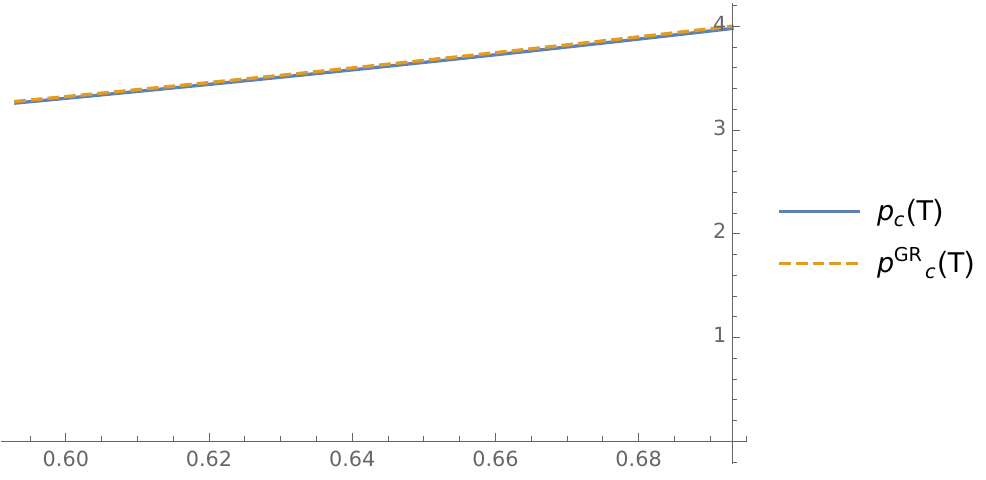}\\
	(g) & (h)  \\[6pt]
\end{tabular}}
\caption{Plots of  the four physical variables $\left(b, c, p_b, p_c\right)$ in the region $T\in (-10, 1)$, in which the transition surface and the black hole horizon are located. The mass parameter $m$ is chosen as  $m/\ell_{pl}=1$, for which we have  $T_{\cal{T}} \simeq -0.946567$ and $T_H \approx0.693$. The initial time is chosen at $T_i = 0.3$.}
\lb{fig1}
\end{figure*}

 \begin{figure*}[htbp]
 \resizebox{\linewidth}{!}{\begin{tabular}{cc}
 \includegraphics[height=4.cm]{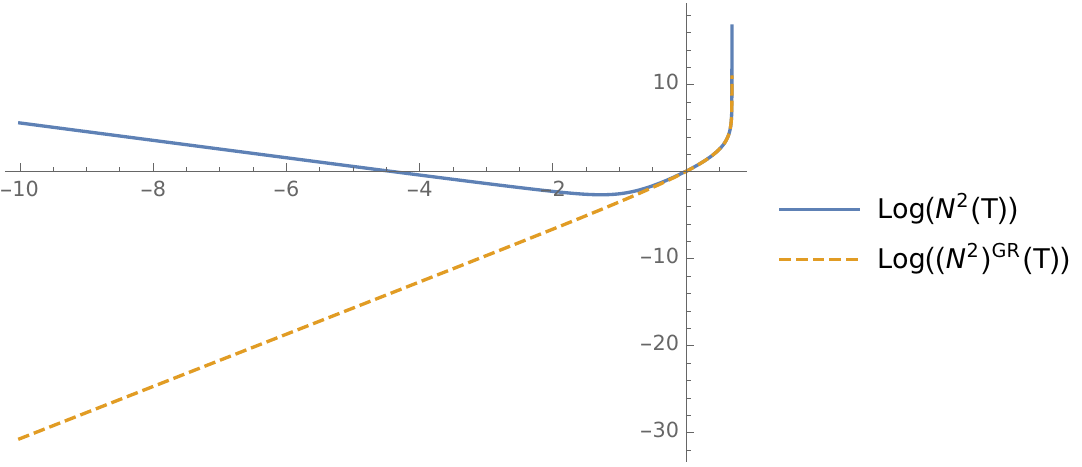}&
\includegraphics[height=4.cm]{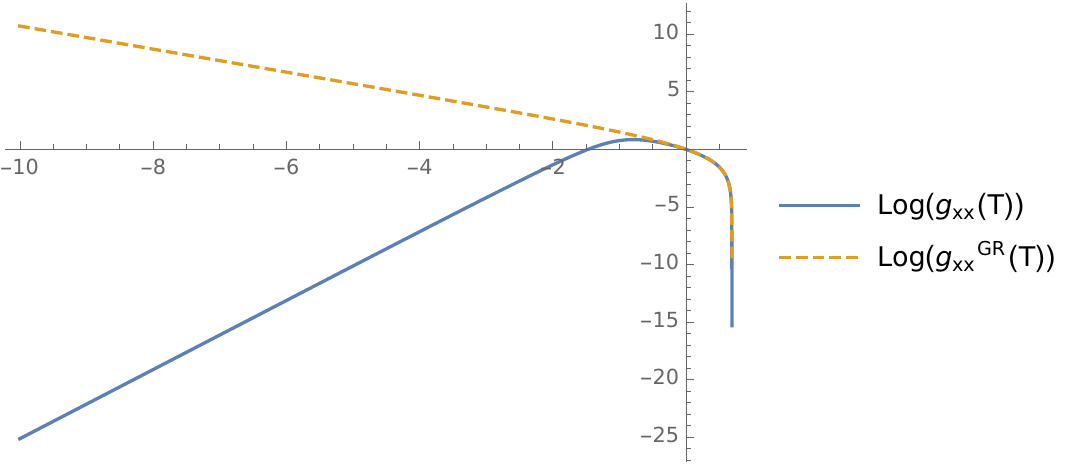}\\
 (a) & (b) \\[6pt]
\\
\includegraphics[height=4cm]{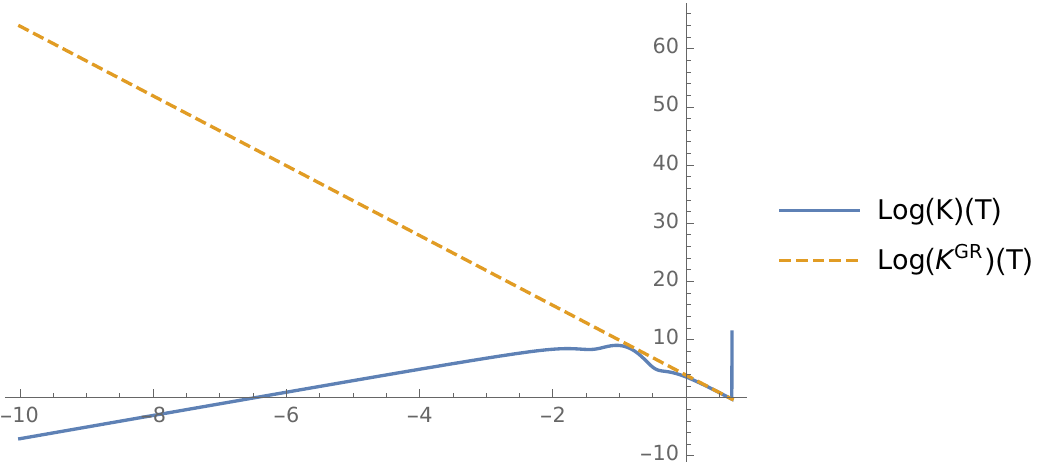}&
\includegraphics[height=4cm]{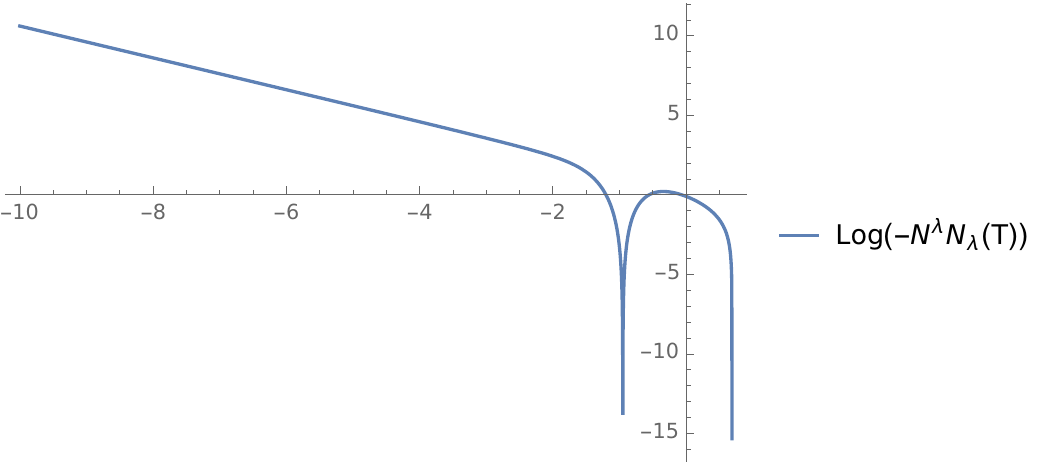}\\
(c) & (d) \\[6pt]
\\
\includegraphics[height=4cm]{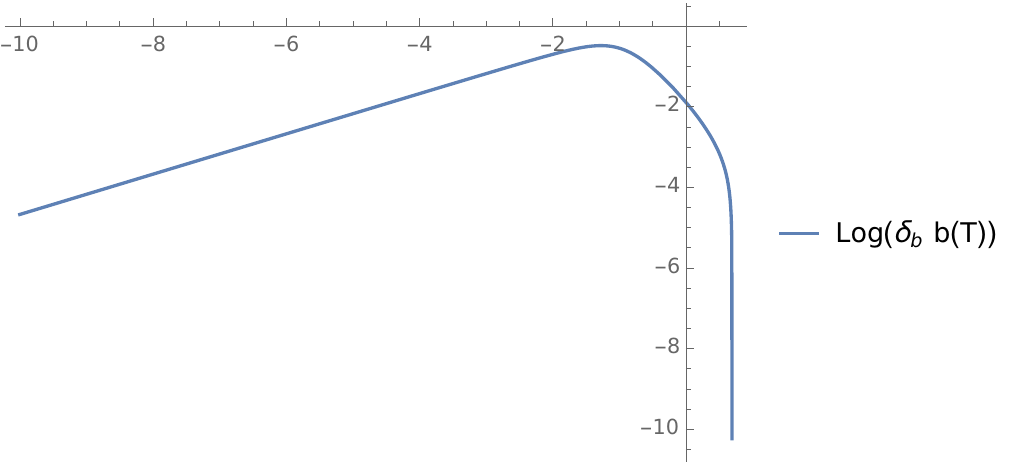}&
\includegraphics[height=4cm]{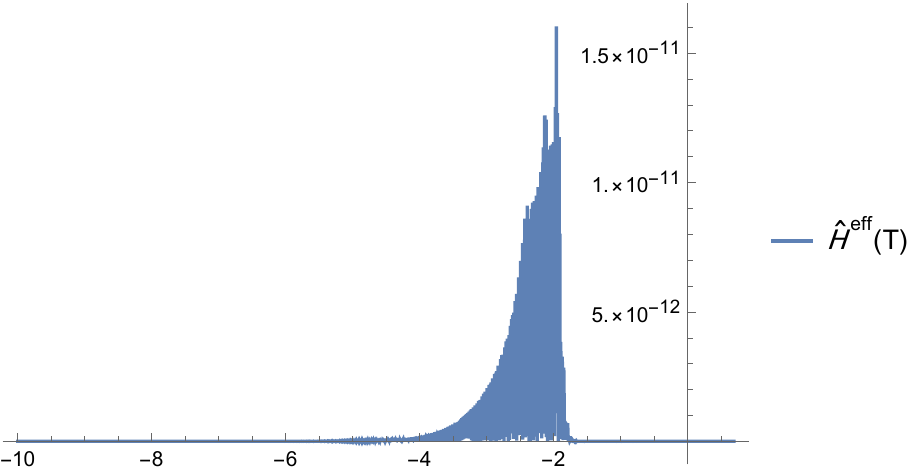}\\
(e) & (f)
	\end{tabular}}
\caption{Plots of  the lapse function $N^2$, the metric component $g_{xx}$, the Kretschmann scalar $K$, the norm $\left(-N^{\lambda}N_{\lambda}\right)$, the quantities 
$\delta_b b$ and $\hat H^{\text{eff}}$ defined by Eq.(\ref{eq2.25kkb}), 
together the classical counterpart $K^{\text{GR}}(T) \equiv 48m^2/p_c^3$
of the Kretschmann scalar. The mass parameter $m$ is chosen as  $m/\ell_{pl}=1$, for which we have  $T_{\cal{T}} \simeq -0.946567$ and $T_H \approx0.693$. The initial time is chosen at $T_i = 0.3$.
}
\lb{fig2}
\end{figure*}

To solve the dynamical equations (\ref{eq2.25s})-(\ref{eq2.25v}), we need to choose the initial time $T_i$ and initial conditions of the four variables
($b, c; p_b, p_c)$. Following \cite{Gan:2022oiy}, we shall choose $T_i$ so that the condition 
\bqn
\lb{eq3.12}
T_{\cal{T}} \ll T_i \ll T_H,
\eqn
holds, where $T_{\cal{T}}$ and $T_H$ denote the locations of the transition surface and the classical black hole horizon, respectively. This will minimize the deviations of the initial conditions with respect to the corresponding ones of the classical Einstein theory. 
Once $T_i$ is chosen, the initial conditions will be chosen as \cite{Gan:2022oiy}
\bqn
\lb{eq3.13}
&& p_b(T_i) = p^{\text{GR}}_b(T_i),  \quad p_c(T_i) = p^{\text{GR}}_c(T_i), \nb\\
&& b(T_i) = b^{\text{GR}}(T_i), \quad 
c(T_i) = c^{\text{eff}}(T_i),
\eqn
where $p^{\text{GR}}_b(T_i),
\; p^{\text{GR}}_c(T_i)$ and $b^{\text{GR}}(T_i)$
are the corresponding  values of the classical theory given by  Eq.(\ref{eq6d}), and
$c^{\text{eff}}(T_i)$ is obtained from the 
Hamiltonian constraint at $T = T_i$
\bqn
\lb{eq3.14}
H^{\text{eff}}(T_i) =0.
\eqn
For more details on the choices of the initial time and conditions, we refer readers to \cite{Gan:2022oiy}.

With the above chosen initial time and conditions,   Eqs.\eqref{eq2.25s}-\eqref{eq2.25v} will 
uniquely determine the trajectories of the four physical variables $\left(p_b, b; p_c, c\right)$ in the pahse space at any given time $T$ for both $T > T_i$ and $T < T_i$. 

 In Figs. \ref{fig1} and \ref{fig2}, we plot several physical quantities for $ m =  \ell_{pl}$ with $T_i = 0.3$, including the Kretchmann scalar, $K(T) \equiv R_{\alpha\beta\mu\nu} R^{\alpha\beta\mu\nu}$.
Note that for  $m = \ell_{pl}$, we find $T_{\cal{T}} \simeq -0.946567$ and $T_H \approx T_H^{\text{GR}}  - 0.0000763718$,  where $T_H^{\text{GR}} \equiv \log\left(2m/\ell_{pl}\right) \simeq 0.693$.
In these figures we also plot out the corresponding quantities of the classical theory, given by 
Eq.(\ref{eq6d}). In these plots, we focus ourselves to the regions near the transition surface $T \simeq T_{\cal{T}}$ and the black hole horizon
$T \simeq T_H$.

From Fig. \ref{fig1} we can see clearly that both $b$ and $p_b$ behave like their classical counterparts as $T \rightarrow - \infty$, while $c$ and $p_c$ behave quite different from their classical counterparts. In particular, from Fig. \ref{fig1} (g) it can be seen that $p_c$ first deceases until it attends a minimal value at $T_{\cal{T}} \simeq -0.947$, and then starts to increase, whereby a transition surface is formed.
On the other hand, at $T = T_H$, a marginally trapped surface (a black hole like horizon) is developed, at which we numerically find \footnote{\lb{f2}To make sure that our numerical calculations are reliable, and our physical conclusions will not depend on numerical errors, we run our {\em Mathematica} code in supercomputers with high precision. In particular, in all calculations we require that the Working Precision and Precision Goal be respectively 250 and 245, where Working Precision specifies how many digits of precision should be maintained in internal computations of {\em Mathematica}, and Precision Goal specifies how many effective digits of precision should be sought in the final result.} 
\bqn
\lb{eq3.17}
b\left(T_H\right) &\simeq& 1.51575 \times 10^{-125},\nb\\
p_b\left(T_H\right) &\simeq& 1.27668 \times 10^{-124},\nb\\
N^{\mu}N_{\mu} \left(T_H\right) &\simeq & -4.0492 \times 10^{-249},\nb\\
\delta T_{H} &\simeq& 0.0000763718,
\eqn
where $\delta T_{H} \equiv T^{\text{GR}}_H - T_{H}$,  
and $N_{\mu}$ is the normal vector to the two-spheres, defined as 
\bqn
\lb{nvector}
N_{\mu} \equiv \frac{\partial (r - r_0)}{\partial x^{\mu}} = \frac{p_{c,T}}{2\sqrt{p_c}}\delta^T_{\mu},
\eqn
with $r_0$ being a constant. A marginally trapped surface will be developed when 
\cite{Hawking:1973uf,Wang:2003bt,Wang:2003xa,Gong:2007md}
\bqn
\lb{eq3.19}
N^{\mu}N_{\mu} &=& - \frac{p_{c, T}^2}{4N^2 p_c} = -\frac{\pi \sin^2\left(\delta_b b\right)}{\gamma^2\Delta p_c} p_{c, T}^2 = 0. ~~~~~~~
\eqn
From Fig. \ref{fig2} (d)
we can see that $N_{\mu}$ becomes null at $T_H \simeq 0.693$ within the numerical errors of our simulations, while Fig. \ref{fig2} (e) tells us that at this point we have $\delta_b b = 0$, which can be also seen from Eq.(\ref{eq3.19}). 
It is interesting to note that $T_H$ is different from $T^{\text{GR}}_H$ very slightly,
even with the Planck mass of a black hole, $M = M_{pl}$. In addition, $N_{\mu}$ also becomes null at the transition surface $T_{\cal{T}} \simeq -0.947$, as can be seen clearly from Fig. \ref{fig2} (d),
at which we have $p_{c, T} = 0$ and $\delta_b b \not= 0$, as can be seen from Figs.
 \ref{fig1} (g) and  \ref{fig2} (e).
On the other hand, 
Fig. \ref{fig2} (f) shows clearly that in our simulations the numerical errors are
well under control. The maximal errors happen around $T \simeq T_{\cal{T}}$, where the curvature reaches the maximum [cf. Fig. \ref{fig2} (c)] and   the metric coefficients  change dramatically. So,  larger numerical errors are expected in this region. However, even in this region, the errors are no larger than $1.5 \times 10^{-11}$, while apart from it, we have $\left|\hat H^{\text{eff}}\right| \ll 10^{-11}$.

 \begin{figure*}[htbp]
 \resizebox{\linewidth}{!}{\begin{tabular}{cc}
 \includegraphics[height=4.cm]{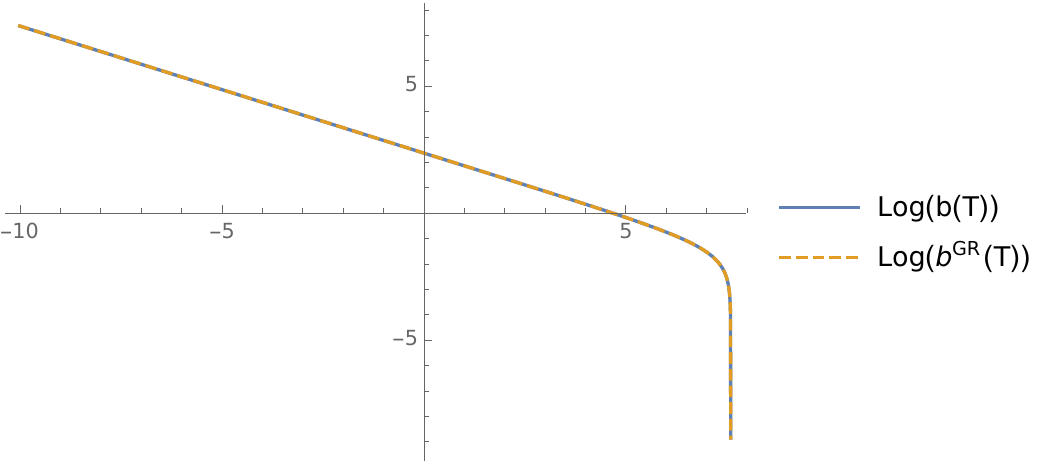}&
\includegraphics[height=4.cm]{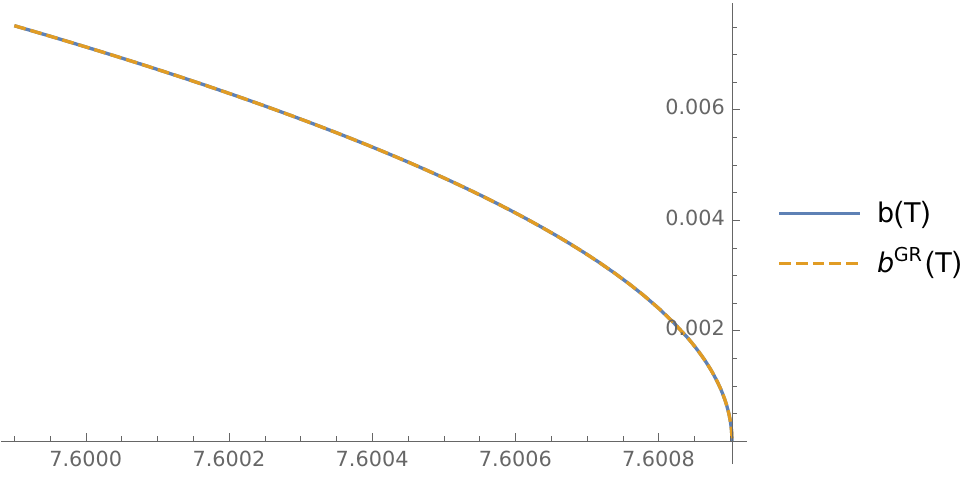}\\
 (a) & (b) \\[6pt]
\\
\includegraphics[height=4cm]{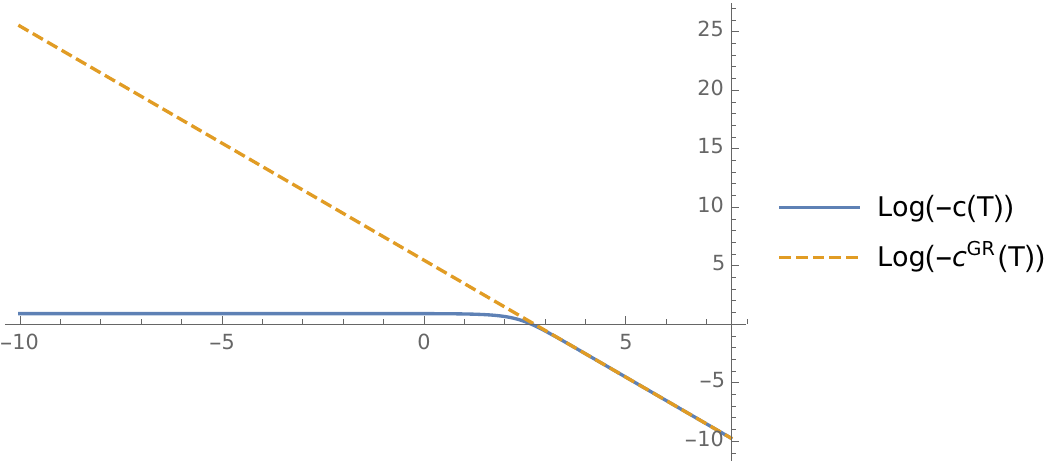}&
\includegraphics[height=4cm]{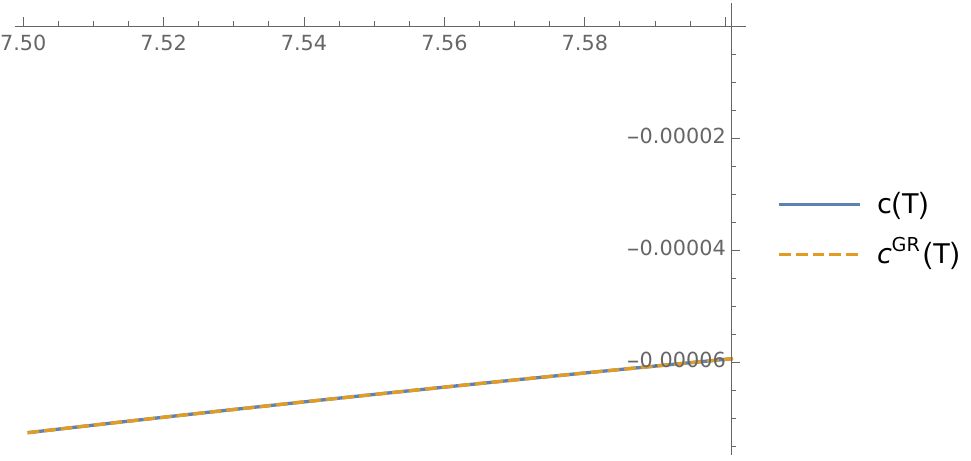}\\
(c) & (d) \\[6pt]
\\
\includegraphics[height=4cm]{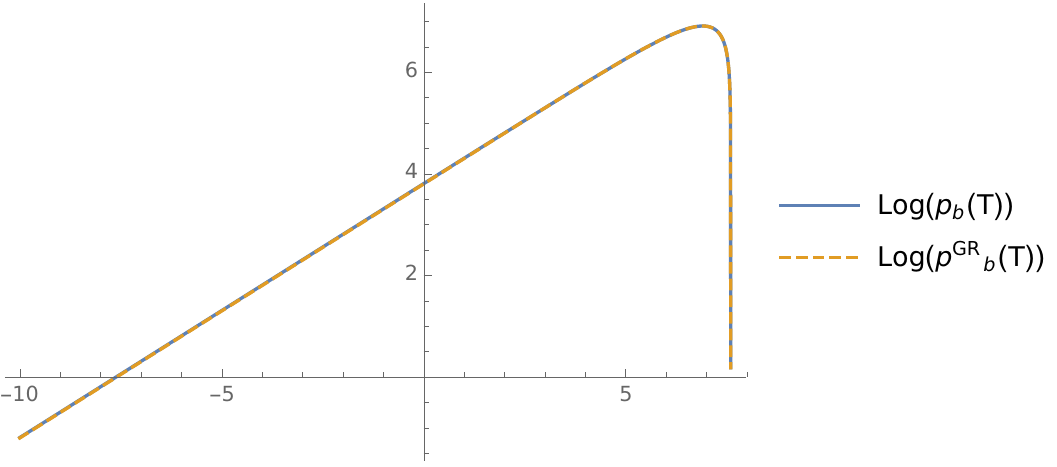}&
\includegraphics[height=4cm]{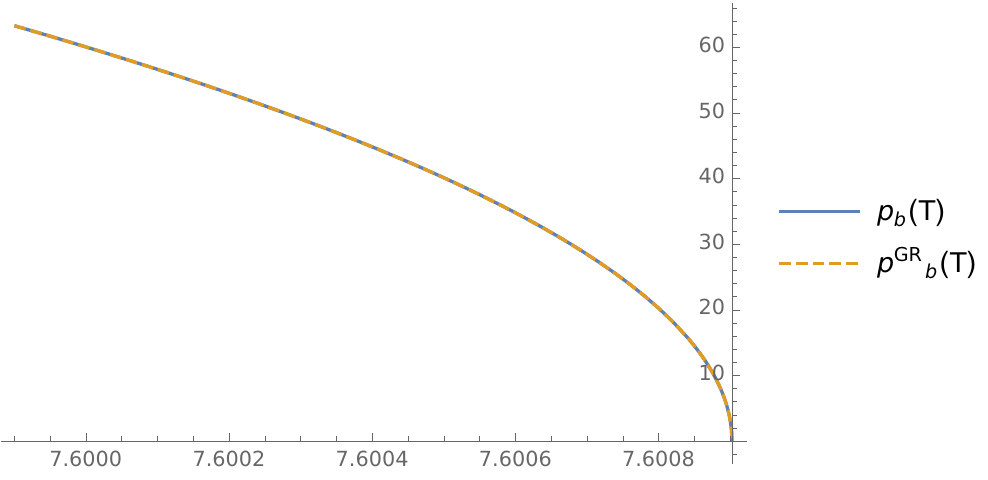}\\
(e) & (f)   \\[6pt]
\\
\includegraphics[height=4cm]{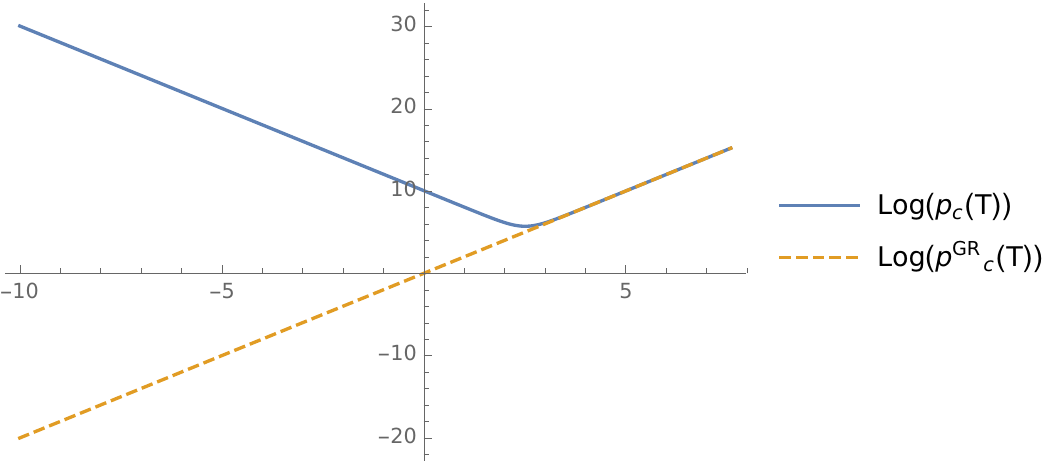}&
\includegraphics[height=4cm]{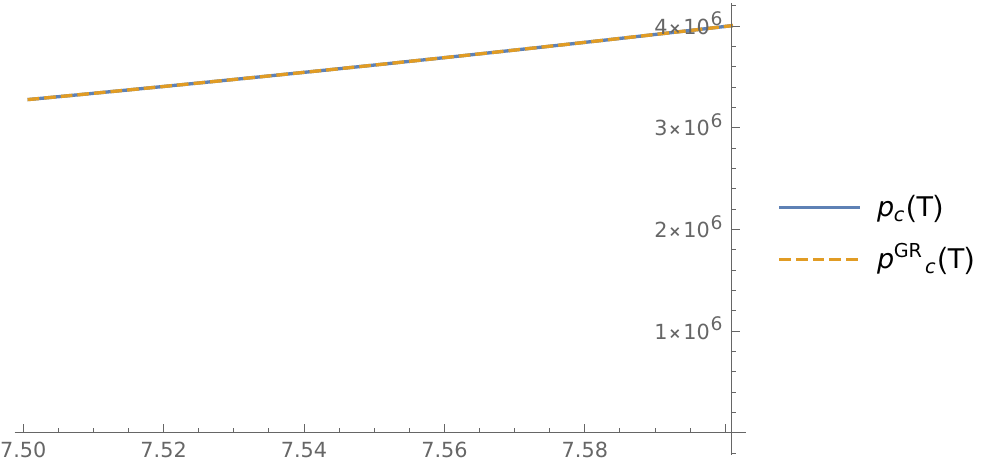}\\
	(g) & (h)  \\[6pt]
\end{tabular}}
\caption{Plots of  the four physical variables $\left(b, c, p_b, p_c\right)$ in the region $T\in (-10, 8)$, in which the transition surface and the black hole horizon are located. The mass parameter $m$ is chosen as  $m/\ell_{pl}=10^3$, for which we have  $T_{\cal{T}} \simeq 2.51204$ and $T_H \approx 7.6009$. The initial time is chosen at $T_i = 7$.}
\lb{fig3}
\end{figure*}

\begin{figure*}[htbp]
 \resizebox{\linewidth}{!}{\begin{tabular}{cc}
 \includegraphics[height=4.cm]{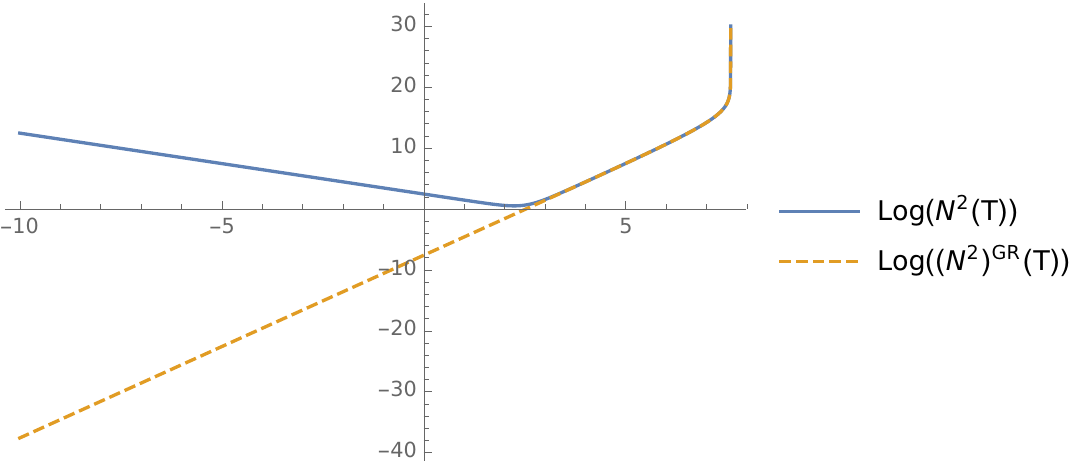}&
\includegraphics[height=4.cm]{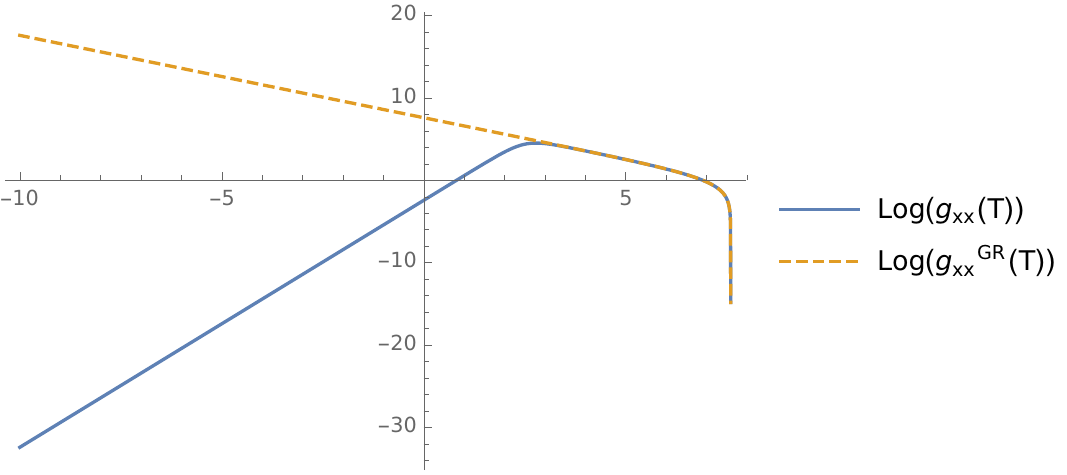}\\
 (a) & (b) \\[6pt]
\\
\includegraphics[height=4cm]{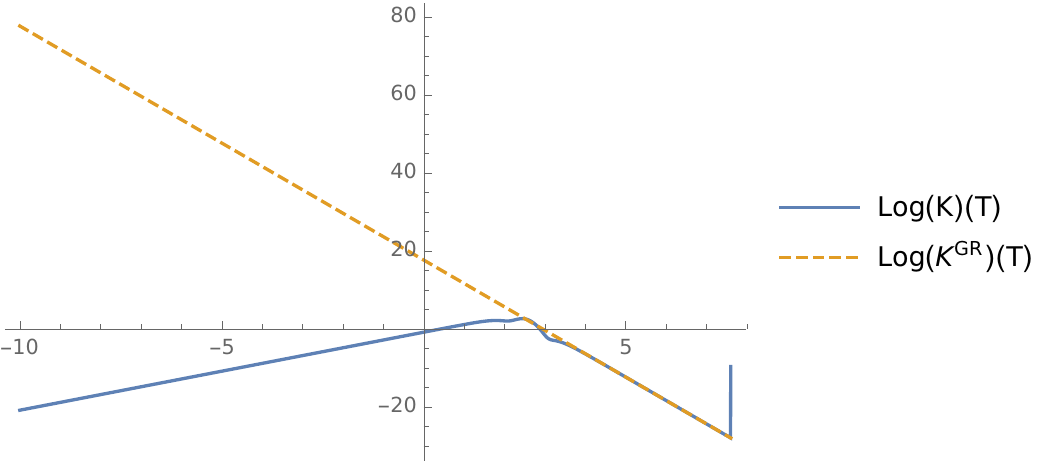}&
\includegraphics[height=4cm]{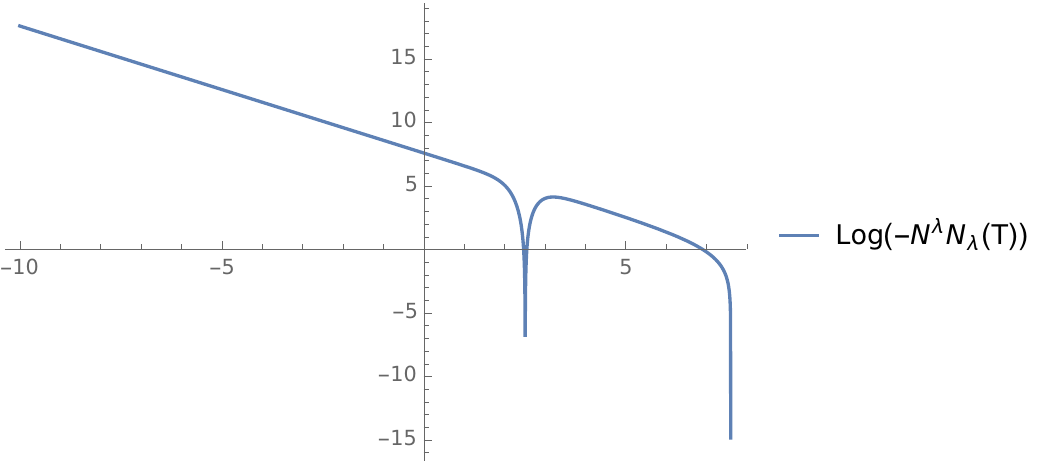}\\
(c) & (d) \\[6pt]
\\
\includegraphics[height=4cm]{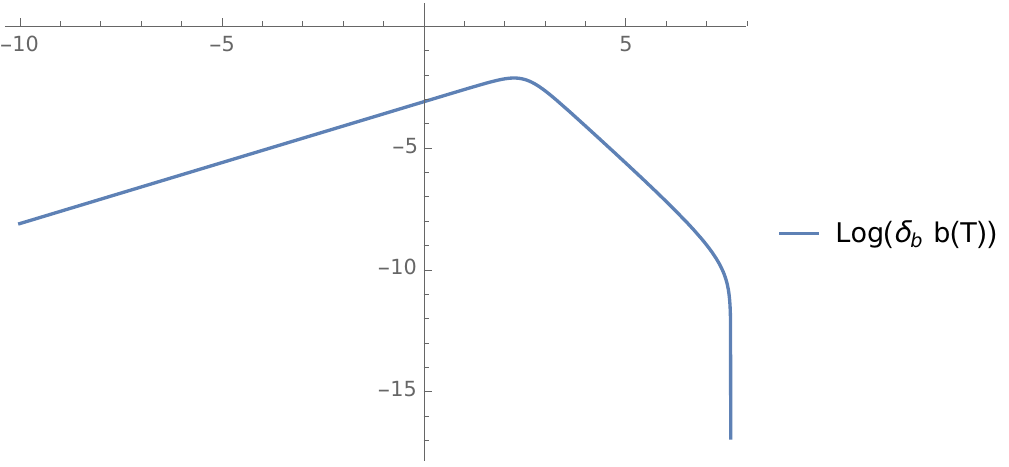}&
\includegraphics[height=4cm]{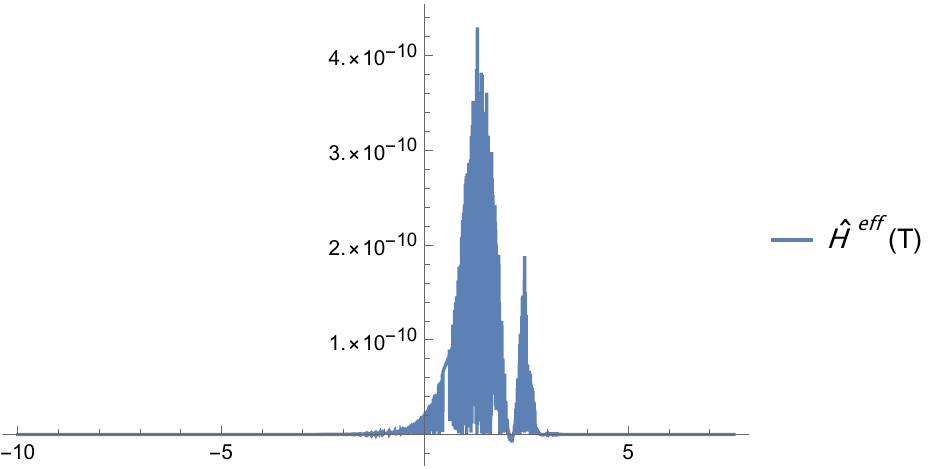}\\
(e) & (f)
	\end{tabular}}
\caption{Plots of  the lapse function $N^2$, the metric component $g_{xx}$, the Kretschmann scalar $K$, the norm $\left(-N^{\lambda}N_{\lambda}\right)$, the quantities 
$\delta_b b$ and $\hat H^{\text{eff}}$, 
together with the classical counterpart $K^{\text{GR}}(T) \equiv 48m^2/p_c^3$
of the Kretschmann scalar. The mass parameter $m$ is chosen as  $m/\ell_{pl}=10^3$, for which we have  $T_{\cal{T}} \simeq 2.51204$ and $T_H \approx 7.6009$. The initial time is chosen at $T_i = 7$.}
\lb{fig4}
\end{figure*}

\begin{figure*}[htbp]
 \resizebox{\linewidth}{!}{\begin{tabular}{cc}
 \includegraphics[height=4.cm]{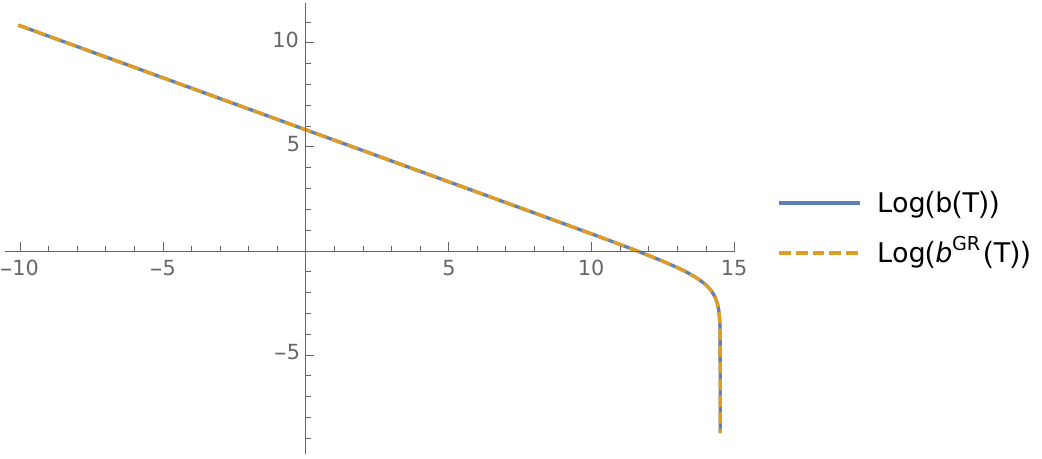}&
\includegraphics[height=4.cm]{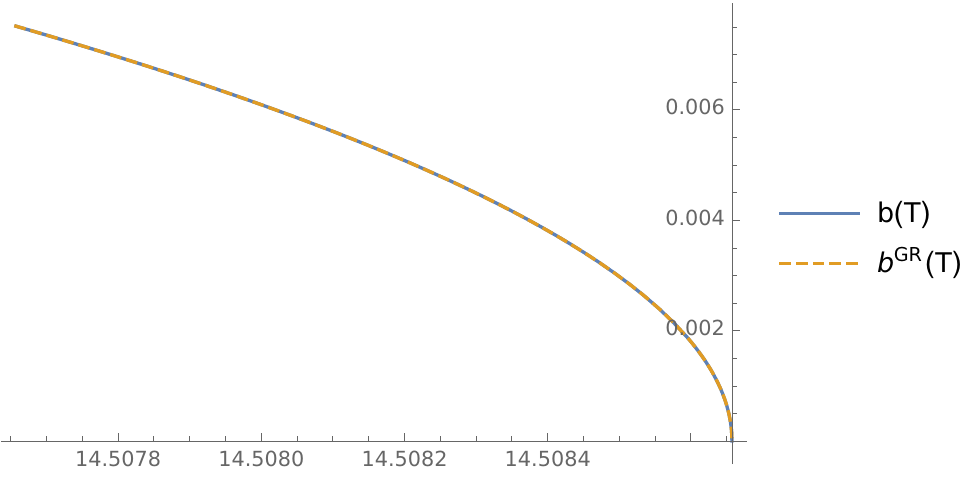}\\
 (a) & (b) \\[6pt]
\\
\includegraphics[height=4cm]{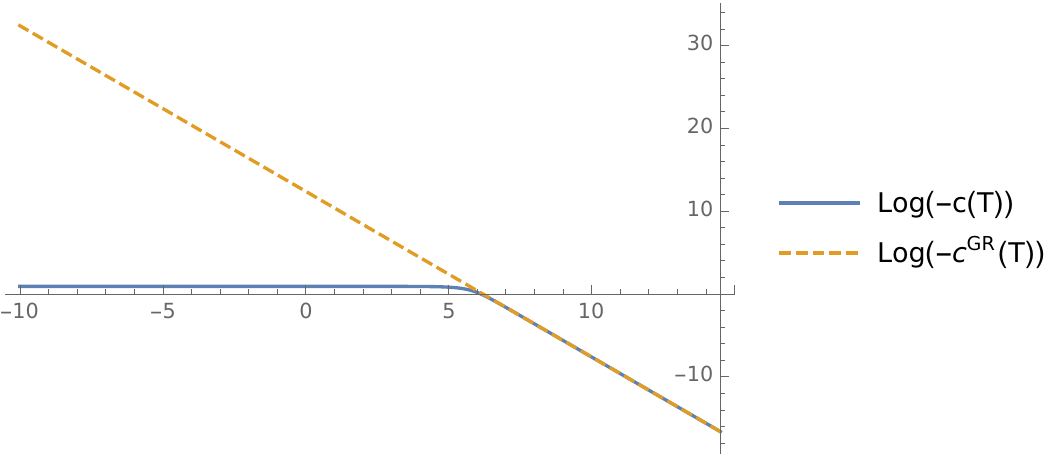}&
\includegraphics[height=4cm]{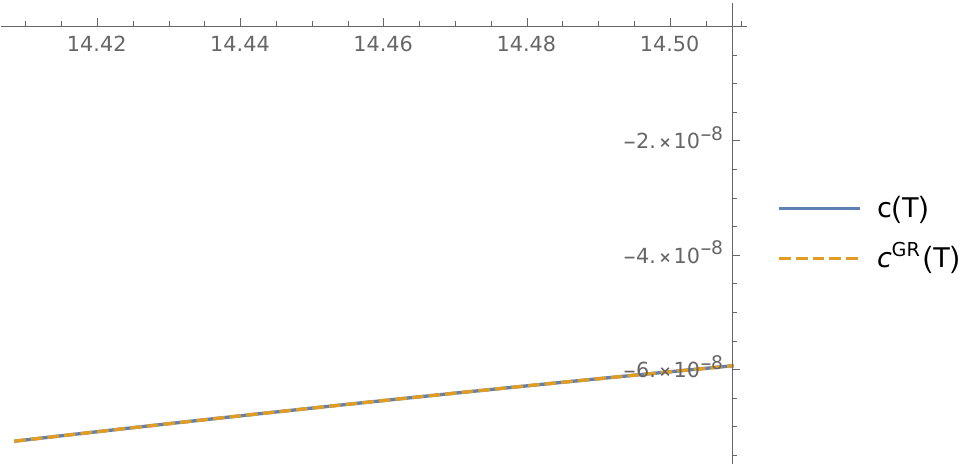}\\
(c) & (d) \\[6pt]
\\
\includegraphics[height=4cm]{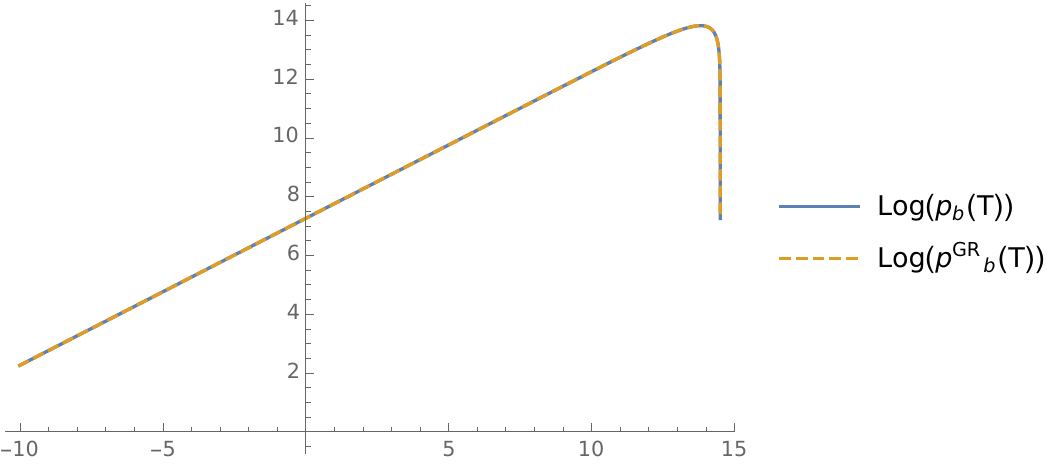}&
\includegraphics[height=4cm]{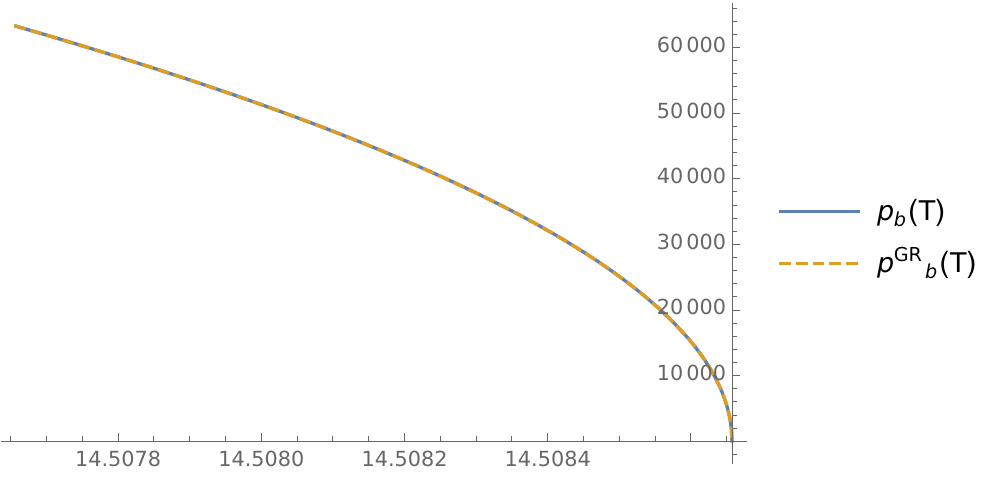}\\
(e) & (f)   \\[6pt]
\\
\includegraphics[height=4cm]{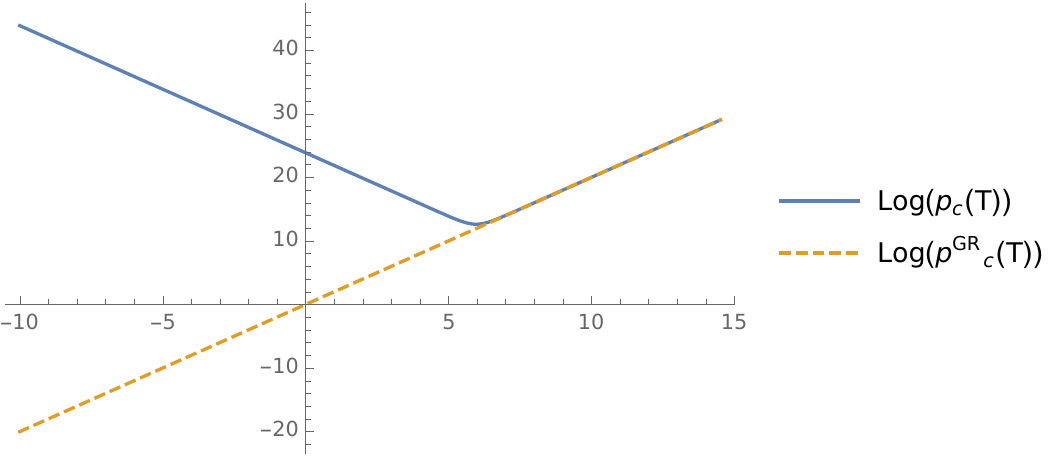}&
\includegraphics[height=4cm]{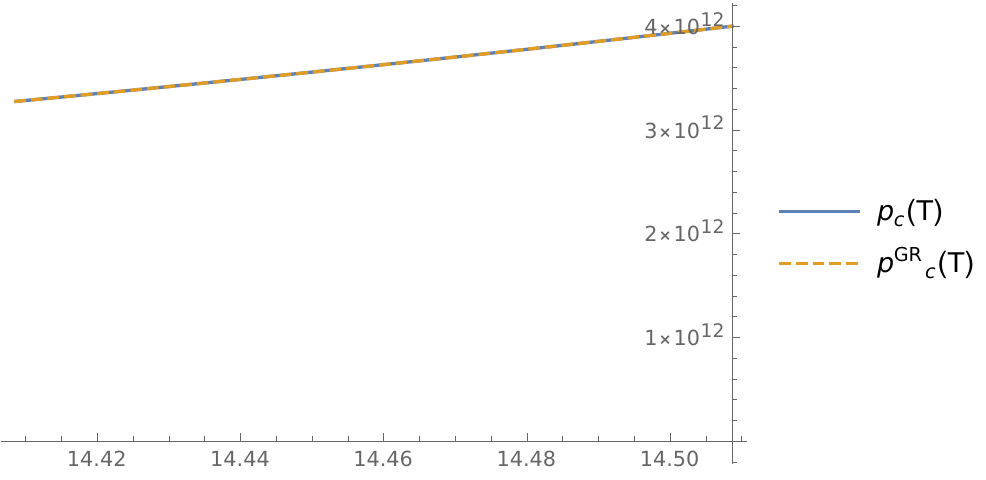}\\
	(g) & (h)  \\[6pt]
\end{tabular}}
\caption{Plots of  the four physical variables $\left(b, c, p_b, p_c\right)$, in the region $T\in (-10, 15)$, in which the transition surface and the black hole horizon are located. The mass parameter $m$ is chosen as  $m/\ell_{pl}=10^6$, for which we have  $T_{\cal{T}} \simeq 5.96686$ and $T_H \approx 14.5087$. The initial time is chosen at $T_i = 14$.}
\lb{fig5}
\end{figure*} 

\begin{figure*}[htbp]
 \resizebox{\linewidth}{!}{\begin{tabular}{cc}
 \includegraphics[height=4.cm]{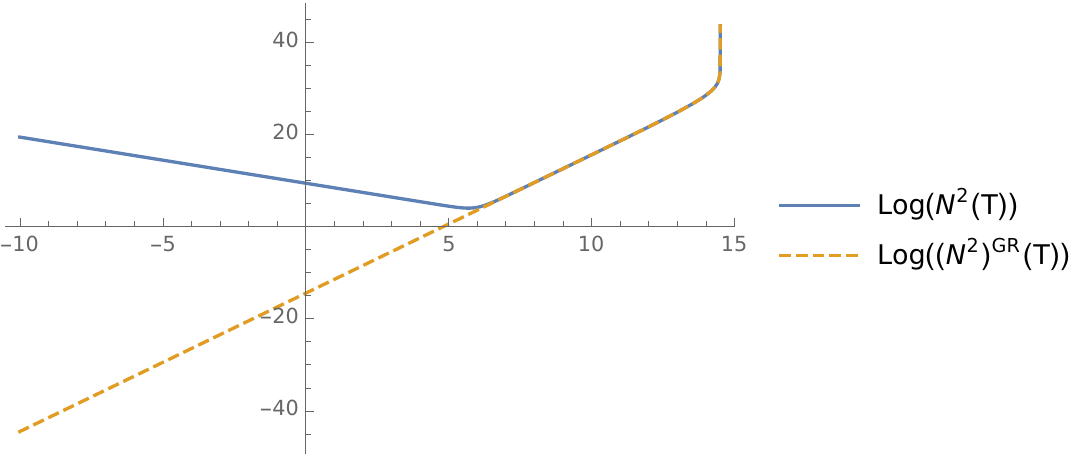}&
\includegraphics[height=4.cm]{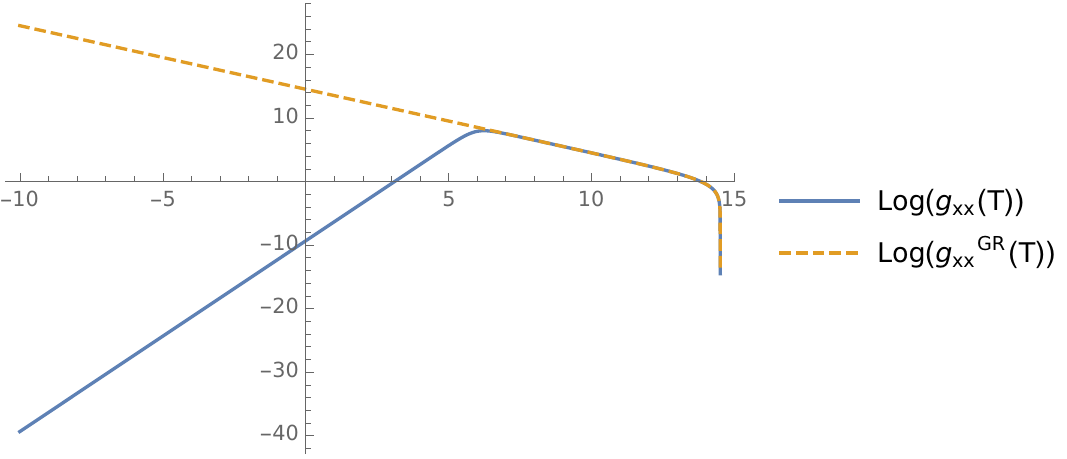}\\
 (a) & (b) \\[6pt]
\\
\includegraphics[height=4cm]{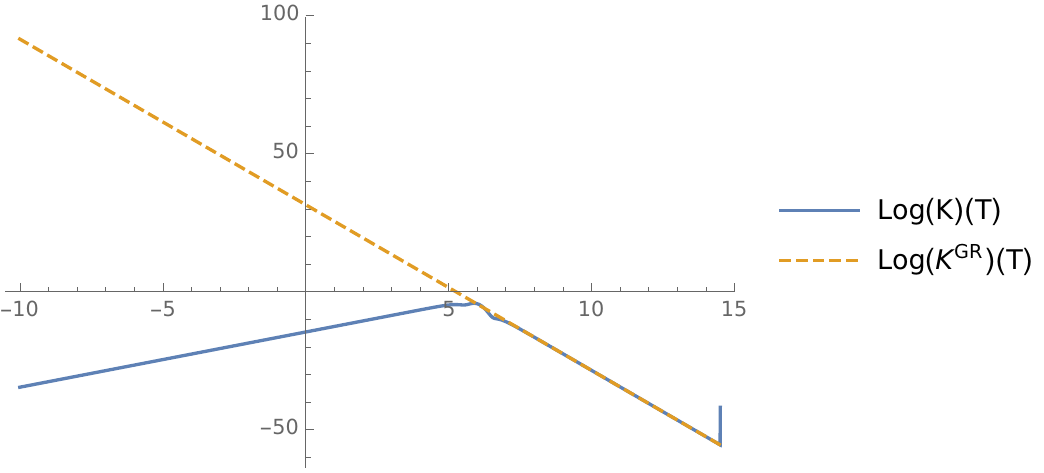}&
\includegraphics[height=4cm]{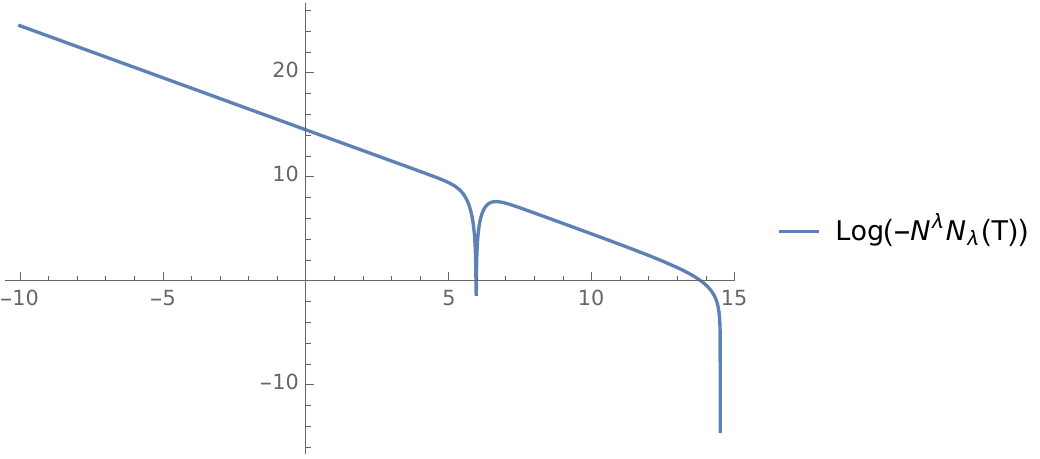}\\
(c) & (d) \\[6pt]
\\
\includegraphics[height=4cm]{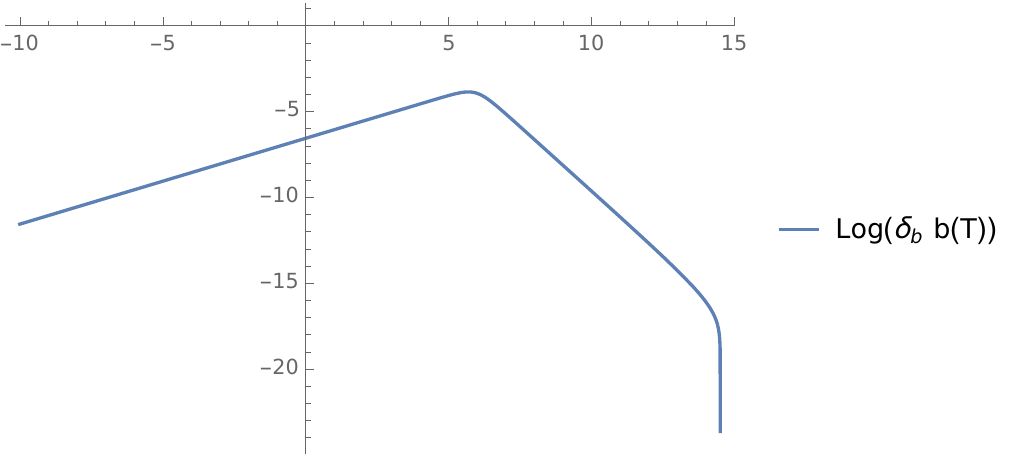}&
\includegraphics[height=4cm]{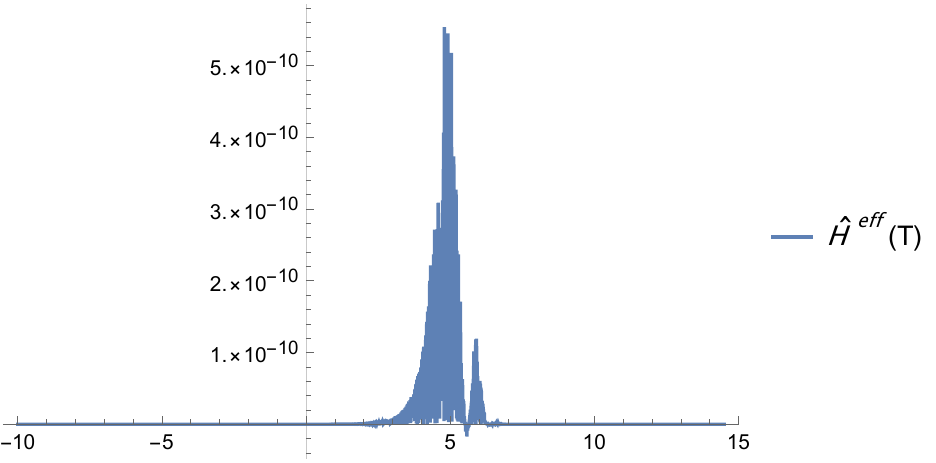}\\
(e) & (f)
	\end{tabular}}
\caption{Plots of  the lapse function $N^2$, the metric component $g_{xx}$, the Kretschmann scalar $K$, the norm $\left(-N^{\lambda}N_{\lambda}\right)$, the quantities 
$\delta_b b$ and $\hat H^{\text{eff}}$, 
together with the classical counterpart $K^{\text{GR}}(T) \equiv 48m^2/p_c^3$
of the Kretschmann scalar.
 The mass parameter $m$ is chosen as  $m/\ell_{pl}=10^6$, for which we have  $T_{\cal{T}} \simeq 5.96686$ and $T_H \approx 14.5087$. The initial time is chosen at $T_i = 14$.}
\lb{fig6}
\end{figure*}

Similar behaviors are also observed for $m/\ell_{pl} = 10^{3},\; 10^{6}$, as shown in Figs. \ref{fig3} - \ref{fig6}. In particular, for $m=10^3 \ell_{pl}$, we find that 
the locations of the transition surface and black hole horizon are respectively at 
$T_{\cal{T}} \simeq 2.51204$ and $T_H \approx 7.6009$, and at the latter we have
\bqn
\lb{eq3.20}
b\left(T_H\right) &\simeq& 1.95882 \times 10^{-124},\nb\\
p_b\left(T_H\right) &\simeq& 1.64953 \times 10^{-120},\nb\\
N^{\mu}N_{\mu} \left(T_H\right) &\simeq & -6.8024 \times 10^{-247},\nb\\
\delta T_{H} &\simeq& 1.47885 \times 10^{-10}.
\eqn
With the high  precision that we use in our numerical simulations  [cf. Footnote \ref{f2}], all the numerical values given in Eq.(\ref{eq3.20}) are physically reliable and meaningful. In particular, it clearly shows the existence of a marginally trapped surface at $T_H \approx 7.6009$. 

On the other hand, for  $m=10^6 \ell_{pl}$ we find that 
the locations of the transition surface and black hole horizon are respectively at 
$T_{\cal{T}} \simeq 5.96686$ and $T_H \approx 14.5087$. At the horizon, from our numerical simulations now we find
\bqn
\lb{eq3.21}
b\left(T_H\right) &\simeq& 1.88451 \times 10^{-124},\nb\\
p_b\left(T_H\right) &\simeq& 1.58696 \times 10^{-117},\nb\\
N^{\mu}N_{\mu} \left(T_H\right) &\simeq & -6.29609 \times 10^{-247},\nb\\
\delta T_{H} &\simeq& -1.18581 \times 10^{-16}.
\eqn
From Eqs.(\ref{eq3.17}), (\ref{eq3.20}) and (\ref{eq3.21}), we can see that the
differences of the locations of the quantum black hole horizons and the corresponding classical ones become smaller and smaller as the black hole mass increases, which indicates the quantum geometric effects become negligible for massive black holes.

In addition, sharply in contrast to the BV model \cite{Boehmer:2007ket}, across the transition surface 
$T_{\cal{T}}$ the geometric radius $r \equiv \sqrt{p_c}$ increases monotonically and no multiple transition surfaces exist.  More interesting, no white-hole-like horizons exist in the future of the transition surface and the spacetime becomes geodesically complete. 

To show our above claims, let us study the asymptotic behavior of the LQBHs of the above model and their properties near the transition surface and black hole horizon in detail.

\subsection{Asymptotic Structure of Spacetimes as $T \rightarrow -\infty$}

For $T \ll T_{\cal{T}}$, we find that the variables $b$ and $p_b$ are approaching to their classical values asmptotically, while the variables $c$ and $p_c$ are deviating significantly from their classical values, as can be seen from Figs. \ref{fig7} - \ref{fig12} in the region $T \in (-600, -10)$ for $m/\ell_{Pl} = 1, \; 10^3, \; 10^6$, respectively. In particular, now $c(T)$ remains almost constant, while $p_c$ is exponentially increasing, in contrast to their behaviors in the classical theory, in which $c^{\text{GR}}(T)$ is exponentially increasing, while $p_c^{\text{GR}}(T)$ is exponentially decreasing, as can be seen from Eq.(\ref{eq6d}). These differences lead to significantly differences in the spacetime properties, as can be seen from Figs. \ref{fig8}, \ref{fig10} and \ref{fig12}, from which we can see that the Kretchmann curvature now is exponentially decreasing as $T \rightarrow -\infty$, instead of exponentially increasing  as $T \rightarrow -\infty$ as in the classical case.
The normal vector $N_{\mu}$ remains timelike in the whole region $T < T_{\cal{T}}$, while the metric coefficients $N^2, \; g_{xx}$ and $p_c$ have exactly the opposite behaviors, that is, now $N^2$ and $p_c$ are all exponentially increasing, while $g_{xx}$ is exponentially decreasing. Setting
\bqn
\lb{eq3.22}
{\cal{F}}(T) = {\cal{F}}_0\left(\frac{m}{\ell_{pl}}\right)^{{\cal{F}}_1}
\exp\left\{{\cal{F}}_2\left(\frac{m}{\ell_{pl}}\right)^{{\cal{F}}_3} T\right\},
\eqn
where ${\cal{F}} \equiv \left(N^2, g_{xx}, p_c\right)$, we find that $N^2, \; g_{xx}, \; p_c$ can be well approximated by
\bqn
\lb{eq3.23}
N^2(T) &\simeq& 0.012\left(\frac{m}{\ell_{pl}}\right)e^{-T},\nb\\
g_{xx}(T) &\simeq& 87.77\left(\frac{m}{\ell_{pl}}\right)^{-1} e^{3T},\nb\\
p_c(T) &\simeq& 0.023\left(\frac{m}{\ell_{pl}}\right)^{2}  e^{-2 T},
\eqn
as $T \rightarrow -\infty$.
 Then, corresponding to the solution of Eq.(\ref{eq3.23}), we find that the effective energy-momentum tensor defined as $ T_{\mu\nu} \equiv \kappa^{-2} G_{\mu\nu}$ can be written as
\bqn
\lb{eq3.24}
\kappa^{2} T_{\mu\nu} = \rho u_{\mu}u_{\nu} + p_x x_{\mu}x_{\nu} + 
p_{\bot}\left(\theta_{\mu}\theta_{\nu} + \phi_{\mu} \phi_{\nu}\right),
\eqn
where $\kappa^2 \equiv 8\pi G$, $(u_{\mu},  x_{\mu}, \theta_{\mu},  \phi_{\mu})$ are the unit vectors along 
the coordinates $T, \; x, \; \theta,\; \phi$ directions, respectively, and 
\begin{widetext}
\bqn
\lb{eqA1}
\rho &=& \frac{g_{xx} \left(4 N^2 p_c+\Dot{p}_c{}^2\right)+2 p_c \Dot{p}_c \Dot{g}_{xx}}{4 N^2 p_c{}^2 g_{xx}}, \quad
p_x = \frac{4 p_c \Dot{N} \Dot{p}_c+N \left(\Dot{p}_c{}^2-4 p_c \Ddot{p}_c\right)-4 N^3 p_c}{4 N^3 p_c{}^2}, \nb\\
p_{\bot} &=& \frac{1}{4 N^3 p_c{}^2 g_{xx}{}^2} \Bigg( 
2 p_c g_{xx} \Dot{N} \left(p_c \Dot{g}_{xx}+g_{xx} \Dot{p}_c\right)+N \Big(p_c{}^2 \left(\Dot{g}_{xx}{}^2-2 g_{xx} \Ddot{g}_{xx}\right)\nb\\
&&-p_c g_{xx} \left(\Dot{p}_c \Dot{g}_{xx}+2 g_{xx} \Ddot{p}_c\right)+g_{xx}{}^2 \Dot{p}_c{}^2\Big)
\Bigg).
\eqn
\end{widetext}
Then, as $T \ll T_{\cal{T}}$, we find that
\bqn
\lb{eq3.25}
\rho &\simeq& -1.4\times 10^4\left(\frac{m}{\ell_{pl}}\right)^{-2}e^{2T}, \nb\\
p_x &\simeq& -7.0\times 10^3\left(\frac{m}{\ell_{pl}}\right)^{-2}e^{2T}, \nb\\
p_{\bot} &\simeq& -1.5\times 10^4\left(\frac{m}{\ell_{pl}}\right)^{-2}e^{2T},
\eqn
which shows clearly that the effective fluid does not satisfy any of the energy conditions \cite{Hawking:1973uf}, although $\rho, p_x$ and $p_{\bot}$ all approach to zero exponentially. On the other hand,  we also find 
\bqn
K(T) \simeq 4.1 \times 10^5 \left(\frac{m}{\ell_{pl}}\right)^{-2}e^{2T}.
\eqn
 Thus, the Krestchmann scalar decreases exponentially to zero as $T \rightarrow -\infty$.
 Combining it  with Eq.\eqref{eq3.23}, we find that asymptotically ($T \rightarrow -\infty$) the Kretschmann scalar takes the form 
 \bqn
 \lb{eq3.25b}
 K \simeq \frac{K_0}{p_c(T)},\; (T \rightarrow -\infty), 
 \eqn
 with $K_0$ being a constant.

\subsection{Spacetime near the Transition Surface}

Near the transition surface $T \simeq T_{\cal{T}}$, 
the quantities, $\rho, p_x, p_{\bot}, 
R^{\mu}_{\mu}, K,  C_{\alpha\beta\mu\nu} C^{\alpha\beta\mu\nu}$
are plotted out in Figs. \ref{fig16}-\ref{fig18}.
From these figures, we can see that all the physical quantities remain finite, and   the classical singularity is replaced by a quantum bounce. Moreover, as mass increases, $\rho, p_x, p_{\bot}, 
R^{\mu}_{\mu}, K,  C_{\alpha\beta\mu\nu} C^{\alpha\beta\mu\nu}$ all decrease.

 \subsection{Quantum Effects Near the Black Hole Horizon}

  The classical black hole horizon is located at
  \bq
  \lb{horizon}
  T_H^{\text{GR}} =\ln(2m),
  \eq
  as can be seen clearly from Eq.(\ref{eq6d}), at which we have $p_b^{\text{GR}} = 0$.
Let us first estimate the quantum effects near $T = T_H$.  
Substituting  the  classical Schwarzschild black hole solution given by Eq.\eqref{eq6d} into Eq.\eqref{eq3}, we find
\bqn\lb{bv-delta_b2}
&& |\delta_b| = \sqrt{\frac{\Delta}{4\pi e^{2T}}} \simeq \frac{1}{4m} \sqrt{\frac{\Delta}{\pi}}, \nb\\
\lb{bv-delta_c2}
&& L_o|\delta_c| = \sqrt{\frac{\Delta}{\pi}} , 
\eqn
as $T \rightarrow T_H$.  Thus, quantum effects near horizon will decrease as $m$ increases, and when $m \gg \frac{1}{4} \sqrt{\frac{\Delta}{\pi}} \approx 0.32 \ell_{pl}$, the quantum effects near $T_H$ is negligibly small.

Requiring that the physical variables be analytic across the black hole horizon, similar to the classical  Schwarzschild black hole, one shall obtain a unique extension to the outside of the black hole horizon. It is not difficult to see that such an extended spacetime is geodesically complete and cover the whole external region, $ p_{c} \in (p_c^H, \infty)$, where $p_c^H \equiv p_c(T_H)$.

\subsection{The Hybrid Scheme with $N=1$}\lb{new-sec-n1}

To study the dynamics of the hybrid scheme further, we consider the gauge 
\bqn
\lb{eq3.26}
N = 1, 
\eqn
that is, the gauge in which the timelike coordinate $T$ becomes   the proper time of a comoving observer with the coordinate system. 
To be distinguishable from the time used in the above, we shall denote it as $\tau$, so that $d\tau = N dT$. 
In this gauge, the effective Hamiltonian is given by
\bqn\lb{eq3.28}
H^{\text{eff}}[N=1]&=& 
-\frac{1}{2G\gamma^2}\bigg(2\frac{\sin(\delta_c c)}{\delta_c}\frac{\sin(\delta_b b)}{\delta_b}\sqrt{p_c}\nb\\
&&+\left(\frac{\sin^2(\delta_b b)}{\delta_b^2}+\gamma^2\right)\frac{p_b}{\sqrt{p_c}}
\bigg).
\eqn
And the corresponding EoMs are given by
\begin{widetext}
\bqn
\lb{eq3.29s}
\dot b &=&-\frac{1}{2 \gamma  \sqrt{p_c(\tau)}}\Bigg(\frac{4 \pi  p_c(\tau) \sin ^2\left(\frac{b(\tau) \sqrt{\frac{\Delta }{p_c(\tau)}}}{2 \sqrt{\pi }}\right)}{\Delta }+\gamma ^2
\Bigg),\\
\lb{eq3.29t}
\dot c &=& \frac{1}{2 \gamma  \left(\Delta  p_c(\tau)\right){}^{3/2}} \Bigg(2 \sqrt{\pi } p_c(\tau) \sin \left(\frac{\sqrt{\Delta } c(\tau)}{\sqrt{\pi }}\right) \left(\Delta  b(\tau) \cos \left(\frac{b(\tau) \sqrt{\frac{\Delta }{p_c(\tau)}}}{2 \sqrt{\pi }}\right)-4 \sqrt{\pi } \sqrt{\Delta  p_c(\tau)} \sin \left(\frac{b(\tau) \sqrt{\frac{\Delta }{p_c(\tau)}}}{2 \sqrt{\pi }}\right)\right) \nb\\
&&+\sqrt{\Delta } p_b(\tau) \left(-4 \pi  p_c(\tau) \sin ^2\left(\frac{b(\tau) \sqrt{\frac{\Delta }{p_c(\tau)}}}{2 \sqrt{\pi }}\right)+2 \sqrt{\pi } b(\tau) \sqrt{\Delta  p_c(\tau)} \sin \left(\frac{b(\tau) \sqrt{\frac{\Delta }{p_c(\tau)}}}{\sqrt{\pi }}\right)+\gamma ^2 \Delta \right)\Bigg), \\
\lb{eq3.29u}
\dot p_c &=& \frac{4 \sqrt{\pi } p_c(\tau) \cos \left(\frac{\sqrt{\Delta } c(\tau)}{\sqrt{\pi }}\right) \sin \left(\frac{b(\tau) \sqrt{\frac{\Delta }{p_c(\tau)}}}{2 \sqrt{\pi }}\right)}{\gamma  \sqrt{\Delta }}, \\ 
\lb{eq3.29v}
\dot p_b &=& \frac{\sqrt{\pi } \left(2 p_b(\tau) \sin \left(\frac{b(\tau) \sqrt{\frac{\Delta }{p_c(\tau)}}}{2 \sqrt{\pi }}\right)+\sqrt{p_c(\tau)} \sin \left(\frac{\sqrt{\Delta } c(\tau)}{\sqrt{\pi }}\right)\right) \cos \left(\frac{b(\tau) \sqrt{\frac{\Delta }{p_c(\tau)}}}{2 \sqrt{\pi }}\right)}{\gamma  \sqrt{\Delta }},
\eqn
\end{widetext}
where now a dot denotes the derivative with respect to $\tau$.
With the same initial conditions and the corresponding initial times $\tau_i \equiv \tau(T_i)$, we integrate the dynamical equations (\ref{eq2.25m}) - (\ref{eq2.25p}), and plot the relevant physical quantities  in Figs. \ref{fig1-new-N1} and \ref{fig3-new-N1} for two representative cases  $m = \ell_{pl}$ and $m = 10^3 \ell_{pl}$, respectively. 
To compare the effective solutions with the classical ones in the gauge $N=1$, in Appendix A, we also present the corresponding  classical Hamiltonian and dynamical equations, which are represented  by orange dashed lines. 

From these figures, we can see that transition surfaces, at which we have $N_{\lambda}N^{\lambda}(\tau) = 0$, replace the classical singularities  used to locate at $p^{\text{GR}}_c = 0$. The transition surfaces are located at $\tau_{\mathcal{T}}\approx -3.064$ for $m= \ell_{pl}$ and  $\tau_{\mathcal{T}}\approx -3141.13$ for $m=10^3 \ell_{pl}$, respectively. After the transition, all the physical variables remain finitie and no singularity is indicated. In particular, $p_c(\tau)$ is monotonically increasing, as it can be seen from Fig. \ref{fig1-new-N1} (d) and Fig. \ref{fig3-new-N1} (d). These are consistent with what we obtained in the previous subsections for the lapse function  chosen as that given by Eq.(\ref{eq8}).

\section{Conclusion}\lb{conclusion}
 \renewcommand{\theequation}{4.\arabic{equation}}\setcounter{equation}{0}

In the studies of quantum black holes in the framework of LQG, various gauges have been used, including the KS gauge \cite{Ashtekar:2023cod},  
Gullstrand-Painlev\'e gauge \cite{Gambini:2022hxr}, and Lema\^itre-Tolman-Bondi gauge \cite{Giesel:2022rxi,han2022improved}. Since the gauge choices usually do not commute with the polymerization (\ref{eq1.2}) \footnote{It should be noted that such ambiguities also exist in quantum field theory and are closely related to the ordering of physical variables.}, it is important to study the robustness of these gauges. In addition, the choices of the two quantum parameters, $\delta_b$ and $\delta_c$, appearing in the polymerization (\ref{eq1.2}), also depend on how to calculate the area in each of the two-dimensional plane   of the quantized three-dimensional spatial spaces, once a foliation of the four-dimensional spacetime is chosen. In particular, in the KS gauge, if one parameterizes $\delta_b$ and $\delta_c$ as those given in \cite{Boehmer:2007ket}, it has been shown recently that   the quantum effects are so strong near the location of the classical black hole horizon so that in the effective theory  black/white hole horizons do not exist any longer, instead, they are replaced by an infinite number of transition surfaces, which always separate trapped regions from anti-trapped ones \cite{Zhang:2023noj,Gan:2022oiy}.  

To understand the effects mentioned above, in  this paper, we propose a new quantization scheme, in which the geometric distances are used along the direction perpendicular to the two-spheres, instead of using its physical distance [cf. Eq.(\ref{eq3.1a})]. This is because in the KS gauge the physical distance along this direction vanishes not only at the classical singularity ($T = -\infty$) but also at the black hole horizon ($T = \ln(2m)$) [cf. Eqs.(\ref{metric}) and (\ref{eq6d})]. As a result, the quantum geometric effects become large not only at the classical singularity but also at the classical black hole horizons. In addition, the physically distance along the $x-$direction is also not gauge-invariant, as shown explicitly by the gauge freedom of Eq.(\ref{GTs}). But  the fraction length $L_o\delta_c$ with respect to the metric ${}^{o}q_{ab}$ is metric independent. Also, the area $4\pi r^2$ of the two-spheres $T, x =$ constant  is gauge-invariant. Therefore, in this paper we have accounted the two areas $A_{(x,\phi)}$ and $A_{(\theta,\phi)}$
as given by Eqs.(\ref{eq3.1a}) and (\ref{eq3.1b}). As a result, $\delta_b$ and $\delta_c$ are uniquely specified and are given by Eq.(\ref{eq3}), i.e., 
\bq
\lb{eq4.1}
\delta_b = \sqrt{\frac{\Delta}{4\pi p_c}}, \quad
L_o \delta_c = \sqrt{\frac{\Delta}{\pi}}.
\eq
With this choice, in Sec. \ref{sec-new} we have systematically studied the effective dynamical system in two different gauges, given, respectively, by Eqs.(\ref{eq8}) and (\ref{eq3.26}). Although the physics does not depend on the choice of the lapse function $N$, in the latter, the time-like coordinate denoted by $\tau$ measures the proper time, and the corresponding geometric meaning of each physical quantity becomes more transparent, as shown explicitly in Sec. \ref{new-sec-n1}. 

In particular,  we have shown that with the choice (\ref{eq4.1}), the quantum gravity effects in the region where the classical black hole horizon used to be present is negligible and in the effective theory the black hole horizon still exists. On the other hand, the classical singularity is replaced by a regular transition surface.  However, in contrast to the models studied in  \cite{Ashtekar:2023cod,Gambini:2022hxr,han2022improved}, white hole horizons are not found.  Instead, we have found that the spacetime in the other side of the transition surface is complete and always anti-trapped. So, no white hole horizons exist in the current model. 


\acknowledgements
W-C.G. was supported by  the Initial Research Foundation of Jiangxi Normal University under the Grant No. 12022827. A.W. is partly supported by the NSF  grant, PHY-2308845.

\section*{Appendix A: The Classical Hamiltonian for $N = 1$}
 \renewcommand{\theequation}{A.\arabic{equation}}\setcounter{equation}{0}

To compare the effective solution with classical result in the gauge $N=1$, we present the classical Hamiltonian here, which is given by
\bqn
H^{\text{GR}}[N=1]= - \frac{1}{2G\gamma^2}\left(2cb\sqrt{p_c}+(b^2+\gamma^2)\frac{p_b}{\sqrt{p_c}}
\right).\nb\\
\eqn
The corresponding EoMs are (here we omitted the superscript ``GR" for simplicity of notation)
\bqn
\dot b &=& -\frac{b(\tau)^2+\gamma ^2}{2 \gamma  \sqrt{p_c(\tau)}},\\
\dot c &=& \frac{(b(\tau)^2+\gamma^2)p_b(\tau)-2b(\tau)c(\tau) p_c(\tau)}{2\gamma p_c(\tau)^{3/2}}, \\
\dot{p_c}&=&\frac{2 b(\tau) \sqrt{p_c(\tau)}}{\gamma },\\
\dot{p_b}&=&\frac{b(\tau) p_b(\tau)+c(\tau) p_c(\tau)}{\gamma  \sqrt{pc(\tau)}}.
\eqn

 \begin{figure*}[htbp]
 \resizebox{\linewidth}{!}{\begin{tabular}{cc}
 \includegraphics[height=4.cm]{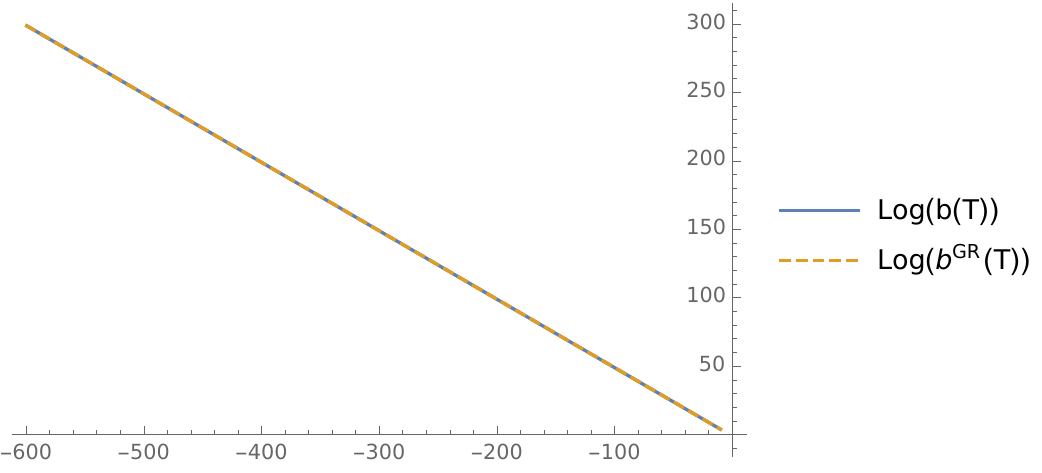}&
\includegraphics[height=4.cm]{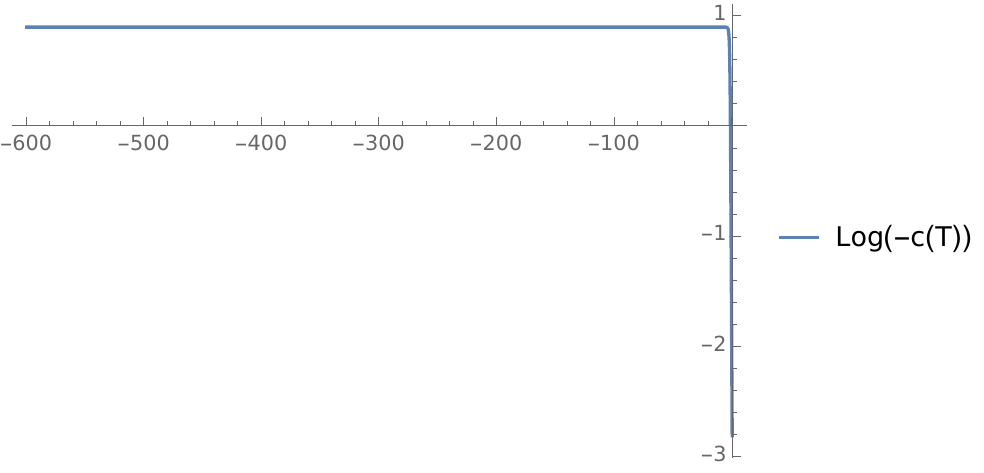}\\
 (a) & (b) \\[6pt]
\\
\includegraphics[height=4cm]{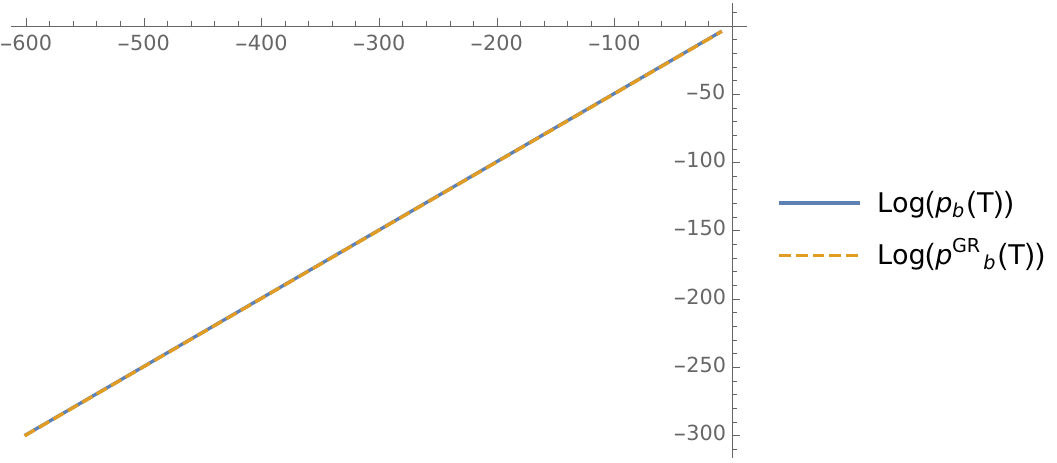}&
\includegraphics[height=4cm]{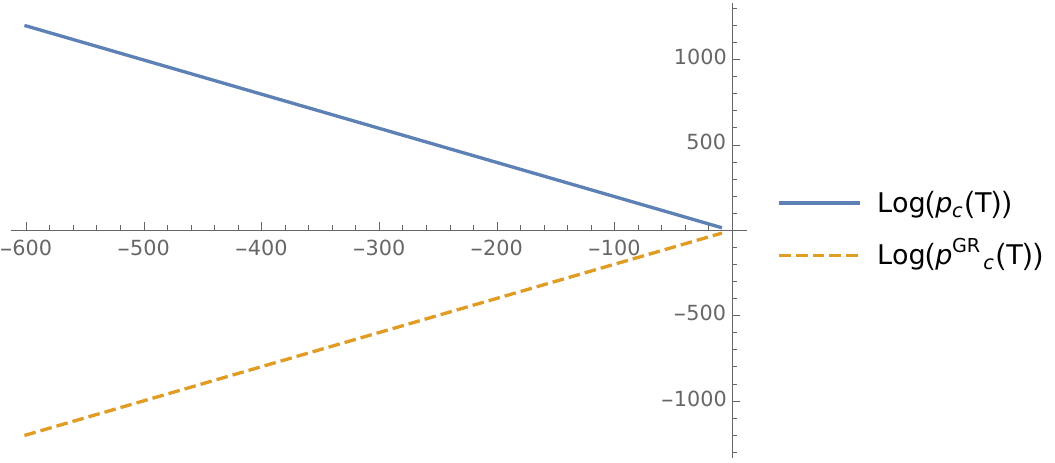}\\
(c) & (d) \\[6pt]
\end{tabular}}
\caption{Plots of  the four physical variables $\left(b, c, p_b, p_c\right)$ in the region $T \in (-600, -10)$, far from the transition surface. The mass parameter $m$ is chosen as  $m/\ell_{pl}=1$, for which we have  $T_{\cal{T}} \simeq -0.946567$ and $T_H \approx0.693$. The initial time is chosen at $T_i = 0.3$. }
\lb{fig7}
\end{figure*}

 \begin{figure*}[htbp]
 \resizebox{\linewidth}{!}{\begin{tabular}{cc}
 \includegraphics[height=4.cm]{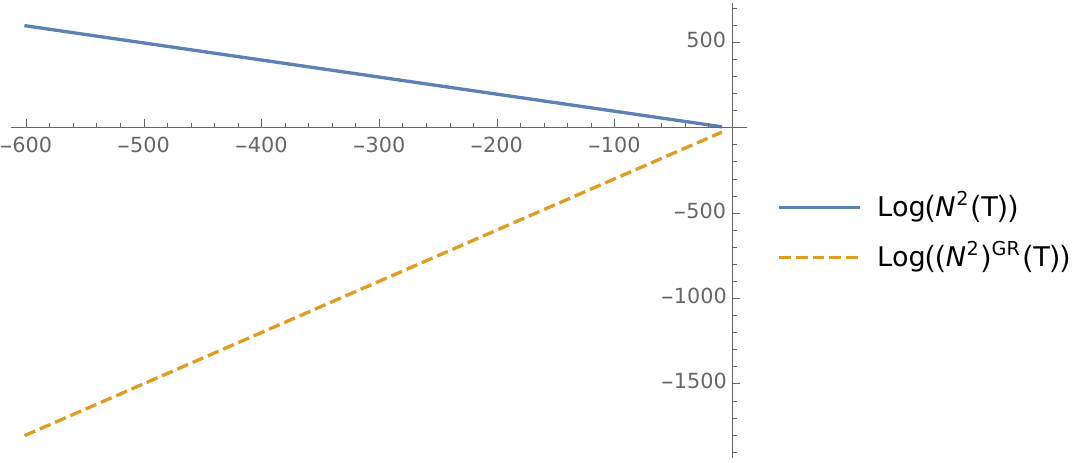}&
\includegraphics[height=4.cm]{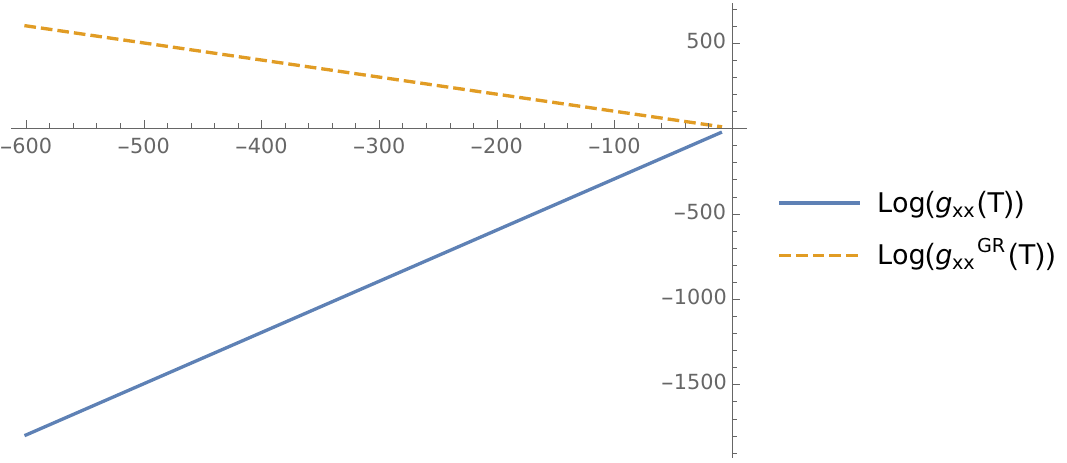}\\
 (a) & (b) \\[6pt]
\\
\includegraphics[height=4cm]{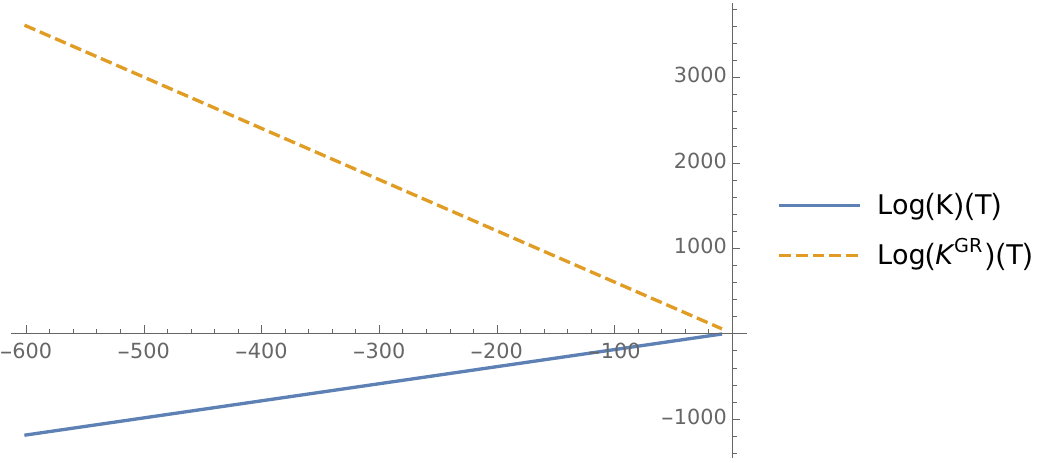}&
\includegraphics[height=4cm]{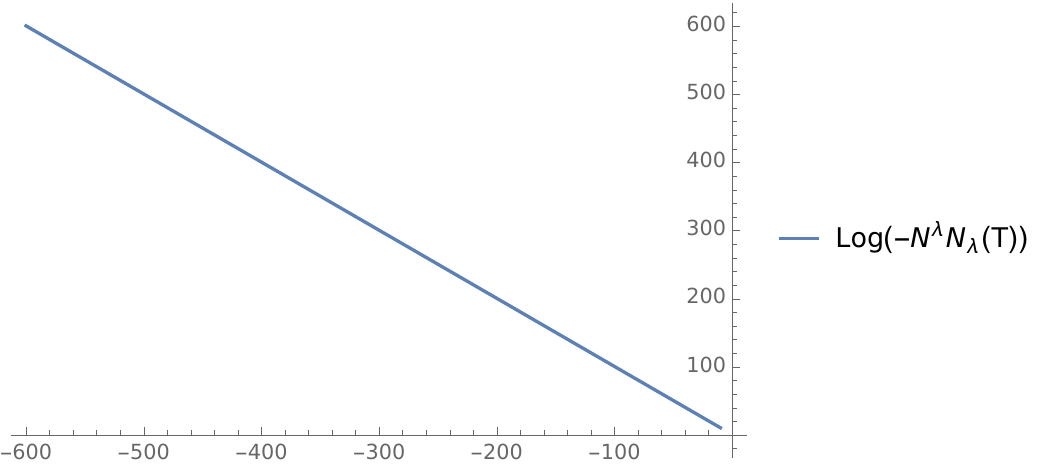} \\
\\
(c) & (d) \\[6pt]
\\
\includegraphics[height=4cm]{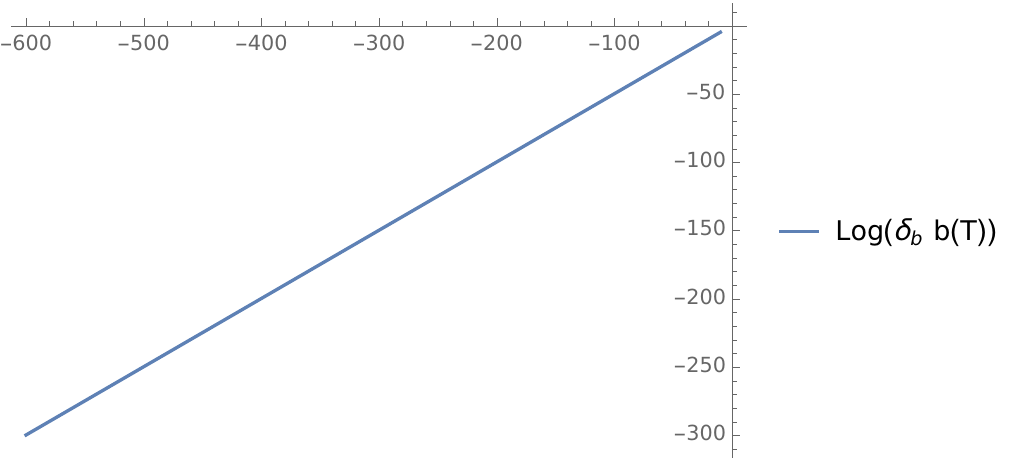}&
\includegraphics[height=4cm]{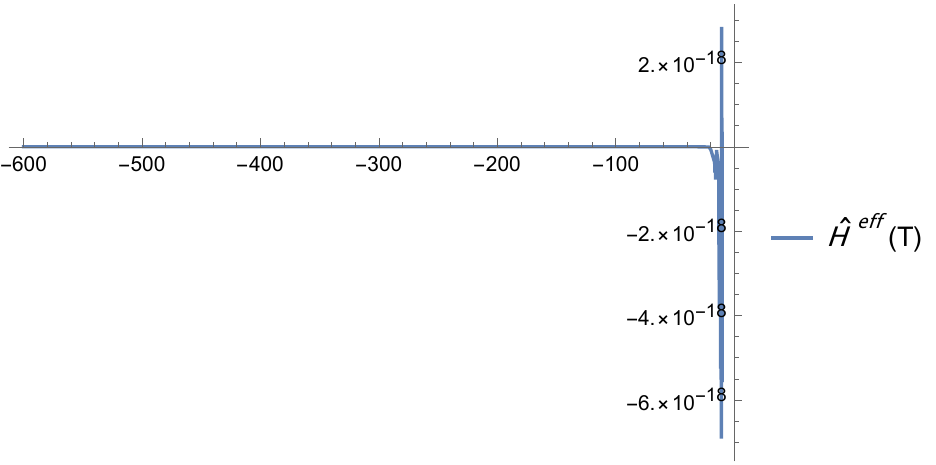}\\
(e) & (f)

 \end{tabular}}
\caption{Plots of  the lapse function $N^2$, the metric component $g_{xx}$, the Kretschmann scalar $K$, the norm $\left(-N^{\lambda}N_{\lambda}\right)$, the quantities 
$\delta_b b$ and $\hat H^{\text{eff}}$, 
together with their classical counterparts  in the range $T\in (-600,-10)$. The mass parameter $m$ is chosen as  $m/\ell_{pl}=1$, for which we have  $T_{\cal{T}} \simeq -0.946567$ and $T_H \approx0.693$. The initial time is chosen at $T_i = 0.3$.
}
\lb{fig8}
\end{figure*} 

\begin{figure*}[htbp]
 \resizebox{\linewidth}{!}{\begin{tabular}{cc}
 \includegraphics[height=4.cm]{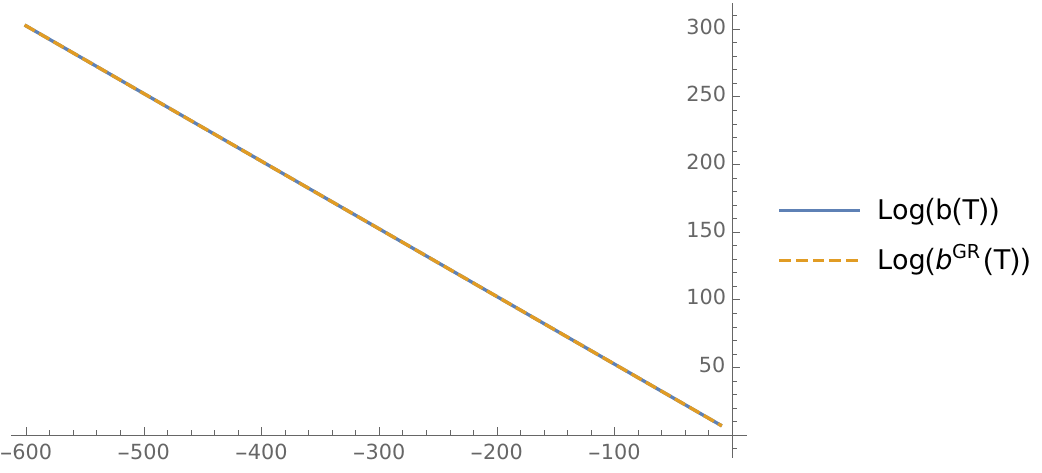}&
\includegraphics[height=4.cm]{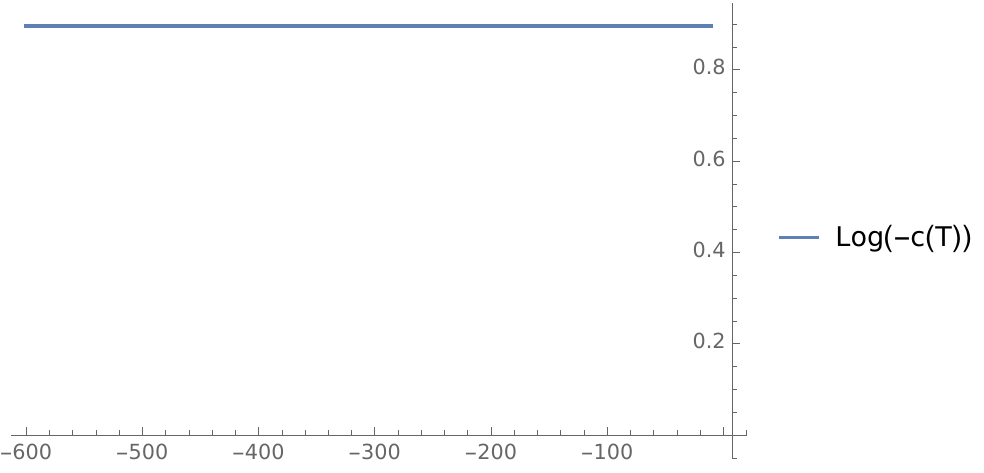}\\
 (a) & (b) \\[6pt]
\\
\includegraphics[height=4cm]{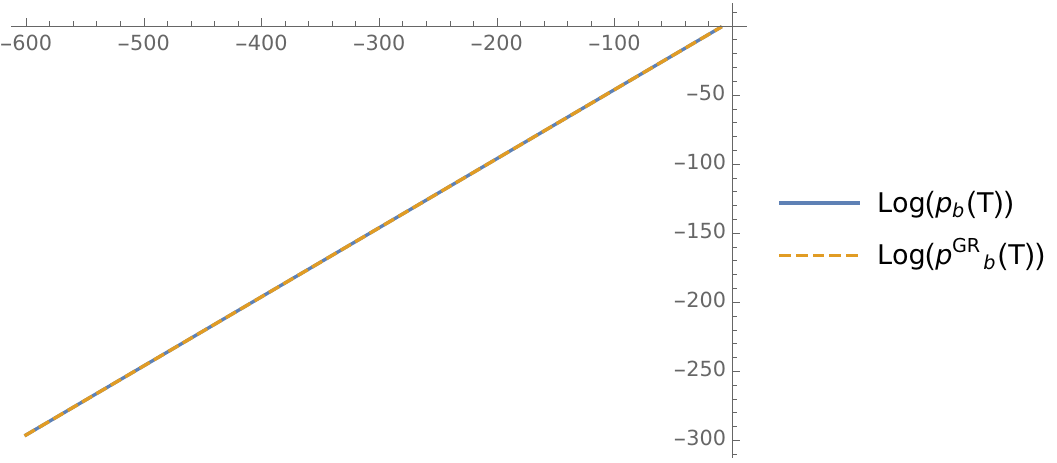} &
\includegraphics[height=4cm]{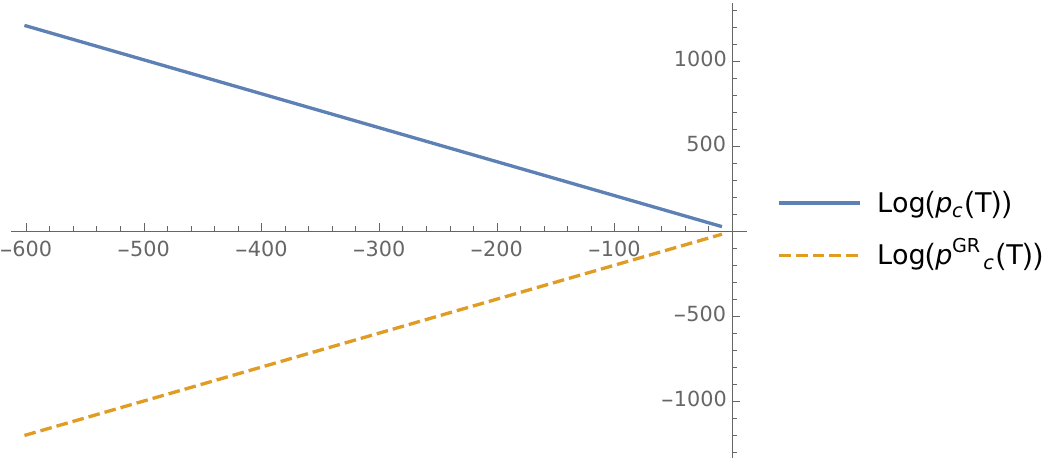}\\
(c) & (d) \\[6pt]
\end{tabular}}
\caption{Plots of  the four physical variables $\left(b, c, p_b, p_c\right)$ for $T \ll T_{\cal{T}}$. The mass parameter $m$ is chosen as  $m/\ell_{pl}=10^3$, for which we have  $T_{\cal{T}} \simeq 2.51204$ and $T_H \approx 7.6009$. The initial time is chosen at $T_i = 7$.}
\lb{fig9}
\end{figure*}

 \begin{figure*}[htbp]
 \resizebox{\linewidth}{!}{\begin{tabular}{cc}
 \includegraphics[height=4.cm]{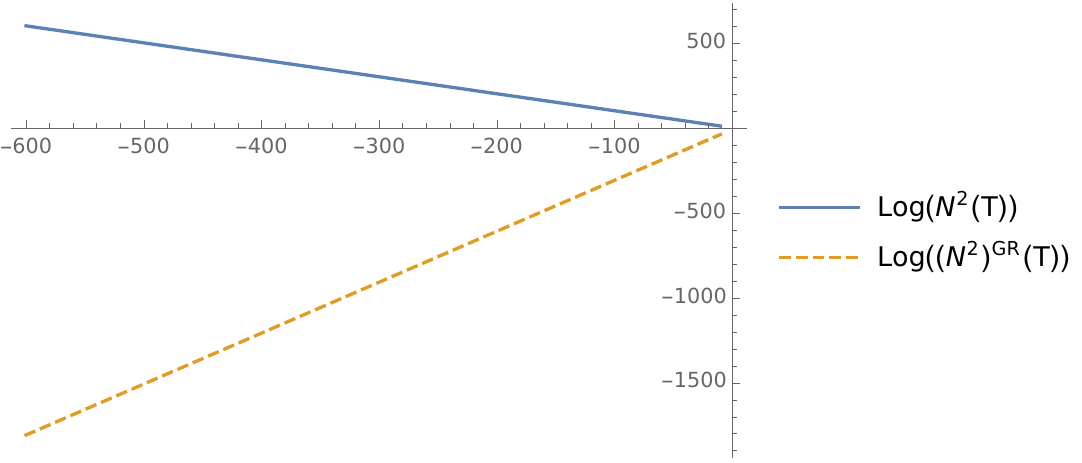}&
\includegraphics[height=4.cm]{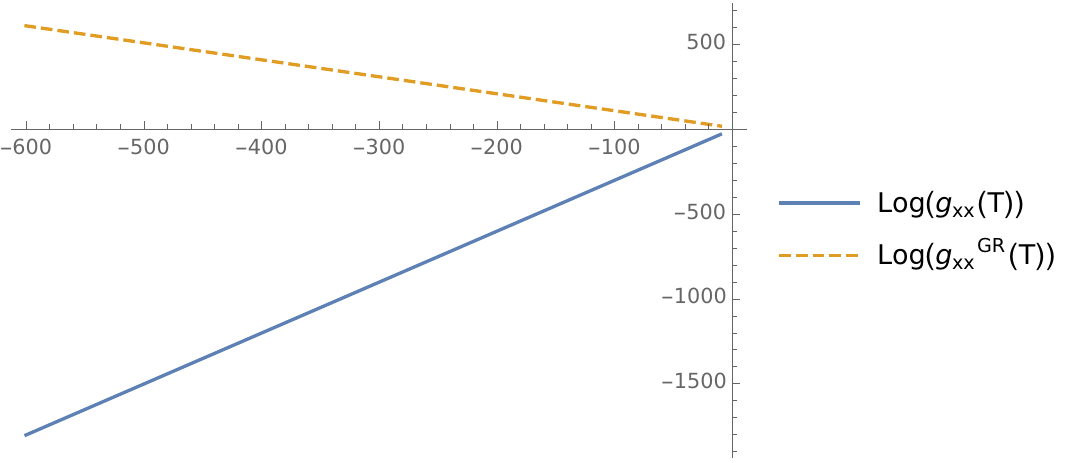}\\
 (a) & (b) \\[6pt]
\\
\includegraphics[height=4cm]{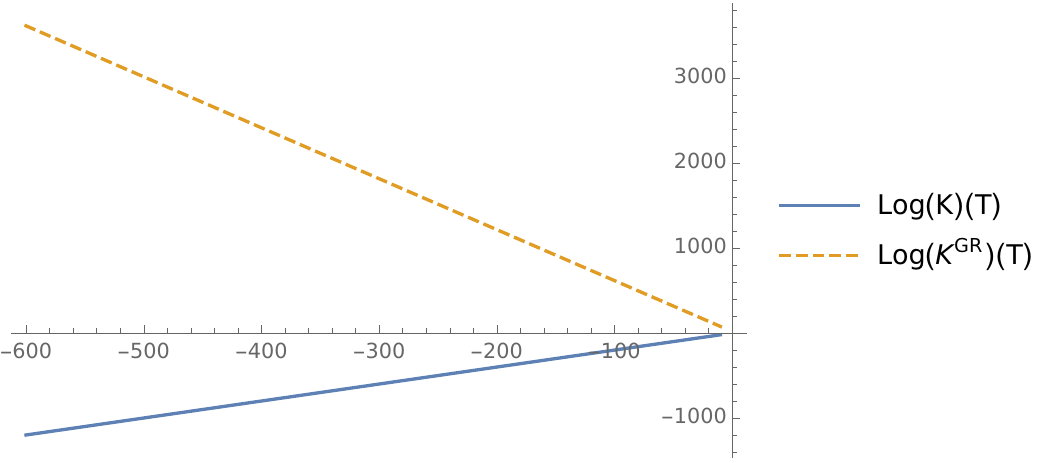}&
\includegraphics[height=4cm]{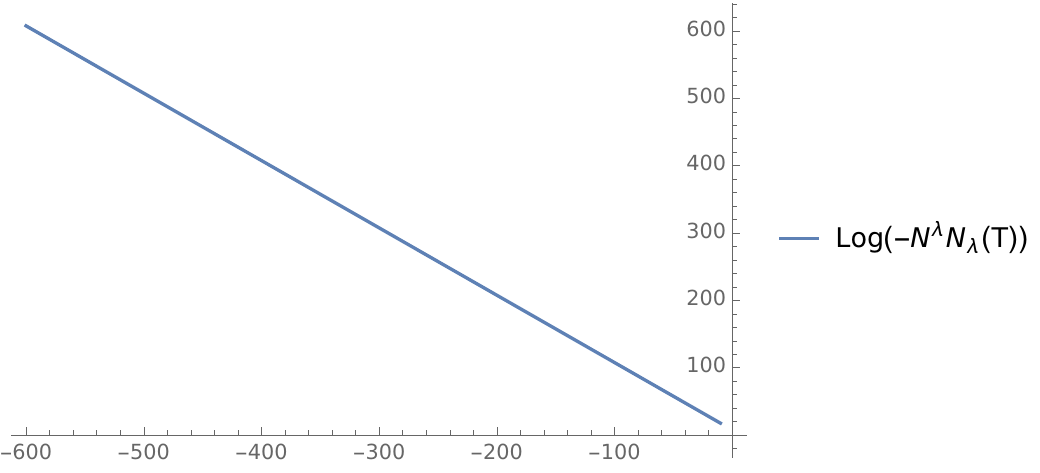}\\
(c) & (d) \\[6pt]
\\
\includegraphics[height=4cm]{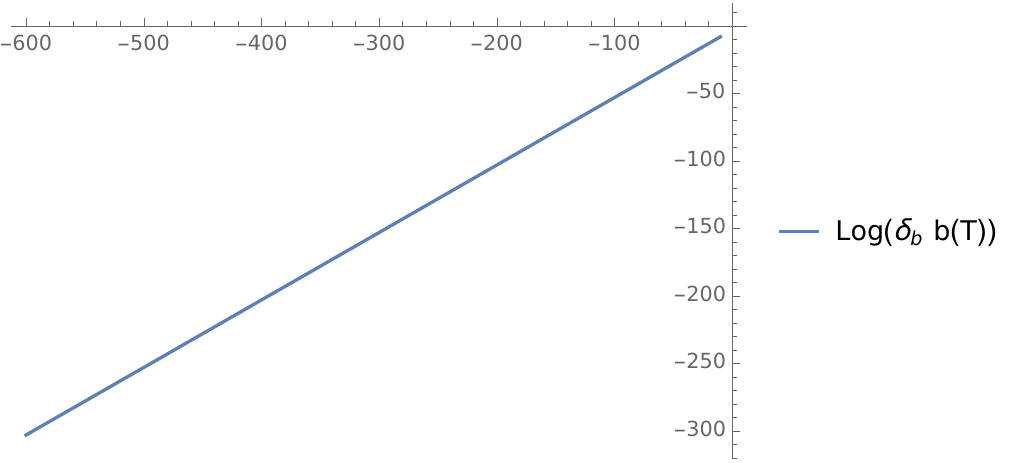}&
\includegraphics[height=4cm]{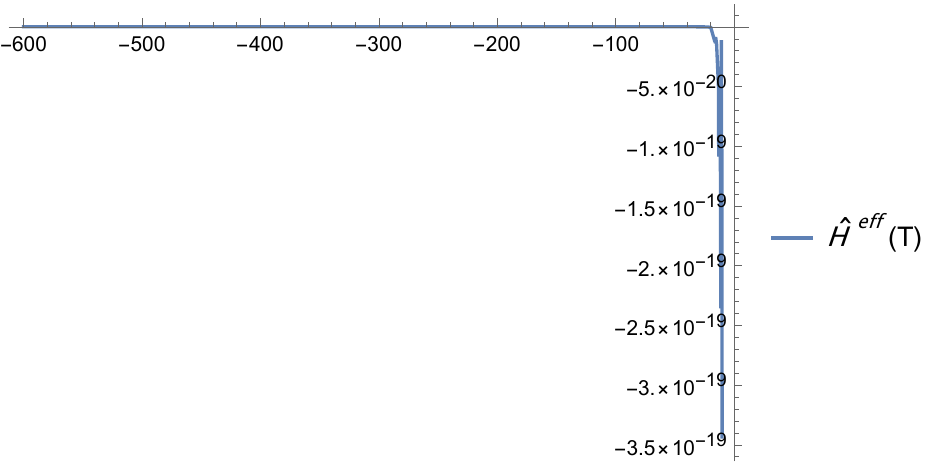}\\
(e) & (f)
	\end{tabular}}
\caption{Plots of  the lapse function $N^2$, the metric component $g_{xx}$, the Kretschmann scalar $K$, the norm $\left(-N^{\lambda}N_{\lambda}\right)$, the quantities 
$\delta_b b$ and $\hat H^{\text{eff}}$, 
together with their classical counterparts  in the range $T\in (-600,-10)$. The mass parameter $m$ is chosen as  $m/\ell_{pl}=10^3$, for which we have  $T_{\cal{T}} \simeq 2.51204$ and $T_H \approx 7.6009$. The initial time is chosen at $T_i = 7$.
}
\lb{fig10}
\end{figure*} 

\begin{figure*}[htbp]
 \resizebox{\linewidth}{!}{\begin{tabular}{cc}
 \includegraphics[height=4.cm]{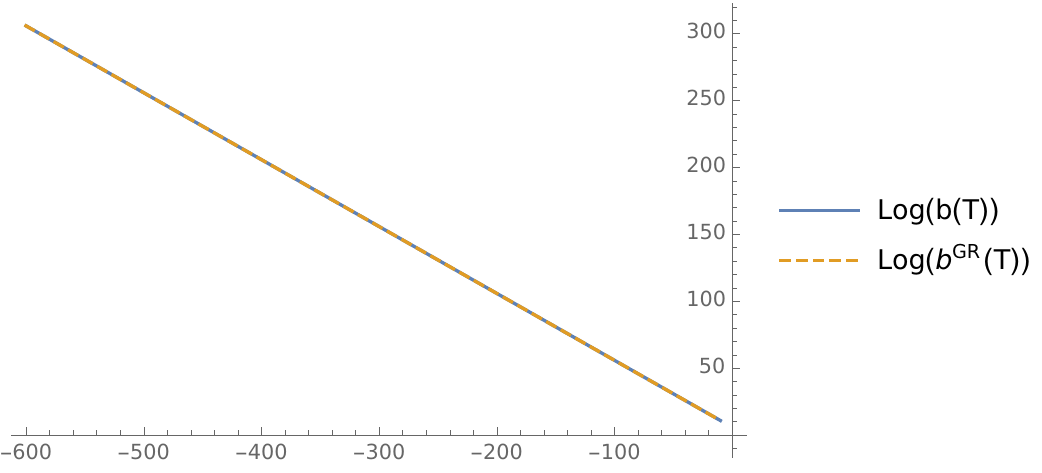}&
\includegraphics[height=4.cm]{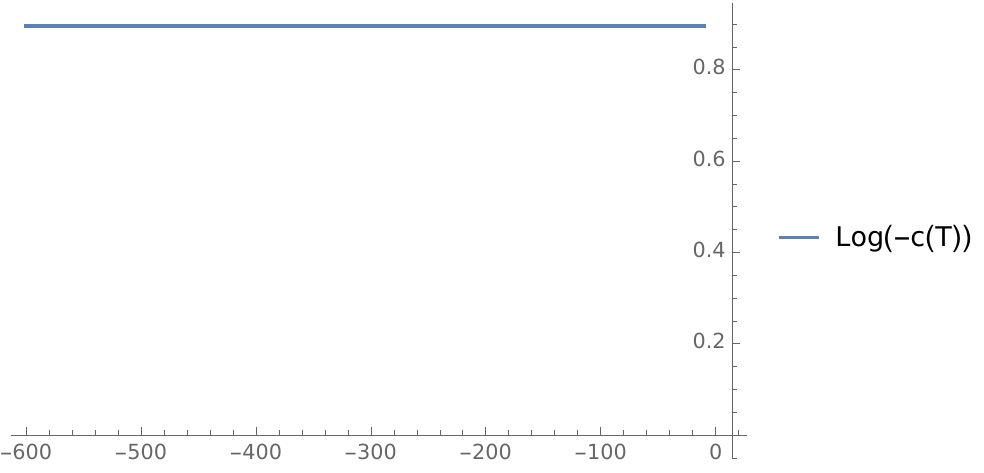}\\
 (a) & (b) \\[6pt]
\\
\includegraphics[height=4cm]{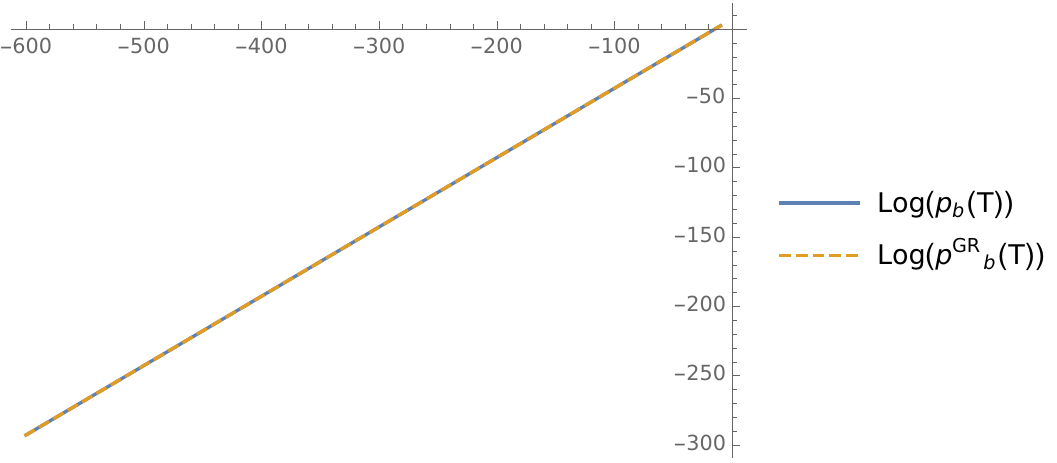}&
\includegraphics[height=4cm]{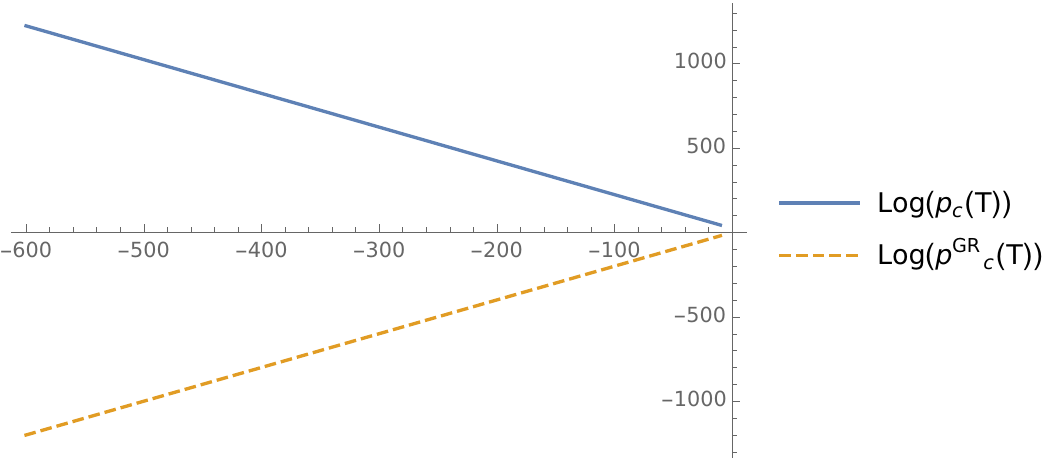}\\
(c) & (d) \\[6pt]
\end{tabular}}
\caption{Plots of  the four physical variables $\left(b, c, p_b, p_c\right)$ for $T \ll T_{\cal{T}}$. The mass parameter $m$ is chosen as  $m/\ell_{pl}=10^6$, for which we have  $T_{\cal{T}} \simeq 5.96686$ and $T_H \approx 14.5087$. The initial time is chosen at $T_i = 14$.}
\lb{fig11}
\end{figure*} 

 \begin{figure*}[htbp]
 \resizebox{\linewidth}{!}{\begin{tabular}{cc}
 \includegraphics[height=4.cm]{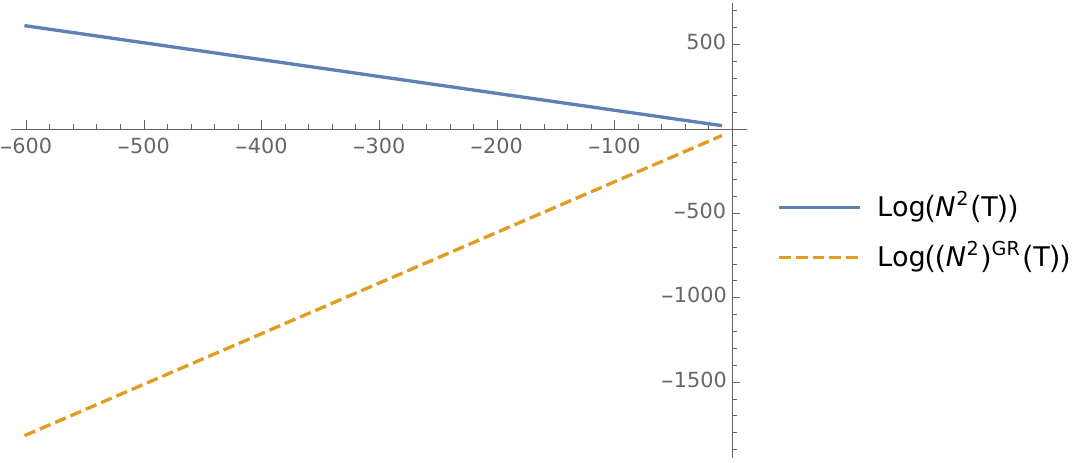}&
\includegraphics[height=4.cm]{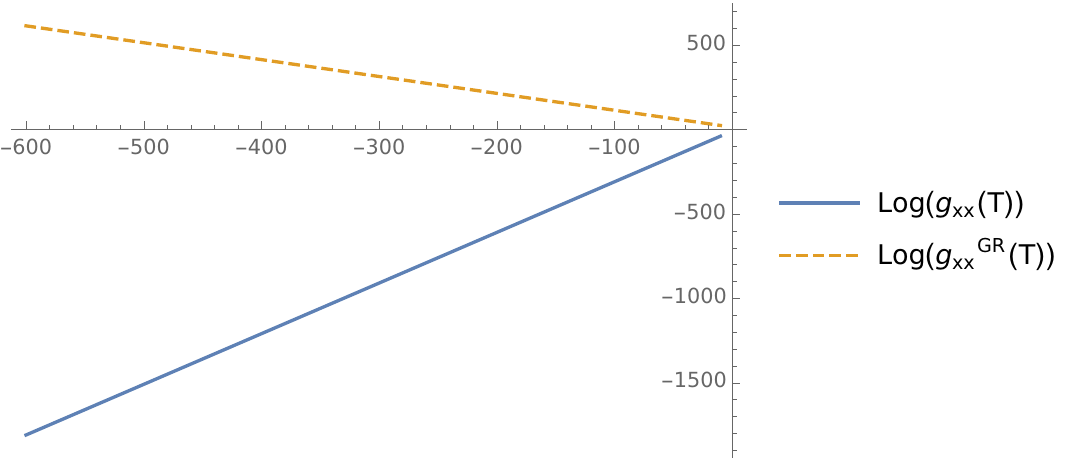}\\
 (a) & (b) \\[6pt]
\\
\includegraphics[height=4cm]{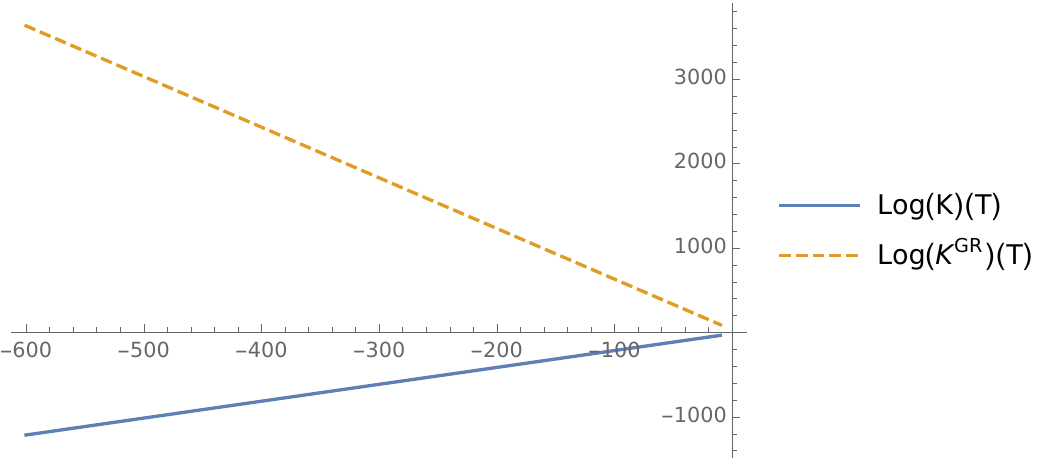}&
\includegraphics[height=4cm]{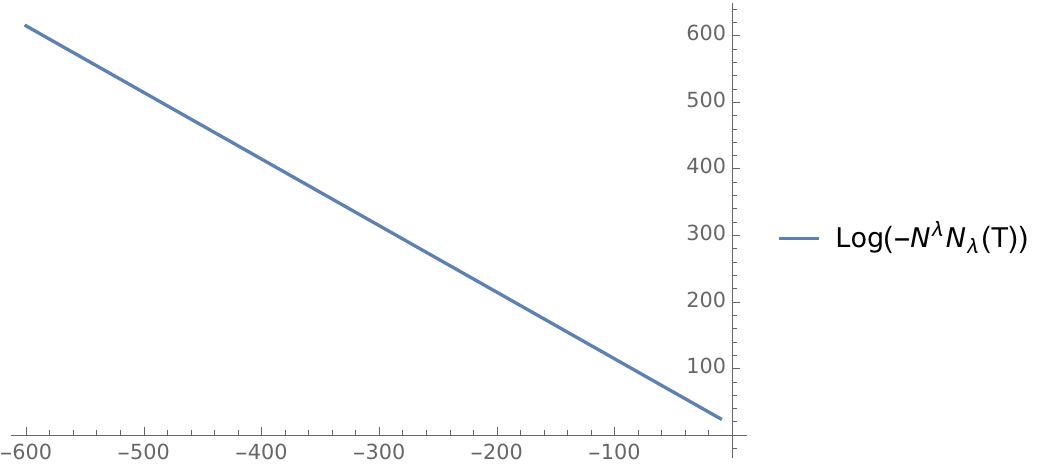}\\
(c) & (d) \\[6pt]
\\
\includegraphics[height=4cm]{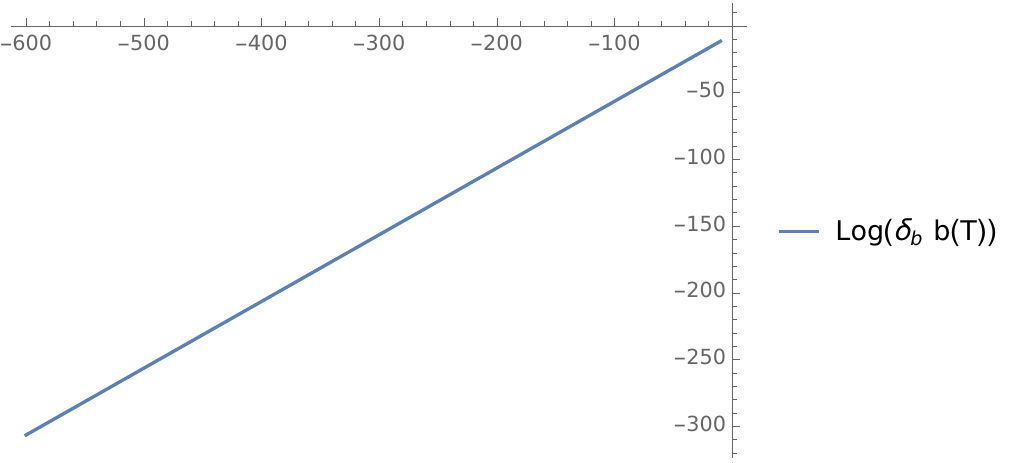}&
\includegraphics[height=4cm]{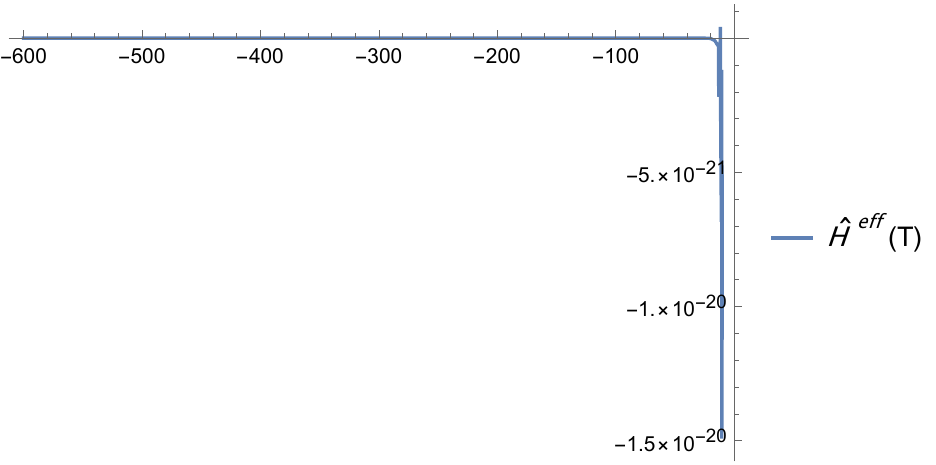}\\
(e) & (f)
	\end{tabular}}
\caption{Plots of  the lapse function $N^2$, the metric component $g_{xx}$, the Kretschmann scalar $K$, the norm $\left(-N^{\lambda}N_{\lambda}\right)$, the quantities 
$\delta_b b$ and $\hat H^{\text{eff}}$, 
together with their classical counterparts  in the range $T\in (-600,-10)$. The mass parameter $m$ is chosen as  $m/\ell_{pl}=10^6$, for which we have  $T_{\cal{T}} \simeq 5.96686$ and $T_H \approx 14.5087$. The initial time is chosen at $T_i = 14$.
}
\lb{fig12}
\end{figure*} 

  \begin{figure*}[htbp]
 \resizebox{\linewidth}{!}{\begin{tabular}{cc}
 \includegraphics[height=4.cm]{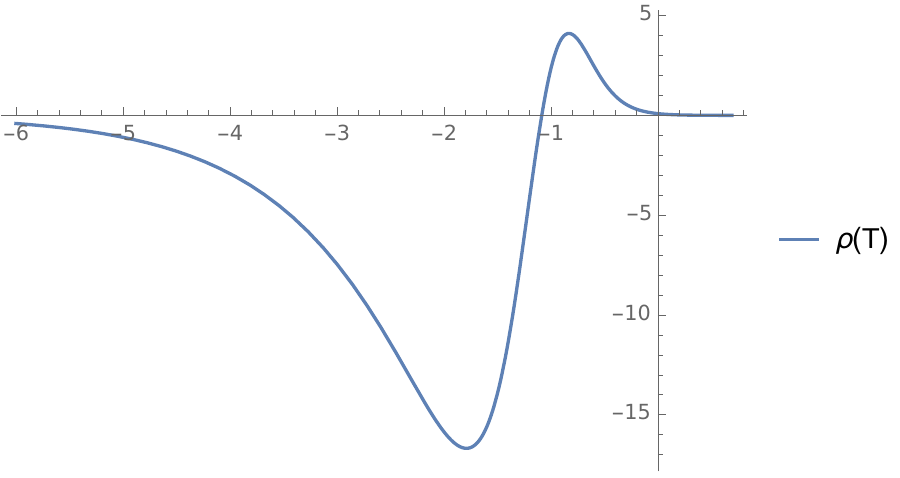}&
\includegraphics[height=4.cm]{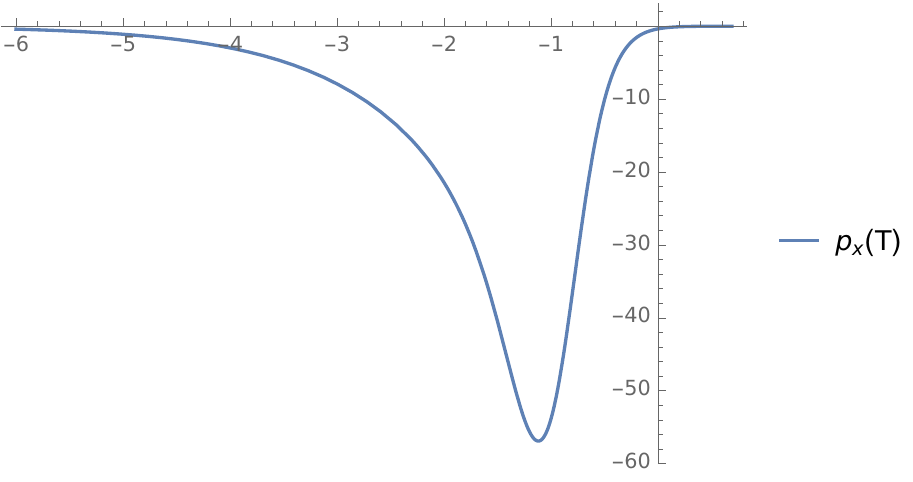}\\
 (a) & (b) \\[6pt]
\\
\includegraphics[height=4cm]{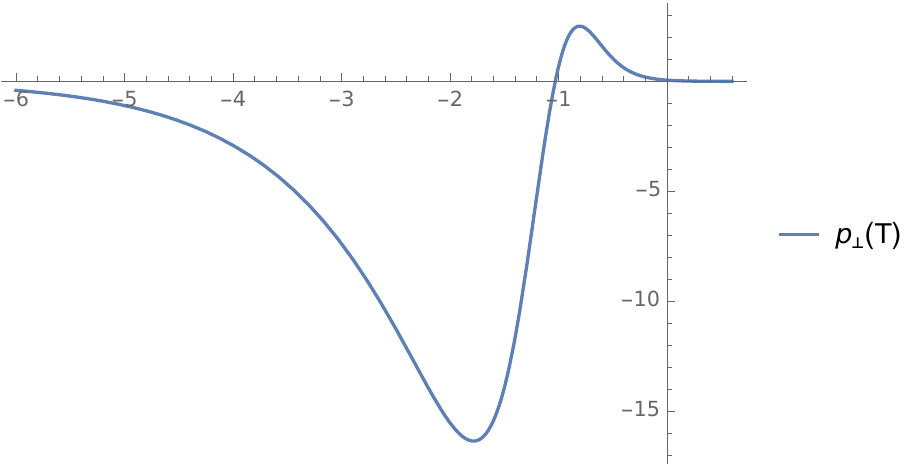}&
\includegraphics[height=4cm]{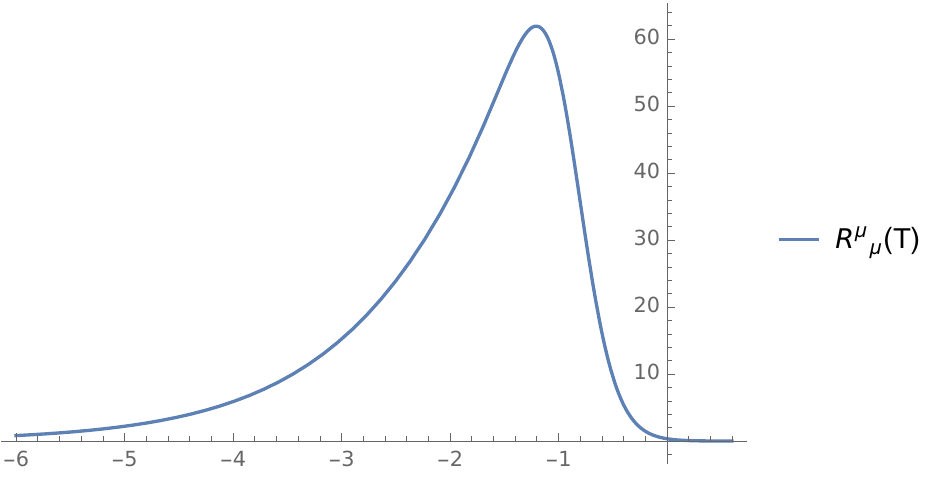} \\
\\
(c) & (d) \\[6pt]
\\
\includegraphics[height=4cm]{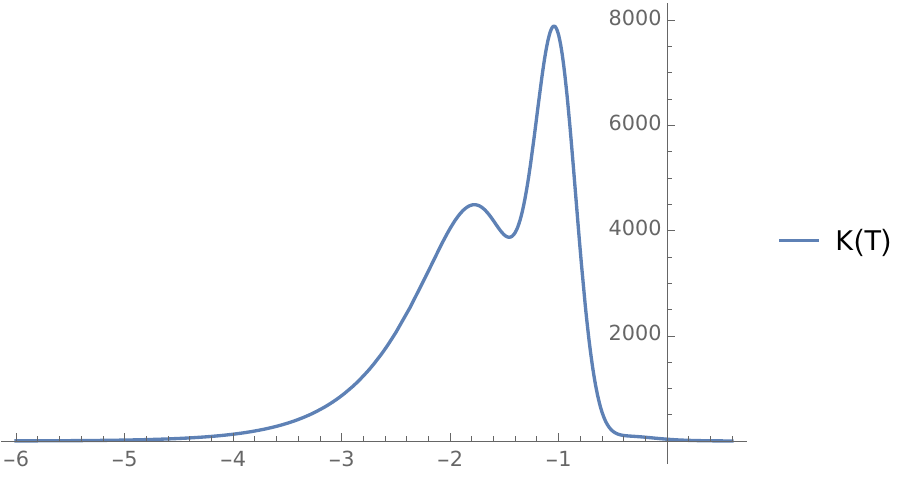}&
\includegraphics[height=4cm]{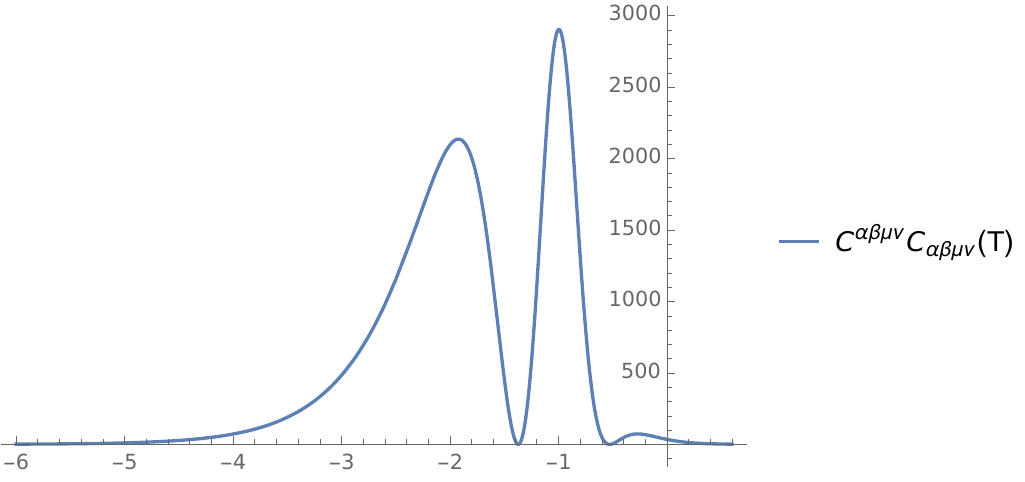}\\
(e) & (f)
	\end{tabular}}
\caption{Plots of the quantities $\rho,\; p_x, \;p_{\bot}, \;
R^{\mu}_{\;\;\mu}, \; K,  \; C_{\alpha\beta\mu\nu} C^{\alpha\beta\mu\nu}$ near the transition surface $T = T_{\cal{T}}$. The mass parameter $m$ is chosen as  $m/\ell_{pl}=1$, for which we have  $T_{\cal{T}} \simeq -0.946567$ and $T_H \approx0.693$. The initial time is chosen at $T_i = 0.3$.}
\lb{fig16}
\end{figure*} 

 \begin{figure*}[htbp]
 \resizebox{\linewidth}{!}{\begin{tabular}{cc}
 \includegraphics[height=4.cm]{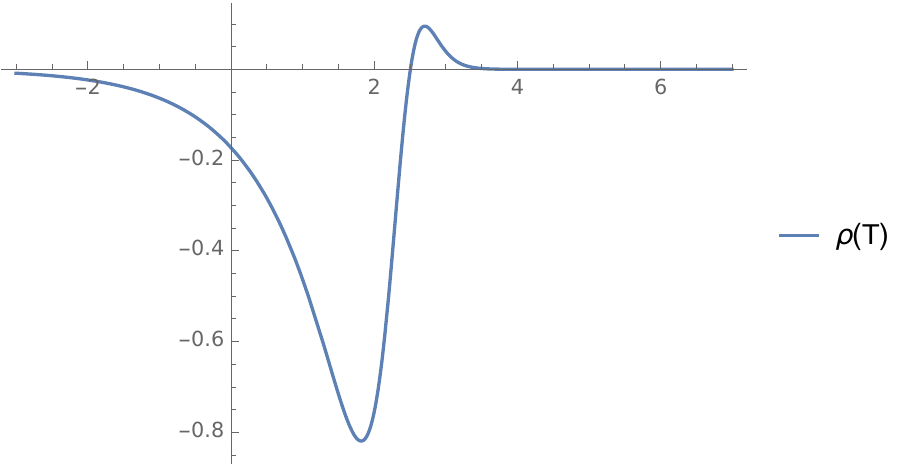}&
\includegraphics[height=4.cm]{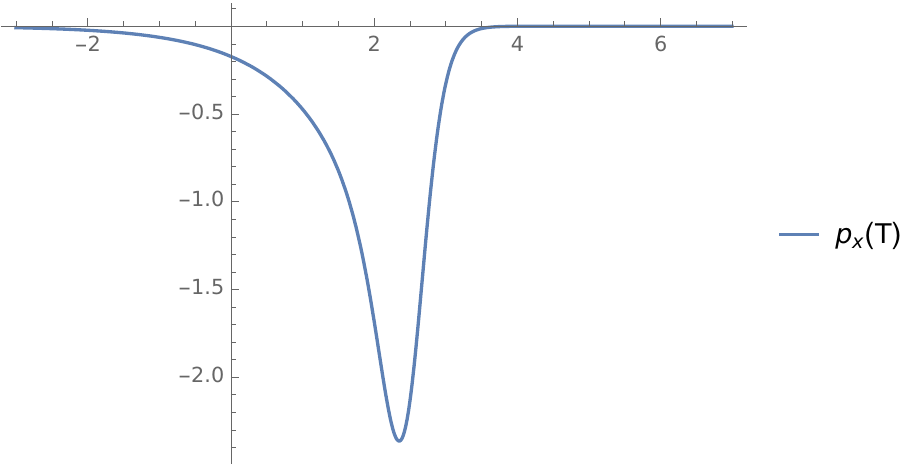}\\
 (a) & (b) \\[6pt]
\\
\includegraphics[height=4cm]{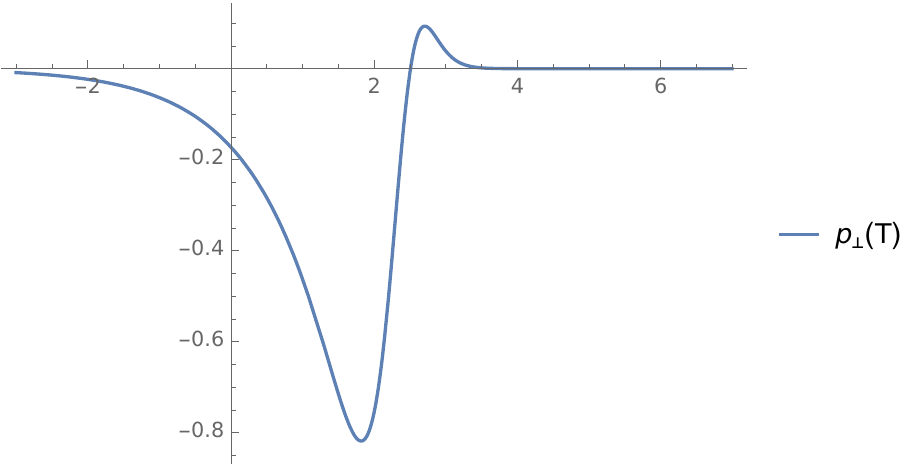}&
\includegraphics[height=4cm]{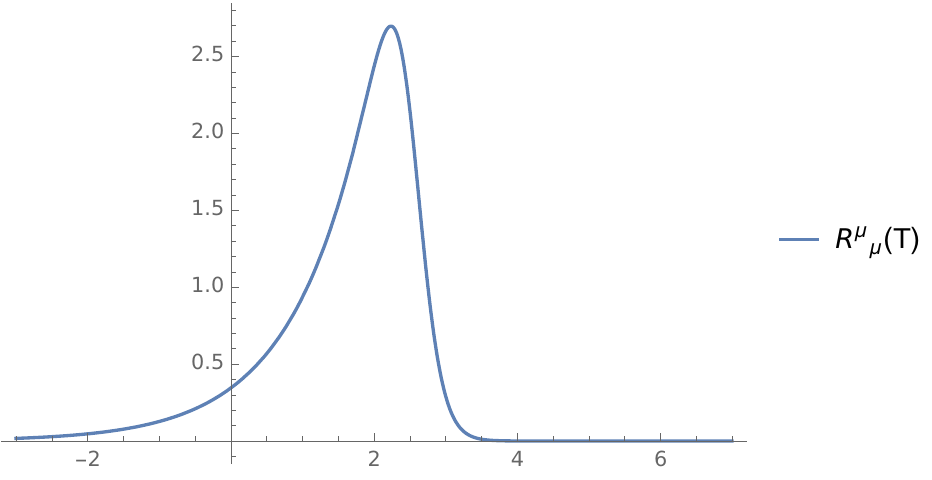} \\
\\
(c) & (d) \\[6pt]
\\
\includegraphics[height=4cm]{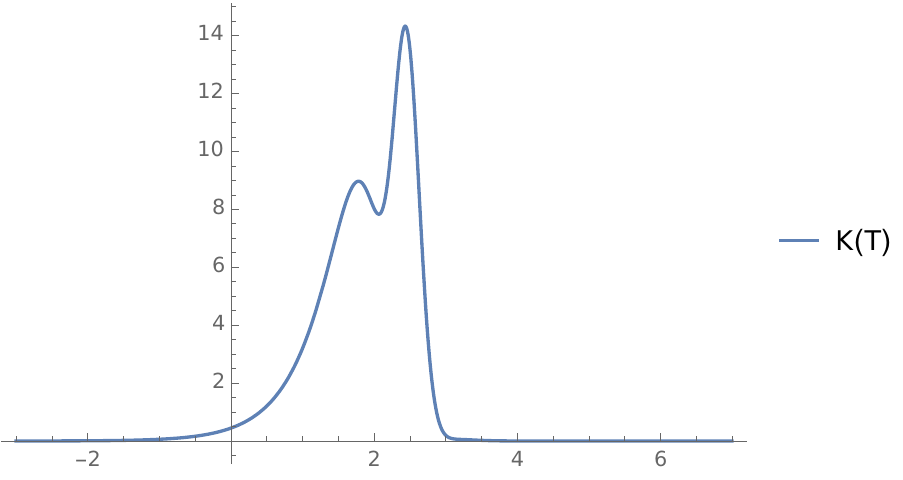}&
\includegraphics[height=4cm]{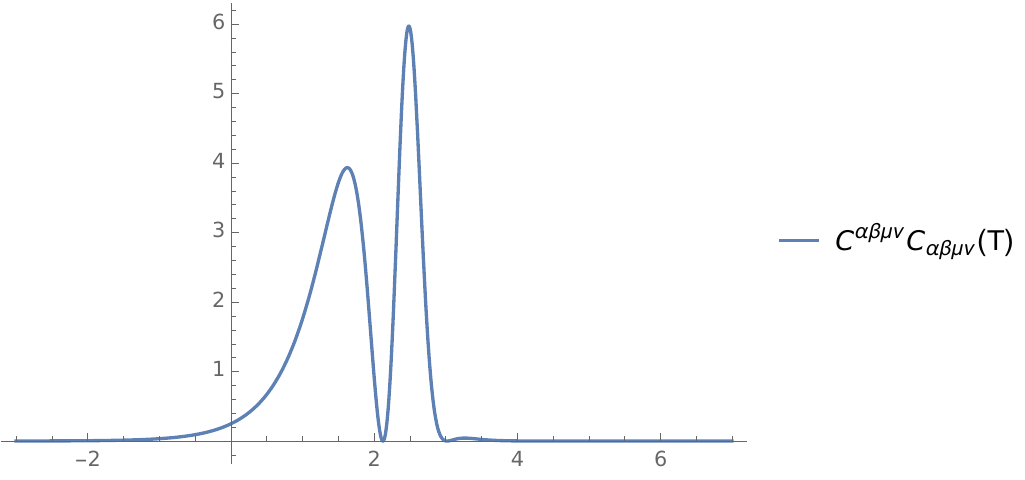}\\
(e) & (f)
	\end{tabular}}
\caption{Plots of the quantities $\rho,\; p_x, \;p_{\bot}, \;
R^{\mu}_{\;\;\mu}, \; K,  \; C_{\alpha\beta\mu\nu} C^{\alpha\beta\mu\nu}$ near the transition surface $T = T_{\cal{T}}$. The mass parameter $m$ is chosen as  $m/\ell_{pl}=10^3$, for which we have  $T_{\cal{T}} \simeq 2.51204$ and $T_H \approx 7.6009$. The initial time is chosen at $T_i = 7$.}
\lb{fig17}
\end{figure*} 

 \begin{figure*}[htbp]
 \resizebox{\linewidth}{!}{\begin{tabular}{cc}
 \includegraphics[height=4.cm]{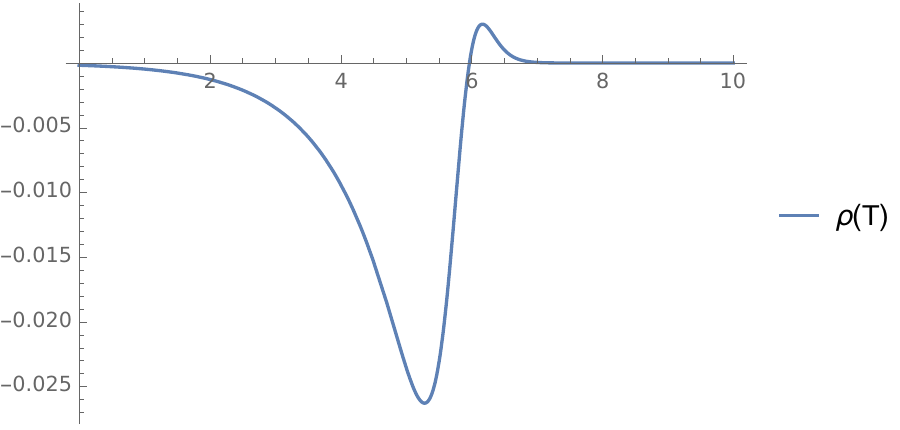}&
\includegraphics[height=4.cm]{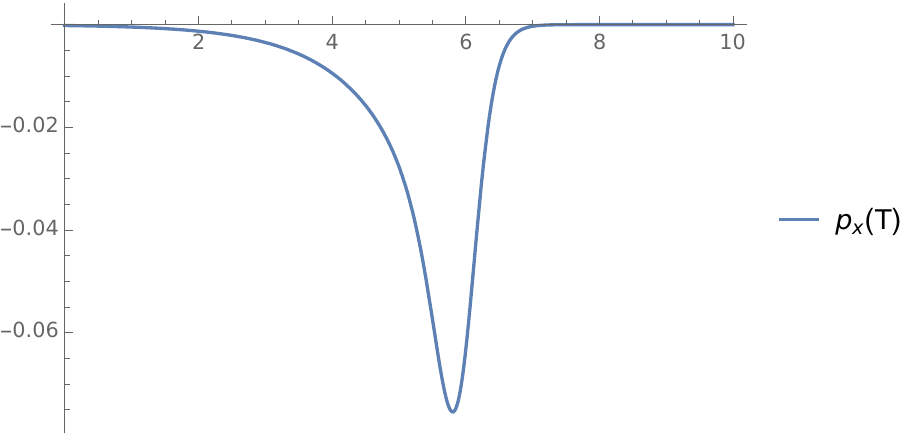}\\
 (a) & (b) \\[6pt]
\\
\includegraphics[height=4cm]{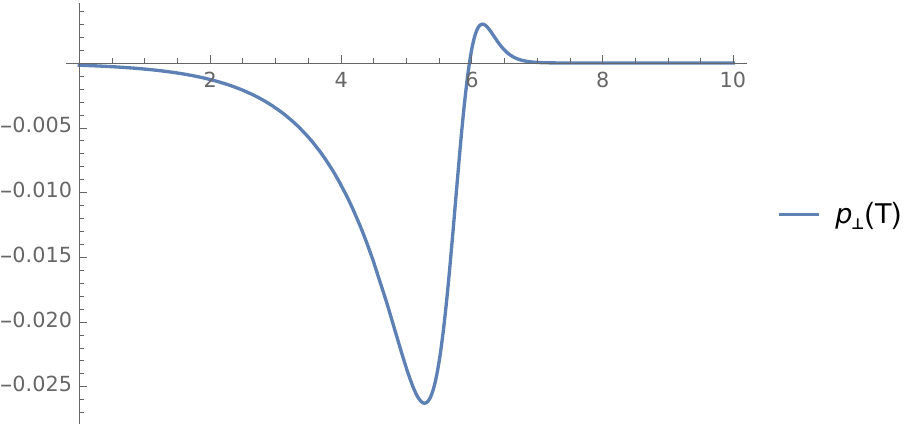}&
\includegraphics[height=4cm]{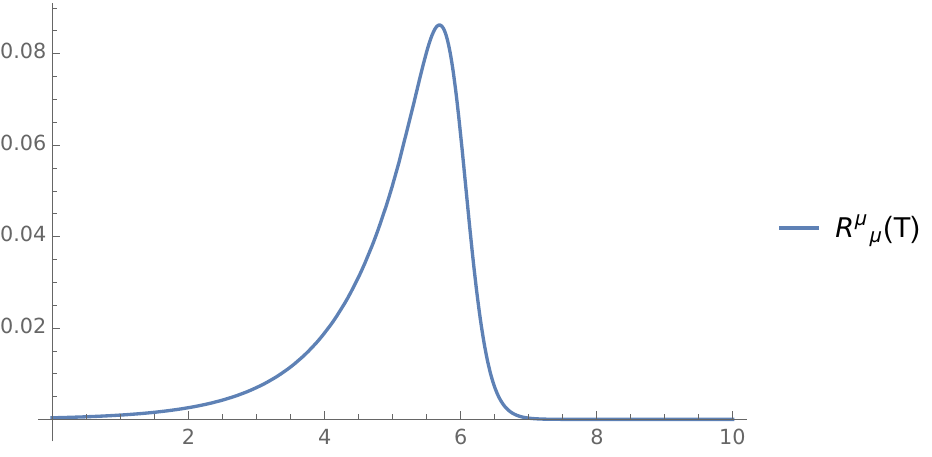} \\
\\
(c) & (d) \\[6pt]
\\
\includegraphics[height=4cm]{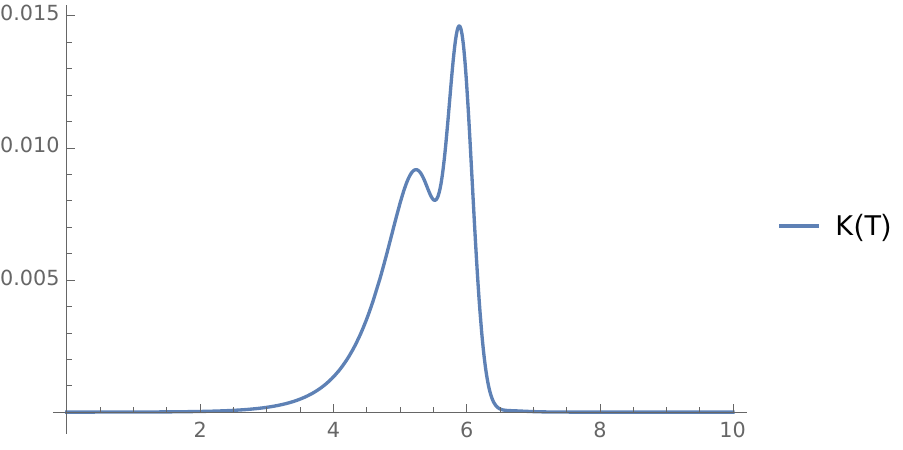}&
\includegraphics[height=4cm]{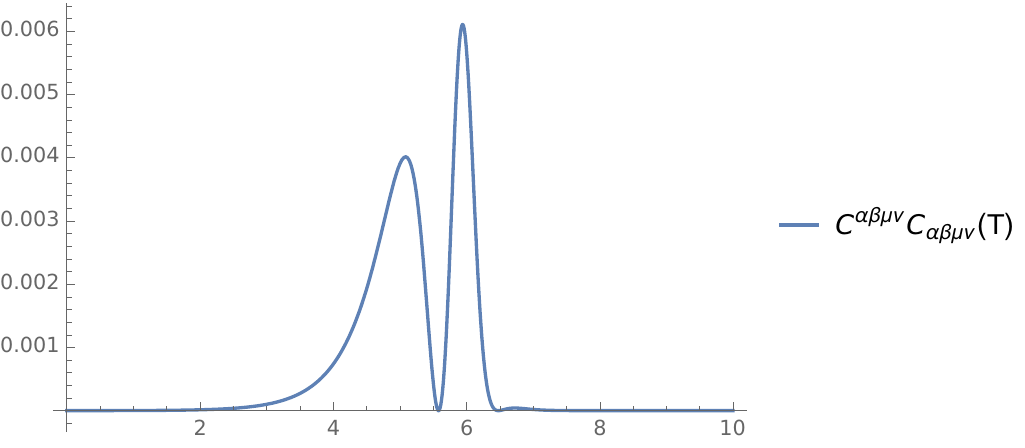}\\
(e) & (f)
	\end{tabular}}
\caption{Plots of the quantities $\rho,\; p_x, \;p_{\bot}, \;
R^{\mu}_{\;\;\mu}, \; K,  \; C_{\alpha\beta\mu\nu} C^{\alpha\beta\mu\nu}$ near the transition surface $T = T_{\cal{T}}$. The mass parameter $m$ is chosen as  $m/\ell_{pl}=10^6$, for which we have  $T_{\cal{T}} \simeq 5.96686$ and $T_H \approx 14.5087$. The initial time is chosen at $T_i = 14$.}
\lb{fig18}
\end{figure*}

\begin{figure*}[htbp]
 \resizebox{\linewidth}{!}{\begin{tabular}{cc}
  \includegraphics[height=4.cm]{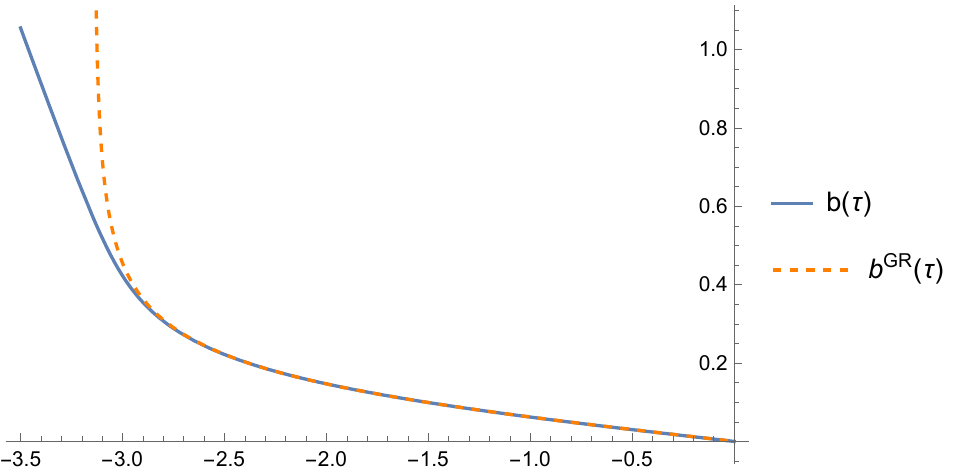}&
\includegraphics[height=4.cm]{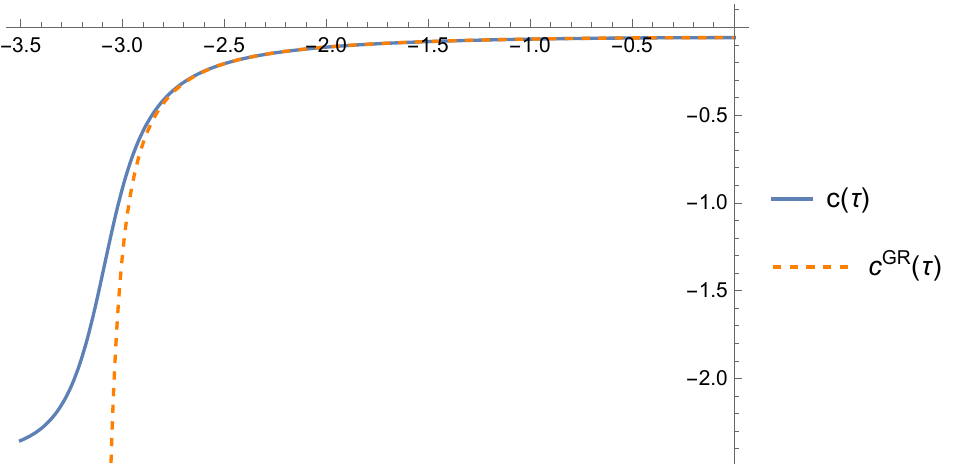}\\
 (a) & (b) \\[6pt]
\\
 \includegraphics[height=4.cm]{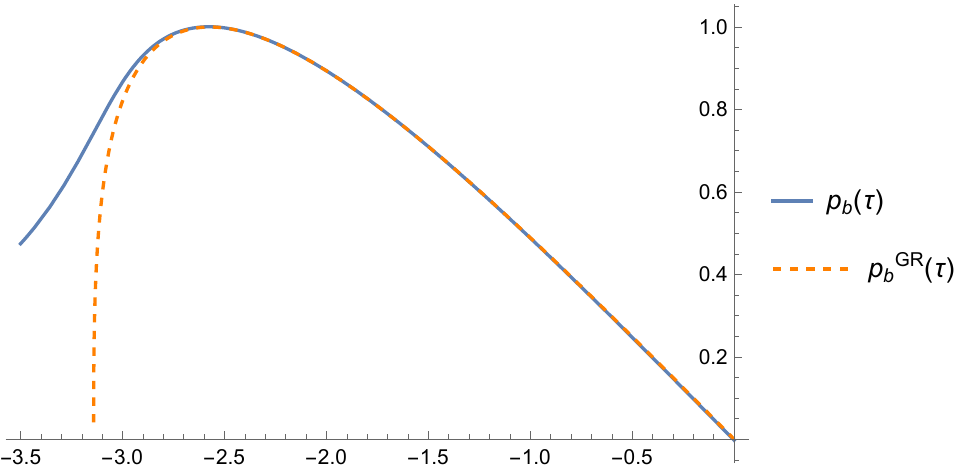}&
\includegraphics[height=4.cm]{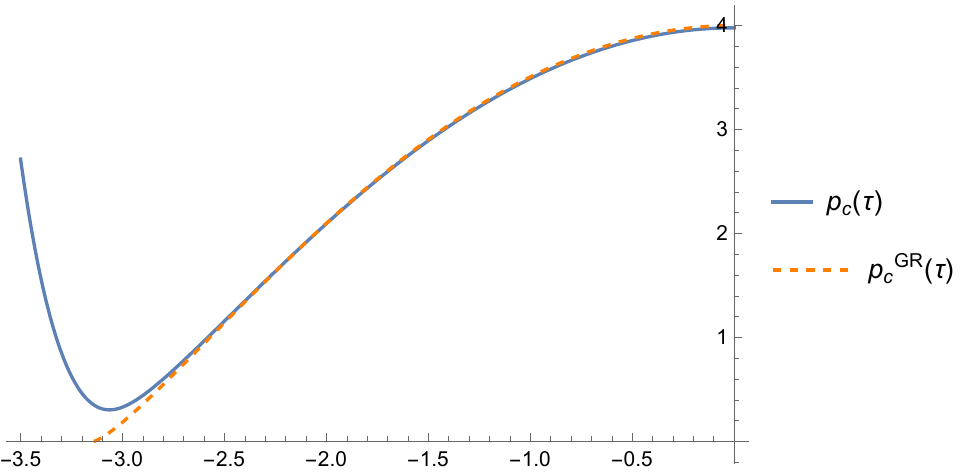}\\
 (c) & (d) \\[6pt]
\\
\includegraphics[height=4cm]{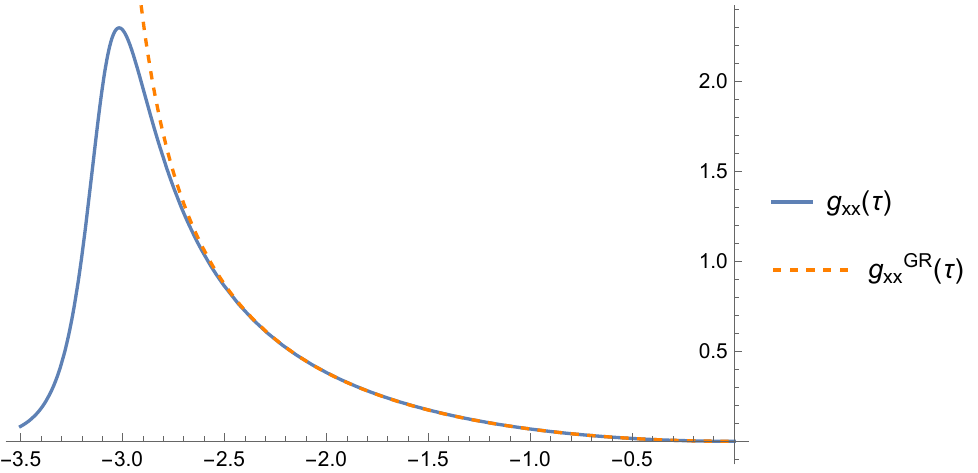}&
\includegraphics[height=4cm]{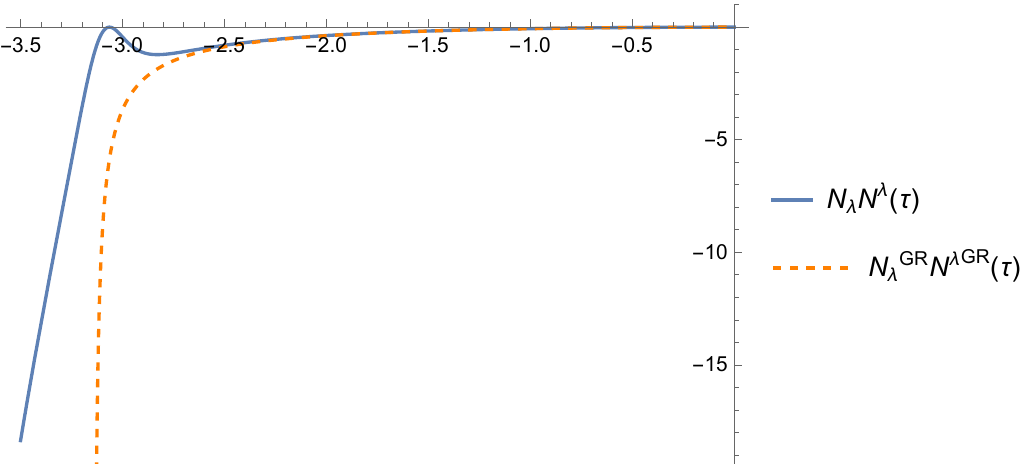}\\
(e) & (f) \\[6pt]
\\
\includegraphics[height=4cm]{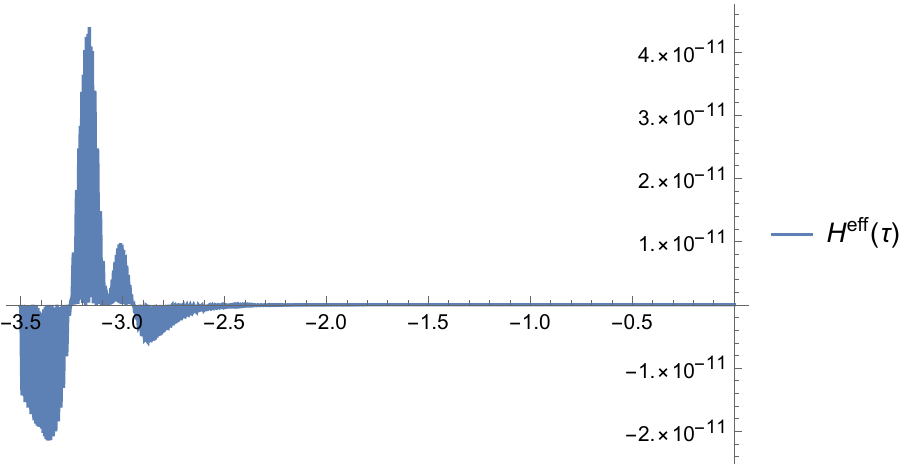}\\
(g)  \\[6pt]
\end{tabular}}
\caption{Plots of the physical quantities $b, \; c, \; p_b, \; p_c,\; g_{xx}$, \; $N^{\lambda}N_{\lambda}$ and $H^{\text{eff}}$ in terms of the proper time $\tau$  with  $m/\ell_{pl}=1$. Transition surface is located at $\tau_{\mathcal{T}}\approx -3.064$.}
\lb{fig1-new-N1}
\end{figure*}

 \begin{figure*}[htbp]
 \resizebox{\linewidth}{!}{\begin{tabular}{cc}
 \includegraphics[height=4.cm]{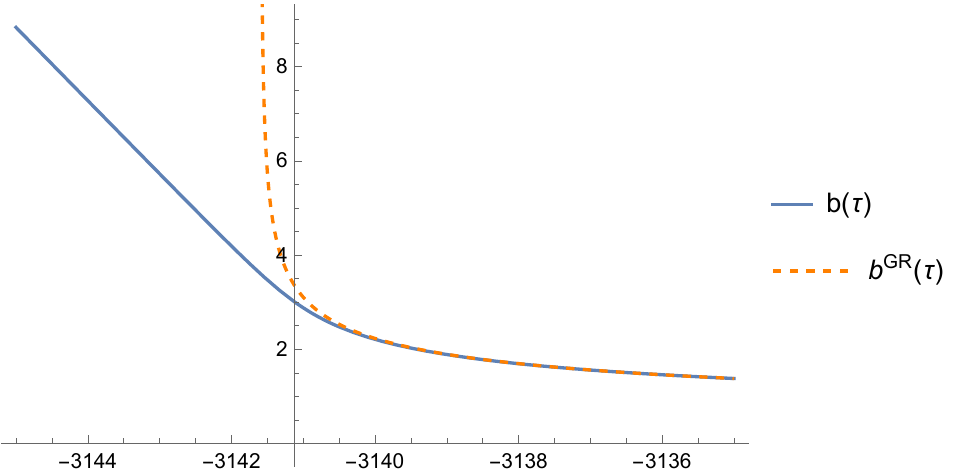}&
\includegraphics[height=4.cm]{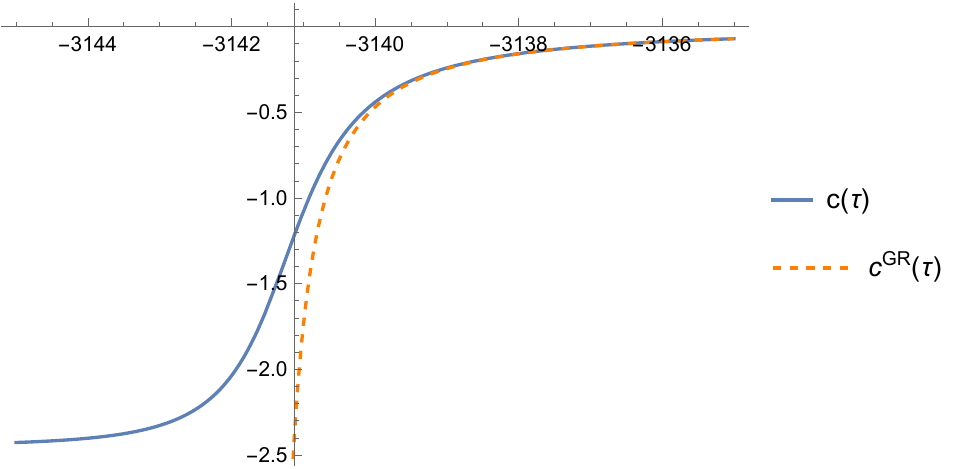}\\
 (a) & (b) \\[6pt]
\\
 \includegraphics[height=4.cm]{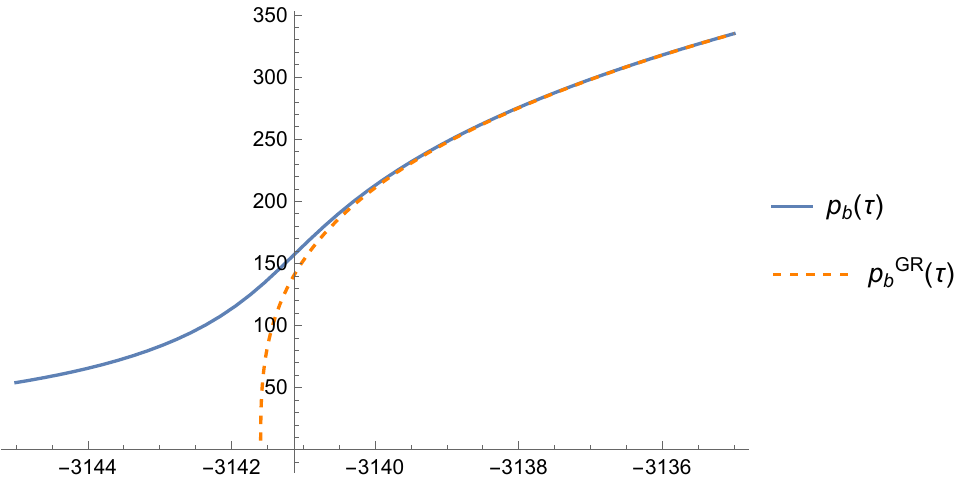}&
\includegraphics[height=4.cm]{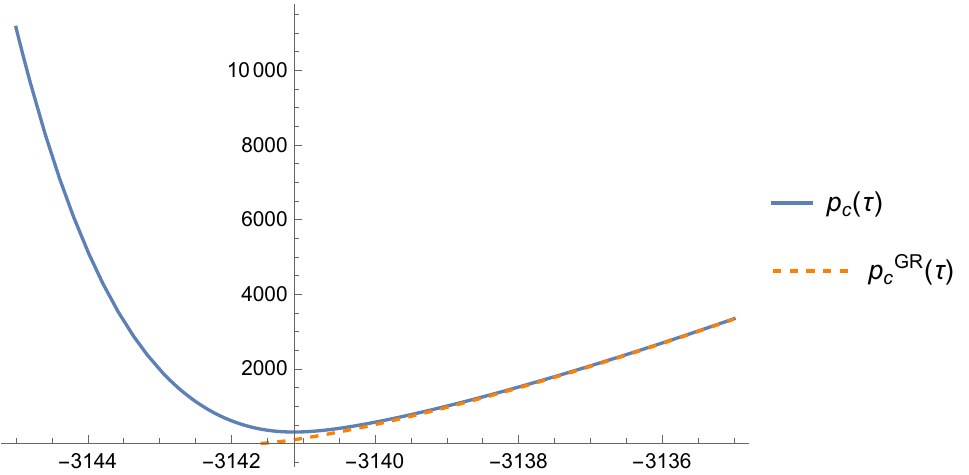}\\
 (c) & (d) \\[6pt]
\\
\includegraphics[height=4cm]{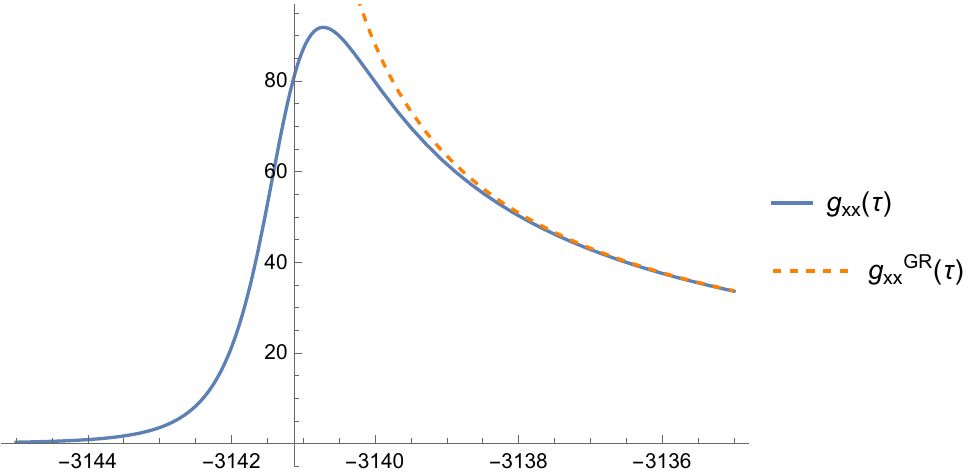}&
\includegraphics[height=4cm]{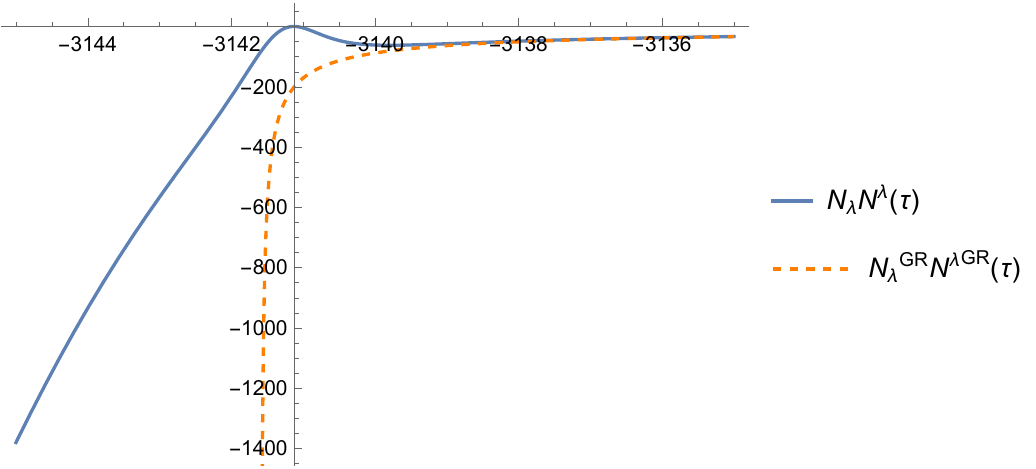}\\
(e) & (f) \\[6pt]
\\
\includegraphics[height=4cm]{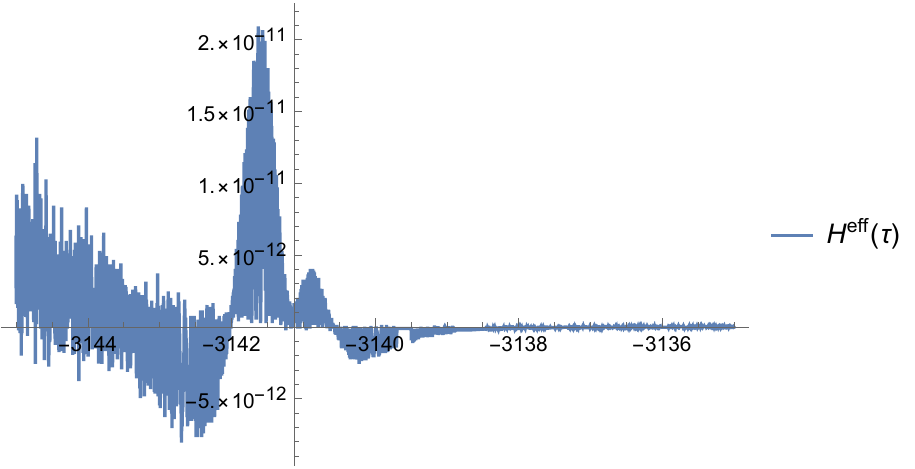}\\
(g)  \\[6pt]
\end{tabular}}
\caption{Plots of the physical quantities $b, \; c, \; p_b, \; p_c,\; g_{xx}$, \; $N^{\lambda}N_{\lambda}$ and $H^{\text{eff}}$ in terms of the proper time $\tau$  with  $m/\ell_{pl}=10^3$ in the vicinity of transition surface $\tau_{\mathcal{T}}\approx -3141.13$.}
\lb{fig3-new-N1}
\end{figure*}

\bibliographystyle{JHEP}
\bibliography{BV}

\end{document}